\documentclass[a4paper,11pt]{article}
\pdfoutput=1 
\usepackage{jcappub}
\usepackage[T1]{fontenc}
\author[a,1]{Asta~Heinesen,\note{Corresponding author.}}
\author[b]{Chris~Blake,}
\author[a]{Yong-Zhuang~Li,}
\author[a]{David~L.~Wiltshire}

\affiliation[a]{School of Physical \& Chemical Sciences, University of Canterbury,\\
Private Bag 4800, Christchurch 8140, New Zealand}
\affiliation[b]{Centre for Astrophysics \& Supercomputing, Swinburne University of Technology,\\ P.O. Box 218, Hawthorn, VIC 3122, Australia}

\emailAdd{asta.heinesen@pg.canterbury.ac.nz}
\emailAdd{cblake@swin.edu.au}
\emailAdd{yong-zhuang.li@pg.canterbury.ac.nz}
\emailAdd{david.wiltshire@canterbury.ac.nz}

\usepackage{braket}
\usepackage{dirtytalk}
\usepackage{changes}
\usepackage{subcaption}
\usepackage[section]{placeins}
\usepackage[titletoc]{appendix}
\usepackage{xcolor}
\usepackage{amssymb}
\usepackage{dsfont}
\usepackage{bm}
\let\mathbb\mathds

\providecommand{\href}[2]{#2}\def\link#1{\href{http://arxiv.org/abs/#1}{{\tt #1}}}
\def\etal{{et al}.}
\def\sgn{\mathop{\rm sgn}\nolimits}
\def\lsim{\mathop{\hbox{${\lower3.8pt\hbox{$<$}}\atop{\raise0.2pt\hbox{$\sim$}}
$}}}
\def\LCDM{$\Lambda$CDM}
\def\RR{{{}^3{\cal R}}}
\def\goesas{\mathop{\sim}\limits}
\def\frn#1#2{{\textstyle{#1\over#2}}}\def\half{\frn12}
\def\fv{f_{\rm v}}\def\fvn{f_{{\rm v}0}}
\def\OMn{{\Omega_{M0}}} \def\OMfid{{\Omega_{M0,{\rm fid}}}}
\def\OLn{{\Omega_{\Lambda{0}}}} \def\Obn{{{\Omega_{b0}}}}
\def\hm{\;\hbox{Mpc}/h} 
\def\transpose{{\raise2pt\hbox{$\scriptstyle\intercal$}}}
\def\fid{{\rm fid}}

\definecolor{MyB}{rgb}{0.1,0.1,1.0}

\definecolor{MyDarkRed}{rgb}{0.71,0.24,0.57}

\title{Baryon acoustic oscillation methods for generic curvature: Application to the SDSS-III Baryon Oscillation Spectroscopic Survey}

\arxivnumber{1811.11963}
\abstract{We develop
methods for investigating baryon acoustic oscillation (BAO) features
in cosmological models with non-trivial (but slowly varying) averaged
spatial curvature: models that are not necessarily flat, close to flat, nor
with constant spatial curvature. The class of models to which our
methods apply include Lema\^{\i}tre-Tolman-Bondi models, modified
gravity cosmologies, and inhomogeneous cosmologies with backreaction
-- in which we do not have a prediction of the shape of the spatial
2-point correlation function, but where we nevertheless expect to
see a BAO feature in the present-day galaxy distribution, in form of
an excess in the galaxy 2-point correlation function.

We apply our methods to the Baryon Oscillation Spectroscopic Survey
(BOSS) dataset, investigating both the Lambda Cold Dark Matter (\LCDM)
and timescape cosmological models as case studies. The correlation functions
measured in the two fiducial models contain a similarly-pronounced BAO
feature. We use the relative tangential and radial BAO scales to
measure the anisotropic Alcock-Paczy\'nski distortion parameter,
$\epsilon$, which is independent of the underlying BAO preferred
scale. We find that $\epsilon$ is consistent with zero in both
fiducial cosmologies, indicating that models with a different spatial
curvature evolution can account for the relative positions of the
tangential and radial BAO scale. We validate our methods using
\LCDM\ mocks.
}
\catcode`\@=11
\gdef\@fpheader{Published:\ \href{https://doi.org/10.1088/1475-7516/2019/03/003}{JCAP 03 (2019) 003}\hfill \href{https://doi.org/10.1088/1475-7516/2019/03/003}{doi: 10.1088/1475-7516/2019/03/003}}
\catcode`\@=12
\begin{document}
\maketitle

\def\PRL#1{{\em Phys.\ Rev.\ Lett.}\ {\bf#1}}
\def\JCAP#1{{\em J.\ Cosmol.\ Astropart.\ Phys.}\ {\bf#1}}
\def\ApJ#1{{\em Astrophys.\ J.}\ {\bf#1}}
\def\PR#1{{\em Phys.\ Rev.}\ {\bf#1}}
\def\MNRAS#1{{\em Mon.\ Not.\ R.\ Astr.\ Soc.}\ {\bf#1}}
\def\CQG#1{{\em Class.\ Quantum Grav.}\ {\bf#1}}
\def\GRG#1{{\em Gen.\ Relativ.\ Grav.}\ {\bf#1}}
\def\IJMP#1{{\em Int.\ J.\ Mod.\ Phys.}\ {\bf#1}}
\def\AaA#1{{\em Astron.\ Astrophys}.\ {\bf#1}}
\def\AJ#1{{\em Astron.\ J}.\ {\bf#1}}
\def\ApJs#1{{\em Astrophys.\ J.\ Suppl}.\ {\bf#1}}
\def\PLB#1{{\em Phys.\ Lett.}\ {\bf B #1}}
\def\APSS#1{{\em Astrophys. \ Space \ Sci.}\ {\bf#1}}
\def\NatA#1{{\em Nature Astron.}\ {\bf#1}}
\def\ARNPS#1{{\em Ann.\ Rev.\ Nucl.\ Part.\ Sci.}\ {\bf#1}}

\def\beq{\begin{equation}} \def\eeq{\end{equation}}
\def\bea{\begin{eqnarray}} \def\eea{\end{eqnarray}}
\def\e{\mathop{\rm e}\nolimits}
\def \domain{\mathcal{D}}
\def\doubleunderline#1{\underline{\underline{#1}}}

\DeclareRobustCommand{\orderof}{\ensuremath{\mathcal{O}}}

\tableofcontents
\newpage

\section{Introduction}

The study of baryon acoustic oscillation (BAO) features in the
recent-epoch matter distribution is, together with the cosmic
microwave background (CMB) and supernovae, a cornerstone of
observational cosmology. In the Lambda Cold Dark Matter (\LCDM) cosmology,
sound waves in the primordial plasma, and the subsequent decoupling of photons
from the baryons, produce a characteristic scale in the distribution of the
baryons at the drag epoch \cite{peeplesBAO,SunZeldovich},
which is predicted to be visible in the matter distribution of today.

The BAO feature, in the form of an excess in the spatial 2-point
correlation function of the matter distribution
\cite{EisensteinHu,Matsu} was first detected in the distribution of
galaxies by \cite{EisensteinDetection,Cole} and, since then, more
precisely measured by large-volume galaxy surveys such as the WiggleZ
Dark Energy Survey \cite{BlakeCov} and the Baryon Oscillation
Spectroscopic Survey \cite{Deff}. The BAO feature has also been
detected using the Lyman-$\alpha$ absorption line of hydrogen as a
tracer of the matter distribution \cite{Busca,Delubac}. The
visibility of a characteristic scale in the 2-point matter
distribution, at around the expected acoustic scale from CMB
constraints \cite{planckParams}, is a success of the \LCDM\
cosmology as a self-consistent framework for the interpretation of
cosmological observations.

However, as successful as the \LCDM\ cosmological model might be
for describing available data, it has foundational mysteries -- physically
unexplained dark components must account for 95\% of the energy content of the
universe -- and observational tensions between different probes
\cite{WendyFreedman,clustercounts,lymanalpha,primordialLithium,kids},
that motivate a continued exploration of alternative models. The
statistical and systematic errors in current data, and the
observational degeneracy of different physical phenomena,
makes it difficult to discriminate between \LCDM\ and
alternative cosmologies. With next-generation surveys by facilities
such as the Large Synoptic Survey Telescope (LSST), Gaia, Euclid, the
Dark Energy Spectroscopic Instrument (DESI) and the Square Kilometre
Array (SKA), we will enter a new level of precision in data, that must
be matched by theoretical precision in order to improve our
understanding of the universe.

Most current analyses interpret cosmological data within the (spatially
flat) Friedman-Lema\^{\i}tre-Robertson-Walker (FLRW) class of models. However,
general relativity allows for a much richer curvature behaviour.
The timescape cosmology \cite{clocks,sol} and tardis cosmology
\cite{tardis} are phenomenological examples
of models with non-FLRW spatial curvature evolution, as arises in schemes
with backreaction from small-scale inhomogeneities \cite{buchert00,BRreview}.
The timescape model has significantly different predictions
to the spatially flat \LCDM\ model for several observables,
such as cosmographic relationships and redshift drift
\cite{obstimescape}. Improved data precision will allow us to
discriminate between cosmological models with different spatial curvature
evolution. For example, projections
for the Euclid satellite show that the \LCDM, timescape and
tardis models should be observationally distinguishable: see \cite{smn14},
Fig.~10. Realizing such goals, however, also requires that data is
reduced in as model--independent manner as possible when performing
any tests of galaxy clustering, such as BAO extraction.

BAO analysis is usually performed assuming a fiducial spatially flat
\LCDM\ cosmology to transform data into a \say{comoving grid},
from which the galaxy 2-point correlation function can be estimated
and the BAO scale extracted by fitting a fiducial \LCDM\ power
spectrum \cite{wedgefit,Standardresults}. Additional fiducial
cosmology analysis steps, such as \LCDM\ density-field
reconstruction \cite{reconstruction}, are also often applied. {\em A
priori}, results based on fiducial data-reduction procedures are not
valid beyond the given fiducial model, and any extension of such
results must be carefully examined for the particular class of models
of interest. The extent to which the fiducial \LCDM\ results
can be applied when considering models with non-trivial spatial curvature
is not clear, as the regime of application is usually investigated for FLRW
models close to the original fiducial cosmology.

In this paper we develop methods for using generic metrics to
transform galaxy data into a correlation function. Furthermore, we
propose and test an empirical fitting procedure with no model
assumptions to extract a characteristic scale in the 2-point
correlation function. Our fitting procedure can be applied to
a large class of cosmological models.
We focus on probing a statistical volume-averaged BAO
feature. This does not mean that local environmental effects in the
BAO feature are unexpected (see,
e.g., \cite{environmentBlake,environmentRoukema,environmentNeyrinck}),
but in this paper we probe the volume-averaged BAO
scale for which local effects are marginalised.

We apply our new methods to the CMASS and LOWZ galaxy surveys of the
Baryon Oscillation Spectroscopic Survey 12th Data Release (BOSS DR12).
Testing our empirical procedure on \LCDM\ BOSS mocks,
we recover the BAO scale as the characteristic scale in our
empirical fitting function. Our fits to the data using a \LCDM\
fiducial cosmology also agree with the results of previous fiducial
\LCDM\ analyses \cite{wedgefit,Standardresults}. We then
demonstrate our new methods by self-consistently re-analysing the BOSS
dataset assuming the timescape cosmological model.

We summarize the structure of our paper as follows. In section
\ref{comoving} we extend the notion of FLRW comoving distances to
geodesic distances on preferred spatial hypersurfaces
in generic globally hyperbolic space-times in order to calculate the spatial
2-point correlation function for generic models. A restriction to
spherical symmetry is then made in order
to be able to split small spatial distances into angular and
transverse parts, and to associate the redshift with a radial
coordinate. The class of models we investigate is detailed
in section \ref{models}, and in section \ref{APscaling} we define an
Alcock-Paczy\'nski scaling equivalent to that used in standard BAO
analyses for FLRW models (see e.g., \cite{APscale}). This allows us to
parameterise the model cosmology in terms of an underlying \say{true}
spherically-symmetric metric. The accuracy of the
Alcock-Paczy\'nski scaling depends on the models tested and the size of
the survey domain. In section \ref{GRM} we present the DR12 CMASS and
LOWZ galaxy surveys, random catalogues and simulated mocks used in
this analysis. In section \ref{empiricalBAO} we propose an empirical
fitting function for BAO analysis, and in section \ref{LCDMmocktest}
we use the \LCDM\ mocks to test that we recover the BAO scale
for a \LCDM\ fiducial cosmology. In section \ref{dataanalysis} we
analyse the BOSS DR12 LOWZ and CMASS surveys in both the
timescape and \LCDM\ cosmologies. We discuss our results and
possible extensions in section \ref{discussion}.

\section{Theory}
\label{theory}

\subsection{Generalising the comoving distance definition to non-FLRW space-times}
\label{comoving}

In BAO analysis we consider the spatial 2-point correlation function,
which describes the excess probability of two galaxies being a certain
spatial distance apart as compared to a Poisson point process. We are
thus concerned with the \emph{spatial} separation of galaxies, even
though we are observing galaxies from a wide range of \say{cosmic
times} when creating our galaxy catalogues.

However, if we know the (statistical) extension of the galaxy
world-lines from the cosmic time of observation, we can map the galaxy
distribution on our null cone to a spatial hypersurface of reference.
In FLRW cosmology this is done by tracking the galaxies through their
comoving coordinates. One can then define spatial comoving distances
between the galaxies at the present epoch, and recover the distances
at any other reference hypersurface via multiplication by the
homogeneous scale factor. For general globally hyperbolic space-times
we can also track the galaxy distribution in comoving coordinates to a
reference hypersurface, on which we can compute the shortest spatial
distances between galaxy pairs that are analogous to FLRW comoving separations.

We consider a globally hyperbolic space-time, and assume that the
vorticity of the matter distribution in this space-time can be
ignored\footnote{This assumption is made in order to define reference
hypersurfaces orthogonal to the fluid frame. However, nothing
prevents us from mapping the galaxy distribution to generic spatial
hypersurfaces of the given space-time, allowing for a generalisation
of the present procedure to the case of vorticity in the matter
distribution.}, and that caustics in the matter distribution can be
ignored at the coarse-graining level and over the timescale
considered. The metric can then in general be written in
Gaussian normal coordinates, $x^{\mu} = (t,x^{i})$,
\begin{align} \label{eq:metriclagrangian}
&ds^2 = - \alpha^2 c^2\, dt^2 + g_{ij}\, dx^{i}\, dx^{j}
\end{align}
where $x^{i}$ are comoving coordinates labelling the fluid elements of
the matter distribution,
$t$ labels the hypersurfaces
normal to the fluid flow,\footnote{
For simplicity we consider
model universes where all relevant matter is in the same
rest frame. This is never entirely true. The present procedure can
easily be generalised to handle multicomponent fluids by simply
choosing one of the fluids as a reference fluid for constructing
hypersurfaces of reference.} $g_{ij}$ is the metric adapted
to the hypersurfaces defined by $t =$ constant, and
$\alpha\, dt$ is the proper time measure
on the particle worldlines.

Consider two particles (galaxies) at space-time events $P_1$ and $P_2$
with coordinates $x_1^{\mu}(P_1) = (t_1,x_1^{i})$ and $x_2^{\mu}(P_2)
= (t_2,x_2^{i})$. We would like to define the shortest spatial
distance between the two particles on a reference hypersurface $t =
T$. Since the particles are by construction moving on lines of
constant comoving coordinates, we can extend the particles to the
reference hypersurface $t = T$. We keep the comoving coordinates
$x_1^{i}$ and $x_2^{i}$ fixed, and consider the new space-time events
$P_{1,T}$ and $P_{2,T}$ with coordinates $x_{1,T}^{\mu} (P_{1,T}) =
(T,x_1^{i})$ and $x_{2,T}^{\mu}(P_{2,T}) = (T,x_2^{i})$. From the
metric eq.~(\ref{eq:metriclagrangian}) we
can compute the shortest spatial distance between $P_{1,T}$ and
$P_{2,T}$ on the surface $t=T$ from the geodesic equation of the
adapted metric \begin{align}
\label{eq:metriclagrangianspatial}
&ds_T^2 = g_{ij}(t=T,x^{k}) dx^{i} dx^{j} .
\end{align}
We denote the resulting shortest distance, $D_{T}(P_1, P_2)$, the
Lagrangian distance between $P_1$ and $P_2$ at the reference surface
$t=T$. This Lagrangian distance definition reduces to the comoving
distance definition in FLRW cosmology, when the matter frame coincides
with the surfaces of homogeneity and isotropy.

\subsection{Models under investigation}
\label{models}

In this section we outline the assumptions regarding the class of
cosmological models for which the procedures outlined in sections
\ref{APscaling} and \ref{empiricalBAO} apply. The motivation for
restricting the class of models is to be able to parameterise
different cosmological models in terms of each other through an
Alcock-Paczy\'nski scaling, as outlined in section \ref{APscaling} (see
e.g., \cite{APscale}).
We note that the results of the data analysis
in the present paper can be applied \emph{only} to the class of models
discussed here.\footnote{The
standard BAO results such as
\cite{Standardresults, wedgefit} are also limited by the regime of
applicability of the AP-scaling.}

As in section \ref{comoving}, we consider globally hyperbolic average
space-times, in which vorticity and caustics of the matter
distribution can be neglected. We can write the metric in such a
space-time as in eq.~(\ref{eq:metriclagrangian}). We are interested
in using this metric to describe the distances between
galaxies within a given survey in a statistical sense. Thus, we need to
write the metric in terms of coordinates $(z,\theta,\phi)$ of the average model
to which the observed redshifts, and angular positions of galaxies
are mapped.

Suppose that we have a set of comoving coordinates
$(r,\theta,\phi)$, where $\theta$ and $\phi$ are mapped to the
observed angles, and where $r$ is a radial coordinate. For simplicity
we shall assume spherical symmetry in $(\theta, \phi)$ such that the
adapted metric eq.~(\ref{eq:metriclagrangian}) can be written
\begin{align} \label{eq:metriclagrangian2}
&ds^2 = - \alpha(t,r)^2 c^2 dt^2 + g_{rr}(t,r)\, dr^2 + g_{\theta \theta}(t,r) \left(d\theta^2 + \cos^2(\theta) d \phi^2 \right)
\end{align}
where $\cos^2(\theta)$ comes from the convention in the definition of
the declination angle. The redshift $z$ of radially propagating null
rays, $\alpha(t,r)^2 dt^2 = g_{rr}(t,r) dr^2$, can therefore be
considered as a function of either $t$ or $r$ (since $t$ and $r$ are
monotonic functions of each other on radial null lines).
Note that since the metric (\ref{eq:metriclagrangian2}) only applies to average light
propagation over large cosmic distances, $z$ is an {\em average} model
parameter.
Although $z$ is not directly observable, it is assumed to be a good approximation for the \emph{mean} observed redshift.
We consider universes that are overall expanding, and neglect the small scale collapse of structures that can cause the redshift to be multivalued along the null rays\footnote{See section 3 of \cite{lightpropndust} for relevant calculations of mean redshift in statistically homogeneous and isotropic space-times, and section 3.2 in particular for a discussion of
multivaluedness of redshift along light cones in relation to
statistical homogeneity and isotropy.}.
In such model-universes it is reasonable to assume that $z$ is a strict monotonic function in $t$ (and therefore
also in $r$)\footnote{The monotonicity assumption is independent of the exact parameterisation, $t$, of the fluid-adapted foliation.
Since $t$ labels surfaces normal to the averaged fluid flow, we have $\bm u \propto \bm \nabla t$, where $\bm u$ is the averaged fluid 4-velocity,
and $t$ is unique up to transformations $t \rightarrow f(t)$ by a monotonic
function $f$. Any function $z$ that is monotonic in $t$ will be monotonic in
$f(t)$.}.
In this case, we can treat $z$ as a radial
coordinate on the spatial sections $t = T$ and write the adapted
metric
(\ref{eq:metriclagrangianspatial}) as\footnote{Since the redshift, $z$,
is only defined along the radial null geodesics it is important to realise
that (\ref{eq:metriclagrangiansphericalsymmetry}), (\ref{eq:gzz}) is a {\em
projection} from the null cones onto fiducial spatial hypersurfaces,
{\em not} a global coordinate transformation in the original space-time
(\ref{eq:metriclagrangian2}).}
\begin{align} \label{eq:metriclagrangiansphericalsymmetry}
&ds_T^2 = g_{zz}(t=T,r) dz^2 + g_{\theta \theta}(t=T,r) \left(d\theta^2 + \cos^2(\theta) d \phi^2 \right) ,
\end{align}
where\footnote{In any spatially flat FLRW model, with $t=T$ corresponding
to the \say{present time} hypersurface, we have $g_{rr}(t=T,r) = a(t = T)^2
= 1$ and $g_{zz}(t=T,r) = \left( \frac{dr}{dz} \right)^2 = \left(
c/H \right)^2$, where $a(t)$ is the scale factor, and we have used
the convention $a(t=T) = 1$.}
\begin{align} \label{eq:gzz}
g_{zz}(t=T,r) \equiv g_{rr}(t=T,r) \left( \frac{dr}{dz} \right)^2 .
\end{align}
The BAO scale is a statistical standard ruler, and in practice the 2-point correlation function probing the BAO scale is obtained by summing over many galaxy pairs. Thus, it is reasonable to consider models with large smoothing scale compared to galaxy pair separations of order the BAO scale $\sim 100$Mpc/h.
In particular, we only discuss
models in which the typical pair separation of galaxies surveyed is
small compared to variations of the adapted spatial metric
(\ref{eq:metriclagrangianspatial}), as detailed in appendix \ref{expgeodesic}.
In these cases we can approximate the Lagrangian distance
$D_{T}(P_1, P_2)$ for two galaxies with coordinates $(z_1,\theta_1,\phi_1)$ and
$(z_2,\theta_2,\phi_2)$ separated by redshift $\delta z = z_2 - z_1$
and angle $\delta \Theta$
\begin{align} \label{eq:angularsep}
\delta \Theta &= \arccos \left[ \sin(\theta _{1})\sin(\theta
_{2})+\cos(\theta _{1})\cos(\theta _{2})\cos(\phi _{2}-\phi _{1})
\right]\\ &\approx \sqrt{(\theta_2 - \theta_1)^2 +
\cos^2(\bar{\theta} )(\phi_2 - \phi_1)^2 } , \qquad \bar{\theta} =
(\theta_1 + \theta_2)/2 \nonumber
\end{align}
as
\begin{align} \label{eq:lagrangiandistapprox}
D_{T}^2(P_1, P_2)& \approx g_{zz}(t=T,\bar{z}) (\delta z)^2 +
g_{\theta \theta}(t=T,\bar{z}) (\delta \Theta)^2 ,
\end{align}
where $\bar{z} = (z_1 + z_2)/2$ is the intermediate redshift.

The validity of the approximation of eq.~(\ref{eq:lagrangiandistapprox}) is
cosmology-dependent\footnote{{The validity of the approximation relies on second order variations of the metric (curvature degrees of freedom) being small as compared to the metric and its first order variations in the adapted coordinate-system $(z,\theta,\phi)$ over scales of the galaxy pair separations of interest (see appendix \ref{expgeodesic}). Examples of models with significant spatial curvature for which eq.~(\ref{eq:lagrangiandistapprox}) apply to a good approximation for galaxy pair separations of order $\sim 100$Mpc/h are the empty Milne universe (FLRW with $\Omega_M=\Omega_\Lambda=0$, $\Omega_k=1$) and the timescape model, which have significant metric variations only on scales $\RR^{-1/2}\goesas c/H_0\goesas3\;\hbox{Gpc}/h$ at the present epoch, where $\RR$ is the spatial Ricci scalar of the given model.}},
and must be assessed for the particular class of model cosmologies of interest.
In appendix \ref{expgeodesic} we give the explicit expansion of the
geodesic path integral up to third order, and in appendix
\ref{sphericalsymmDist} we apply our results to spherically-symmetric
metrics. For the FLRW and timescape models with reasonable model
parameters, we find that higher-order corrections to
eq.~(\ref{eq:lagrangiandistapprox}) are of order $\lsim 10^{-3}$ for
Lagrangian galaxy separations of order $100\hm$.

It will be convenient to define
\begin{align} \label{eq:mu}
\mu_T(P_1,P_2) = \frac{ \sqrt{ g_{zz}(t=T,\bar{z}) (\delta z)^2} }{D_{T}(P_1, P_2)}
\end{align}
as the \say{radial fraction} of the separation. Note that such a
splitting into the radial and transverse components of a geodesic
distance is not meaningful for general metrics. However, when the
approximation of eq.~(\ref{eq:angularsep}) is valid, such an
Euclidean notion still applies.

Conventionally, the surface of evaluation $t = T$ is taken to be the
present day. Whenever we refer to evaluation at the present day we
shall omit the $T$ subscript on eq.~(\ref{eq:lagrangiandistapprox})
and (\ref{eq:mu}). For ease of notation the dependence on the points
of the galaxies will also be implicit, and we will just write $D$ and
$\mu$ respectively.

\subsection{Alcock-Paczy\'nski scaling}
\label{APscaling}

In the later analysis it will be convenient to parameterise the model
cosmology in terms of an unknown \say{true} cosmology. We will assume
that the universe is well-described by a \say{true} metric of the form
in section \ref{models}, and that we have a model cosmology also of
the form outlined in section \ref{models}, but not necessarily with
the same adapted metric.

We can write the model Lagrangian distance between two galaxies
eq.~(\ref{eq:lagrangiandistapprox}) at mean redshift $\bar{z}_i$ and
separation $\delta z_i$, $\delta \Theta_i$ on the sky in terms of the
\say{true} distance measures
\begin{align} \label{eq:modelDintermsoftrue}
(D_{T,i})^2 & \approx g_{zz}(t=T,\bar{z}_i) (\delta z_i)^2 + g_{\theta \theta}(t=T,\bar{z}_i) (\delta \Theta_i)^2 \\
& = \frac{1}{\alpha^2_{\parallel,i} }g^{\rm tr}_{zz}(t^{\rm tr}=T^{\rm tr},\bar{z}_i) (\delta z_i)^2 + \frac{1}{\alpha^2_{\perp,i} } g^{\rm tr}_{\theta \theta}(t^{\rm tr}=T^{\rm tr},\bar{z}_i) (\delta \Theta_i)^2 \nonumber
\end{align}
where \say{tr} stands for the \say{true} cosmology, the index $i$ labels
the galaxy pair, and
\begin{align} \label{eq:APdef}
\alpha_{\parallel,i} & \equiv \sqrt{ \frac{g^{\rm tr}_{zz}(t^{\rm tr}=T^{\rm tr},\bar{z}_i) }{g_{zz}(t=T,\bar{z}_i)} } , \qquad \alpha_{\perp,i} \equiv \sqrt{ \frac{ g^{\rm tr}_{\theta \theta}(t^{\rm tr}=T^{\rm tr},\bar{z}_i) }{ g_{\theta \theta}(t=T,\bar{z}_i) }}
\end{align}
are the Alcock-Paczy\'nski (AP) scaling parameters. Note that we are
comparing a reference hypersurface of the \say{true} cosmology
$t^{\rm tr}=T^{\rm tr}$ to the reference hypersurface $t =T$ of the model
cosmology, by associating points of the same observational
coordinates $(z,\theta,\phi)$.

Each galaxy pair will be associated with its own unique scalings of
eq.~(\ref{eq:APdef}). For sufficiently small volumes of the galaxy survey
considered, we might approximate the individual distortion parameters
by one global scaling $\alpha_{\parallel}$, $\alpha_{\perp}$ to lowest
order. This is a reasonable approximation if the survey volume has a
relatively narrow redshift distribution, and if both the \say{true}
and the model metric are slowly changing in redshift. As a rule of
thumb, the narrower the redshift distribution, and the larger the
curvature scales of the models of interest, the better the global scaling
approximation is. In the present work we use the
global AP-scaling as a rough tool for testing consistency of the
investigated fiducial cosmologies, keeping in mind the limitations of
this approximation.

We can define the \say{isotropic scaling} $\alpha$ and the
\say{anisotropic scaling} $\epsilon$
\begin{align} \label{eq:alphaepsilon}
\alpha \equiv (\alpha^2_{\perp} \alpha_{\parallel})^{1/3} , \qquad (1+\epsilon)^3 \equiv \frac{\alpha_{\parallel}}{\alpha_{\perp}} .
\end{align}
Such a decomposition will be useful in the following analysis, since in
an isotropically-sampled galaxy distribution we expect the BAO
feature to be degenerate with $\alpha$ and not $\epsilon$. (See section
\ref{empiricalBAO} for explicit expressions in the context of the
particular fitting function used in this analysis.) We note that
$\alpha$ and $\epsilon$ as defined in eq.~(\ref{eq:alphaepsilon}) are
analogous to the AP-scaling parameters outlined in,
e.g., \cite{APscale}, when associating $g_{zz}$ with the inverse Hubble
parameter multiplied by the speed of light $c/H$ and $g_{\theta
\theta}$ with the angular diameter distance $D_A$.

The isotropic scaling $\alpha$ describes how the volume measure of a
small coordinate volume $\delta z \, \cos(\theta) \, \delta \theta \,
\delta \phi$ differs to lowest order
between the \say{true} and the model cosmology,
\begin{align} \label{eq:APalphaepsilon}
\alpha & \approx \left( \frac{\delta V^{\rm tr}_{i}(t^{\rm tr}=T^{\rm tr},\bar{z}_i) }{\delta V_{i}(t=T,\bar{z}_i) } \right)^{1/3} = \left( \frac{g^{\rm tr}_{zz}(t^{\rm tr}=T^{\rm tr},\bar{z}_i) \left( g^{\rm tr}_{\theta \theta}(t^{\rm tr}=T^{\rm tr},\bar{z}_i) \right)^2 }{ g_{zz}(t=T,\bar{z}_i) \left(g_{\theta \theta}(t=T,\bar{z}_i) \right)^2 } \right)^{1/6}
\end{align}
with
\begin{align} \label{eq:dVol}
\delta V_{i}(t=T,\bar{z}_i) \equiv \sqrt{ \det( g )(t=T,\bar{z}_i) }\, \delta z \,\delta \theta \, \delta \phi ,
\end{align}
where $\det( g )$ is the determinant of the spatial metric
(\ref{eq:metriclagrangiansphericalsymmetry}) in the coordinate
basis $(z,\theta, \phi)$.

It will prove convenient to parameterise $\alpha$ and $\epsilon$ of
two model cosmologies in terms of the relative transverse and radial
distance measures of the models
\begin{align} \label{eq:alphaepsilontrans}
\alpha_1 &= \alpha_2 \left( \frac{g_{2, zz}(t_2 =T_2,\bar{z}) }{g_{1, zz}(t_1=T_1,\bar{z})} \right)^{1/6} \left( \frac{g_{2, \theta \theta}(t_2 =T_2,\bar{z}) }{g_{1, \theta \theta}(t_1=T_1,\bar{z})} \right)^{1/3} \\
\epsilon_1 &= (1 + \epsilon_2) \left( \frac{g_{2, zz}(t_2 =T_2,\bar{z}) }{g_{1, zz}(t_1=T_1,\bar{z})} \right)^{1/6} \left( \frac{g_{2, \theta \theta}(t_2 =T_2,\bar{z}) }{g_{1, \theta \theta}(t_1=T_1,\bar{z})} \right)^{-1/6} - 1 \nonumber .
\end{align}
Knowing $\alpha$ ($\epsilon$) within a reference/fiducial cosmology,
we can calculate $\alpha$ ($\epsilon$) within a different cosmology
from the known model distance measures using the identity in
eq.~(\ref{eq:alphaepsilontrans}).

From the assumption of slowly varying $\alpha_{\perp}$ and
$\alpha_{\parallel}$ over the survey volume we can approximate
\begin{align} \label{eq:DTapprox}
(D_{T,i})^2 & \approx \frac{1}{\alpha^2_{\parallel} }g^{\rm tr}_{zz}(t^{\rm tr}=T^{\rm tr},\bar{z}_i) (\delta z_i)^2 + \frac{1}{\alpha^2_{\perp} } g^{\rm tr}_{\theta \theta}(t^{\rm tr}=T^{\rm tr},\bar{z}_i) (\delta \Theta_i)^2 ,
\end{align}
which we can invert to $D^{\rm tr}_{T^{\rm tr}}$ approximated in terms of
$D_T$, $\mu_T$, and the global Alcock-Paczy\'nski scaling parameters
$\alpha_{\parallel}$, $\alpha_{\perp}$.
\begin{align} \label{eq:DTapproxinverse}
(D^{\rm tr}_{T^{\rm tr},i})^2 & \approx g^{\rm tr}_{zz}(t^{\rm tr}=T^{\rm tr},\bar{z}_i) (\delta z_i)^2 + g^{\rm tr}_{\theta \theta}(t^{\rm tr}=T^{\rm tr},\bar{z}_i) (\delta \Theta_i)^2 \\
& \approx \alpha^2_{\parallel} g_{zz}({t=T},\bar{z}_i) (\delta z_i)^2 + \alpha^2_{\perp} g_{\theta \theta}({t=T},\bar{z}_i) (\delta \Theta_i)^2 \nonumber\\
& = \alpha^2_{\perp} (D_{T,i})^2 \left( 1 + \left( \frac{\alpha^2_{\parallel} }{ \alpha^2_{\perp} } - 1 \right) \mu_{T,i} \right) \nonumber \\
& = \alpha^2 (D_{T,i})^2 \left( \frac{\alpha_{\perp}}{\alpha_{\parallel}} \right)^{2/3} \left( 1 + \left( \frac{\alpha^2_{\parallel} }{ \alpha^2_{\perp} } - 1 \right) \mu^2_{T,i} \right) , \nonumber
\end{align}
where the definition of $\mu_T$ in eq.~(\ref{eq:mu}) has been used.
Similarly we have for $\mu^{\rm tr}$
\begin{align} \label{eq:muTapproxinverse}
\mu^{\rm tr}_{T^{\rm tr},i} = \frac{ \sqrt{ g^{\rm tr}_{zz}(t^{\rm tr}=T^{\rm tr},\bar{z}_i) (\delta z_i)^2} }{D^{\rm tr}_{T,i}} & \approx \frac{\alpha_{\parallel}}{\alpha_{\perp}} \mu_{T,i} \frac{1}{\sqrt{ 1 + \left( \frac{\alpha^2_{\parallel} }{ \alpha^2_{\perp} } - 1 \right) \mu^2_{T,i} }} .
\end{align}

\subsection{Overview of the timescape model}

In the present analysis we apply our methods to the spatially flat \LCDM\
and the timescape cosmologies. Both
models are part of the class described in section \ref{models}, and we
can therefore test them with the procedures outlined in this paper.

The timescape cosmology \cite{clocks,sol} is a model which invokes
non-trivial backreaction of inhomogeneous structures on $\lsim100\hm$
scales on the average expansion of the
universe. In particular, Einstein's equations are not taken to govern
a global background metric, rather matter and geometry couple
\emph{locally}, allowing for non-trivial curvature evolution. In the
timescape model the early universe is close to critical density (and
well-approximated by a spatially flat FLRW geometry) and evolves into a
void-dominated present day universe of average negative spatial curvature.
Typical observers in the timescape model have a mass-biased view of
the universe, as they are located in gravitationally-bound structures
which are overdense with respect to the average density of the
universe. For a review of the timescape model and its observable
consequences, see \cite{TSreview}.

The timescape model and spatially \LCDM\ model have the same number of
free parameters. In both cases, the main features of the late epoch universe
are determined by two independent parameters, which can be thought of as the
Hubble parameter and a matter density parameter. However, in the timescape
case, on account of inhomogeneities, not every observer is the same average
observer with identical clocks and rulers. There are ``bare'' and ``dressed''
versions of each parameter -- the bare ones referring to volume
averages of the small scale Einstein equations which best describes average
cosmic evolution \cite{buchert00}, and the
dressed ones to observers like ourselves in gravitational bound systems
and the measurements on our past light cone in terms of our rulers and clocks
\cite{clocks}. The dressed parameters are the observationally relevant
ones for model comparisons.

In the present paper we aim to demonstrate feasibility of the method, by
making just one choice of the timescape dressed matter density parameter, equal
to that of the \LCDM\ matter density parameter, $\OMn = 0.3$ at the
present epoch. However, it should be stressed that dressed parameter in
the timescape case does not enter any Friedmann-like Hamiltonian constraint
equation. Furthermore, the value chosen
is a reasonable one in the timescape case \cite{dnw,nw}, but not singled
out as a best fit in other tests \cite{sn1aTS}.

In the timescape \say{tracking limit} \cite{obstimescape} applicable to redshifts $z\lsim 10$, the adapted metric (\ref{eq:metriclagrangiansphericalsymmetry}) of the timescape model is given by an exact solution
\cite{sol}
\begin{align} \label{eq:gthetaTS}
\sqrt{g_{\theta \theta}(t=t_0, r(t))} = (1+z) c t^{2/3} \left( \mathcal{F}(t_0) - \mathcal{F}(t) \right) , \qquad \sqrt{g_{zz}(t=t_0, r(t))} = \frac{c}{H} ,
\end{align}
where the position $r(t)$ of sources on radial null rays is parameterised
in terms of the volume--average time parameter at emission, $t_0$
denotes the present epoch value of $t$,
\begin{align} \label{eq:HtTS}
H(t) = \frac{3(2t^{2} + 3bt + 2b^2)}{t(2t + 3b)^2},
\end{align}
is the dressed Hubble parameter relevant for all observational measures in
this paper,
\begin{align} \label{eq:FTS}
\mathcal{F}(t) = 2\,t^{1/3} + \frac{b^{1/3}}{6} \ln\left(\frac{(t^{1/3} + b^{1/3})^2 }{ t^{2/3} - b^{1/3} t^{1/3} + b^{2/3} }\right) + \frac{b^{1/3}}{\sqrt{3}} \tan^{-1} \left( \frac{2t^{1/3} - b^{1/3} }{\sqrt{3} b^{1/3} } \right),
\end{align}
$b \equiv 2(1-\fvn) (2+\fvn)/[9\fvn \bar{H}_0]$, $\bar{H}_0=
2(2+\fvn)H_0/(4{\fvn}^2+\fvn+4)$ is the ``bare Hubble
constant'', $H_0=H(t_0)$ the ``dressed Hubble constant'', and $\fvn=\fv(t_0)$
the present epoch value of the void volume fraction. In the timescape model
the void volume fraction,
\beq\fv(t)={3\fvn\bar{H}_0 t\over3\fvn\bar{H}_0 t+(1-\fvn)(2+\fvn)}\,,
\eeq
is a parameter which arises in
the Buchert average \cite{buchert00}. The dressed $\Omega_M$ parameter is not
a natural timescape parameter but is constructed to take
values similar to the \LCDM\ case, being given by
$\Omega_M=\half(1-\fv)(2+\fv)$
with inverse $\fv=\half[-1+\sqrt{9-8\,\Omega_M}\,]$. An important feature
of the timescape model, crucial to inferring late--epoch apparent acceleration,
is that
the statistical time parameter $t$ is {\em not} the observed expansion age. Rather, this is given by
\beq
\tau=\frn23 t+{2(1-\fvn)(2+\fvn)\over27\fvn\bar{H}_0}\ln\left(1+{9\fvn\bar{H}_0 t
\over2(1-\fvn)(2+\fvn)}\right)\, \label{tsol}
\eeq
in the tracking limit. The observed redshift then also reads
\begin{align} \label{eq:ztTS}
z + 1 = \frac{2^{4/3} t^{1/3} (t+b)}{f^{1/3}_{v0} \bar{H}_0 t (2t + 3b)^{4/3} } .
\end{align}
For a derivation and discussion of the above results, see \cite{obstimescape}.

\begin{figure}[!htb]
\centering
\includegraphics[scale=0.50]{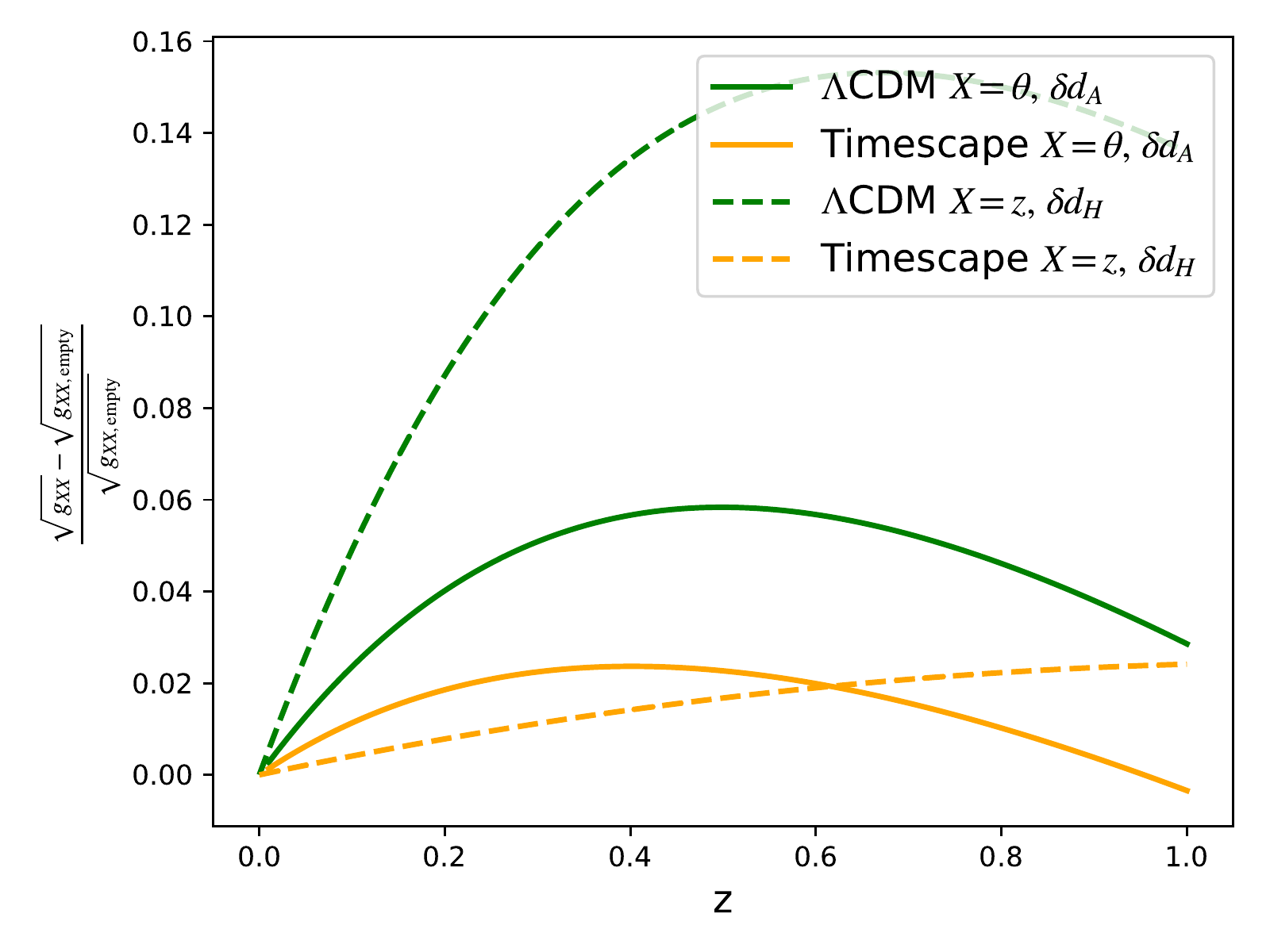}
\caption{\LCDM\ $\OMn = 0.3$ and timescape dressed
$\OMn = 0.3$ radial and transverse distance measures, relative to the
empty universe distance measures, as a function of redshift $z$. The axis
$(\sqrt{g_{XX} }- \sqrt{g_{XX , \text{empty}}})/\sqrt{g_{XX ,
\text{empty}}}$ represents the fractional difference of the angular diameter
distance and radial Hubble distances for \LCDM\ and timescape relative to
an empty universe for an observer at the present epoch, assuming the same value of the Hubble parameter for all three models. I.e., when $X=\theta$ it corresponds to $\delta d_A \equiv(d_A - d_{A,\text{empty}})/d_{A,\text{empty}}$ and when $X=z$ it corresponds to $\delta d_H \equiv(H^{-1}- H^{-1}_{\text{empty}})/H^{-1}_{\text{empty}}$.}
\label{fig:LCDMvsTimescape}
\end{figure}

Figure \ref{fig:LCDMvsTimescape} shows $\sqrt{g_{\theta \theta}}$ and
$\sqrt{g_{zz}}$ of eq.~(\ref{eq:metriclagrangiansphericalsymmetry})
for the timescape and \LCDM\ models with $\OMn = 0.3$
relative to the empty universe.
The same global Hubble parameter $H_0$ is assumed for all three models.
Since ${d_A\equiv}\sqrt{g_{\theta \theta}(t=t_0, r(t))} /(1+z)$ is the angular diameter distance, while ${d_H\equiv}\sqrt{g_{zz}(t=t_0, r(t))}/(1+z)$ represents the projected radial proper distance between two particles separated by a small distance $\delta z$ in redshift,
these quantities represent the standard
angular and radial distance measures.

The timescape model redshift--distance relation is closer to that of
the empty universe than to
\LCDM\ for redshifts $z\lsim1$\footnote{For high
redshifts the timescape model expansion history is closer to a FLRW model
containing the usual matter and radiation species. Its distance--redshift
relation effectively interpolates between those of \LCDM\ models with different
values of $\OMn$ and $\OLn$ at different redshifts
\cite{obstimescape,TSreview}.}. While the timescape model distance measures
are within $\sim 2$\% of the empty universe case,
the \LCDM\ model differs from the empty universe by up to $\sim
15$\% in the redshift range $0.15 \leq z \leq 0.7$.\footnote{These percentage estimates would in general change for distances measured in units of Mpc (rather than units of $\hbox{Mpc}/h$) for reasonable values of $H_0$ of the individual models. Typical values of $H_0$ for the timescape model are around 10\% smaller than for the \LCDM\ model.} The
low-redshift proximity of the timescape model expansion history to
that of the empty universe
reflects the late-epoch volume dominance of voids relative to
gravitationally-bound structures, which in the timescape model gives
rise to a present-day on average negatively-curved universe. Given this
comparison, BOSS large-scale structure data has the potential to
distinguish between these scenarios.

The timescape model is currently much less experimentally constrained
than the \LCDM\ model \cite{dnw,nw}, since a perturbation theory
describing structure formation within the timescape model has yet to be
developed. As a consequence CMB constraints on the BAO scale are much less
precise for timescape as compared to \LCDM. (One can fit the angular positions
of the acoustic peaks
CMB using conservative priors for the baryon-to-photon ratio, following an
equivalent procedure to that described in appendix D of \cite{sn1aTS}.) This
makes the $\epsilon$ parameter the most powerful discriminator between the
timescape model and \LCDM, in the context of the present analysis.

\subsection{The Landy-Szalay estimators}
\label{2PCFestimatorsmain}

The 2-point correlation function in cosmology (see for example
\cite{PeeblesTheory}) describes the excess in correlation between
structure in a spatial section of the universe, relative to the case
in which matter is distributed according to an uncorrelated Poisson
process. Thus the 2-point correlation function describes
characteristic scales in the matter distribution.

The spatial 2-point correlation function is defined as
\begin{align} \label{eq:xidef1}
\xi(X,Y)= \frac{ f(X,Y)}{f(X)f(Y)} - 1
\end{align}
where $f(X,Y)$ is the ensemble probability density of finding two
galaxies at points $X$ and $Y$, and $f(X)$ is the uncorrelated
probability density of finding a galaxy at point $X$. By assuming that the
galaxy distribution is well-described by a homogeneous and isotropic
point process, eq.~(\ref{eq:xidef1}) reduces to
\begin{align} \label{eq:xidef2}
\xi(D)= \frac{f(D) }{ f_{\mathrm{Poisson}}(D) } - 1 ,
\end{align}
where $D$ is the Lagrangian distance of the \say{true} underlying
metric between the points $X,Y$ defined in section \ref{comoving},
$f(D)$ represents the probability density of finding two objects with the
mutual distance $D$, and $f_{\text{Poisson}}(D)$ represents the
analogous probability density in the uncorrelated case. Note that we can
define a correlation function with a similar form to
eq.~(\ref{eq:xidef2}) for an inhomogeneous and anisotropic point
process by marginalising over the position and direction degrees of
freedom in $f(X,Y)$ (see appendix \ref{2PCF}). For a given
spherically-symmetric metric, where in addition to the
Lagrangian distance $D$ we can define the radial fraction of the
separation $\mu$ (see section \ref{models}), it will be convenient to
define the correlation function analogous to eq.~(\ref{eq:xidef2}),
\begin{align} \label{eq:ximudef1}
\xi(D,\mu)= \frac{f(D,\mu)}{f_{\text{Poisson}}(D,\mu) } - 1 ,
\end{align}
parameterised by $\mu$ and $D$. (See appendix \ref{2PCF} for details.)

Various estimators of the 2-point correlation function have been tested within
\LCDM\ \cite{CFestimatorsconvergence}. An efficient estimator is
found to be the Landy-Szalay (LS) estimator \cite{LSestimator}
\begin{align} \label{eq:xiestimatorLS1}
\hat{\xi}_{LS}(D, \mu) = \frac{DD(D, \mu) + RR(D, \mu) - 2DR(D, \mu)}{RR(D, \mu)} ,
\end{align}
where $DD$ is the binned normalised number count
\begin{align} \label{eq:DD1}
DD(D, \mu) = \frac{1}{N_D (N_D - 1)} \sum^{N_D}_{a, b} \mathbb{1}_{D \pm \Delta D } (D(x^i_a, x^i_b) ) \mathbb{1}_{\mu \pm \Delta \mu } (\mu(x^i_a, x^i_b) )
\end{align}
over galaxies in the survey, where $N_D$ is the total number of galaxies, and $\Delta D$ and $\Delta \mu$ are the binning size, and $\mathbb{1}_{A}(y)$ is the indicator function, having the value $1$ for $y \in A$ and 0 for $y \notin A$.
$RR$ is defined in the same way
\begin{align} \label{eq:RR1}
RR(D, \mu) = \frac{1}{N_R (N_R - 1)} \sum^{N_R}_{a, b} \mathbb{1}_{D \pm \Delta D } (D(x^i_a, x^i_b) ) \mathbb{1}_{\mu \pm \Delta \mu } (\mu(x^i_a, x^i_b) ) ,
\end{align}
except that the sum is now over $N_R$ artificial galaxies in a random
Poisson catalogue, designed to match the galaxy density of the galaxy
survey. We also define $DR$, the normalised cross pair-count between the galaxy
catalogue and the random sample, by
\begin{align} \label{eq:DR1}
DR(D, \mu) = \frac{1}{N_D N_R} \sum^{N_D}_{a} \sum^{N_R}_{b} \mathbb{1}_{D \pm \Delta D } (D(x^i_a, x^i_b) ) \mathbb{1}_{\mu \pm \Delta \mu } (\mu(x^i_a, x^i_b) )
\end{align}

We will use the LS estimator (\ref{eq:xiestimatorLS1}) to
estimate the underlying 2-point correlation function in this paper.
It will be convenient to average this estimator in $\mu$ to obtain
the wedge LS estimator,
\begin{align} \label{eq:xiestimatorLSwedge1}
\hat{\xi}_{LS[\mu_{1}, \mu_{2}]} (D) = \frac{1}{\mu_{2}- \mu_{1}} \int_{\mu_{1}}^{\mu_{2}}\, d \mu\, \hat{\xi}_{LS}(D, \mu) .
\end{align}
We define the isotropic wedge $\hat{\xi} (D)$, the transverse wedge $\hat{\xi}_{\perp} (D)$ and radial wedge $\hat{\xi}_{\parallel} (D)$ estimator as respectively
\begin{align} \label{eq:xiestimatorLSwedgeTR1}
\hat{\xi} (D) \equiv \hat{\xi}_{LS [0, 1]} (D) , \qquad \hat{\xi}_{\perp} (D) \equiv \hat{\xi}_{LS [0, 0.5]} (D) , \qquad \hat{\xi}_{\parallel} (D) \equiv \hat{\xi}_{LS [0.5, 1]} (D)
\end{align}
where we have dropped the subscript LS.

\section{Galaxy surveys, random catalogues, and mocks}
\label{GRM}

In this section we describe the datasets (observed and simulated) used
in this analysis. Since the 2-point correlation function is defined as
an excess probability of the correlation of galaxies compared to an
unclustered Poisson distribution, we also use a random catalogue to
construct the Landy-Szalay estimators
(\ref{eq:xiestimatorLSwedgeTR1}). We use mock catalogues to test
our analysis methods in a fiducial \LCDM\ cosmology, and to
estimate the covariance of our measurements.

\subsection{The galaxy surveys}
\label{galaxysurvey}

The Sloan Digital Sky Survey (SDSS) III \cite{SDSS} is a large
spectroscopic redshift survey performed at the Apache Point
Observatory in New Mexico. SDSS contains the Baryon Oscillation
Spectroscopic Survey (BOSS) \cite{BOSS} of Luminous Red Galaxies
(LRGs), which constitutes the current largest-volume map of large-scale
structure, spanning the approximate redshift range $0.1 \leq z \leq
0.7$ across $10{,}000$ deg$^2$ of sky. Different colour and magnitude
cuts are used to select homogeneous galaxy types across redshift
ranges $0.15 \leq z \leq 0.43$ (the LOWZ sample) and $0.43 \leq z \leq
0.7$ (the CMASS sample). The samples are split into disconnected
sub-surveys containing the galaxies from the North Galactic Cap (NGC)
and South Galactic Cap (SGC).

\begin{figure}[!htb]
\centering
\includegraphics[scale = 0.6]{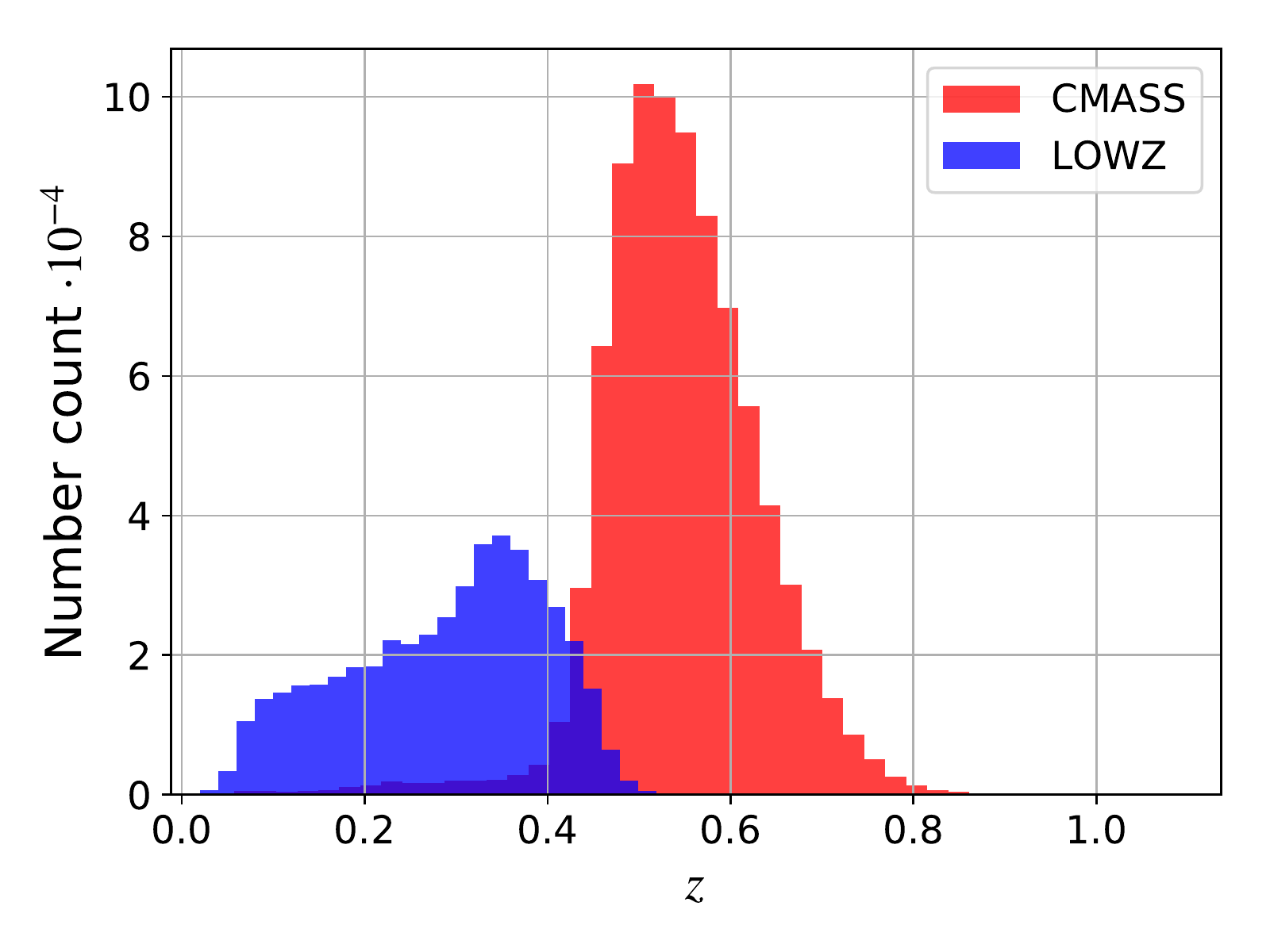}
\caption{Redshift distribution of CMASS (red) and LOWZ (blue).}
\label{fig:zdist}
\end{figure}

We use the BOSS Data Release 12 (DR12) \cite{DR12} in this analysis.
Each of the galaxies is labelled by observed coordinates
$(z,\theta,\phi)$, where $z$ is the observed redshift, $\theta$ is the
angle of declination and $\phi$ is the angle of right ascension. The
redshift distribution of the surveys is shown in figure
\ref{fig:zdist}. The total number of galaxies contained in our
selected redshift intervals is 361,762 for LOWZ and 777,202 for CMASS.

We do \emph{not} use a reconstruction procedure of peculiar motions of
galaxies such as the one described in \cite{reconstruction}. Such a
procedure reconstructs the displacements of galaxies relative to a
\LCDM\ background based on the density field of the survey,
using the relation between the linear density field and velocity
fields in \LCDM\ perturbation theory. Such a perturbation theory has not yet been developed for the
timescape cosmology, so we do not apply it in our
analysis.

In computing the spatial 2-point correlation function, we make use of
the cosmology-independent \say{total galaxy weights} (or completeness
weights) described by \cite{randompoisson}. These weights are designed
to account for observational biases, in order to make the observed
galaxy distribution an unbiased estimate of the underlying galaxy
distribution. For example, neighbours to galaxies for which redshift
determination failed are up-weighted in order to compensate for the
missing galaxy in the sample. We do not use Feldman, Kaiser \&
Peacock (FKP) weights (see \cite{FKP,randompoisson}), since they are
derived in the context of a fiducial cosmological model. However, the
application of FKP weights does not significantly affect acoustic peak
measurements in BOSS.

\subsection{The random catalogues}

We use random catalogues generated from the CMASS and LOWZ galaxy
distributions as described by \cite{randompoisson}. The random
catalogues are generated independently of a cosmological model and are
based solely on the distribution of the galaxies in observed
coordinates $(z,\theta,\phi)$. The random catalogue uniformly samples
the angular coverage of the data, and random redshifts are assigned
from the redshift probability distribution of the survey. We use a
random catalogue 10 times the size of the given galaxy catalogue or
mock.

\subsection{The mocks}
\label{mocks}

The errors in the correlation function used in BAO analysis can be
estimated in the context of a fiducial \LCDM\ cosmology using
theory or simulations. Alternatively, non-parametric methods such as
jack-knife estimation can be applied.

The assumption of a fiducial cosmology in error analysis is not
satisfying from the point of view of investigating a broader class of
models than the fiducial cosmology. However, in practice
non-parametric methods are hard to implement, as the assumptions
underlying them cannot be satisfied for current galaxy surveys. To
apply jack-knife variance estimation we must be able to divide our
sample into a (large) number of subsamples that are well approximated
as resulting from identical and independent probability distributions,
i.e., we must be able to view the regions as realisations of an
ensemble. Furthermore, jack-knife regions must be sufficiently large
to contain enough galaxy pairs separated by the relevant scales, which
conflicts with the requirement that the number of jack-knife regions
must be sufficiently large to allow an accurate inverse covariance
matrix to be constructed.

We instead use the Quick Particle Mesh (QPM) mocks as described in
detail in \cite{QPM} for error analysis. These mocks are based on
\LCDM\ $N$-body simulations, and are generated specifically for
the BOSS clustering analysis. The number density in the mock
catalogues is designed to match the observed galaxy number density of
the BOSS catalogues, and to follow the radial and angular selection
functions of BOSS. The fiducial \LCDM\ cosmology of the QPM
simulations is
\begin{align} \label{eq:qpmparams}
\OMn =0.29, \quad \OLn =0.71, \quad \Obn =0.048, \quad \sigma_8 =0.8, \quad h=0.7 ,
\end{align}
where $\OMn$, $\OLn$ and $\Obn$ are the present
epoch matter density parameter, dark energy density parameter, and baryonic
matter density parameter respectively, $\sigma_8$ is the
root mean square of the linear mass fluctuations at the present epoch
averaged at scales $8\hm$ given by the integral over the
\LCDM\ power spectrum, and $H_0 = 100\, h$ km/s/Mpc is the
Hubble parameter evaluated at the present epoch. The sound horizon at
the drag epoch within this model is $r_s = 103.05\hm$.

There are 1000 QPM mocks available. We use all of these to construct
an approximate covariance matrix of the measured galaxy correlation
function. Furthermore, we use these mocks to test how well our
empirical procedure can recover the input acoustic scale and the
anisotropic distortion in the fiducial \LCDM\ cosmology with
parameters (\ref{eq:qpmparams}).

\section{Empirical model for the correlation function, and extraction of the BAO characteristic scale}
\label{empiricalBAOmain}

Conventional \LCDM\ BAO fitting procedures \cite{Standardresults,wedgefit}
involve the construction of a template power spectrum model motivated
by \LCDM\ perturbation theory. We cannot necessarily apply
these techniques in more general cosmological models. In this section
we therefore develop an empirical approach for fitting the baryon
acoustic oscillation feature in models with non-trivial curvature,
where we do not have a model for the shape of the correlation
function, but where we nevertheless expect a characteristic scale in
the matter distribution to be sourced from early-universe oscillations
of the baryonic plasma.

In our analysis we will leave the Hubble constant free to vary and
extract the BAO scale in units $\hbox{Mpc}/h$, rather than fixing $h$
independently to some particular value. Our key fitted parameter, $\epsilon$, is dimensionless and independent of $H_0$.
In future analysis we aim to obtain independent constraints on both $\OMn$ and $H_0$ from
joint BAO and CMB observations; $\OMn$ is just fixed in the present paper
to develop the methodology.

\subsection{The fitting function}
\label{empiricalBAO}

The simplest model-independent form we might consider for the BAO
correlation function is the superposition of a Gaussian and a
featureless (e.g., polynomial) fitting function. Such empirical models have been considered in e.g., \cite{MotionAccousticScale,sancheztransverse,sanchezradial}. For a universe with statistical homogeneity and
isotropy, we expect the BAO characteristic scale to be statistically
independent of the direction of separation of the galaxies relative to
our position, up to observational biases such as redshift-space
distortions and non-representative sampling of the underlying galaxy
distribution. These considerations motivate the following empirical
model as a function of the Lagrangian separation $D$ and radial
fraction $\mu$:
\begin{align} \label{eq:xifit}
& \xi_{Fit}(D^{\rm tr}, \mu^{\rm tr}) = (D^{\rm tr})^2 A \exp\left[\frac{-\left(D^{\rm tr} - r_{\text{BAO}} \right)^2}{2 \sigma^2}\right] + C_0(\mu^{\rm tr}) + \frac{C_1(\mu^{\rm tr}) }{D^{\rm tr}} + \frac{C_2(\mu^{\rm tr} )}{(D^{\rm tr})^2}
\end{align}
where the superscript \say{tr} refers to the underlying \say{true}
cosmology. The polynomial terms model the underlying featureless
shape of the correlation function without the BAO feature and are
equivalent in form to those of \cite{wedgefit}. The scaled Gaussian
empirically models the BAO feature, and replaces the \LCDM\
power spectrum model of \cite{wedgefit}.

We note that the local maximum of the 2-point correlation function at
the BAO peak does not in general correspond to the BAO scale in a
particular cosmological model (for example, these two characteristic
scales differ by $\sim 2-3\%$ in \LCDM\ cosmology, a systematic
difference which is significantly larger than the statistical
measurement error in the scale). This is a significant issue for
empirical modelling, if we wish to incorporate predictions of the
underlying BAO scale.

To partially address this issue, we include a factor $(D^{\rm tr})^2$
multiplying the Gaussian term in eq.~(\ref{eq:xifit}), which changes
the position of the local maximum in order to produce a closer match
to the expected fiducial characteristic scale $r_{\text{BAO}}$ of the
\LCDM\ mock catalogues, within the current level of statistical
precision. This calibration would need to be re-assessed in the
context of other cosmological models.\footnote{Models that are not
developed with respect to perturbation theory cannot be tested
against the full information in the CMB anisotropies, and are consequently more
weakly constrained than \LCDM\ scenarios.} Furthermore, we do
not assume any calibration of $r_{\text{BAO}}$ in this study, instead
quoting results for $r_{\text{BAO}}/\alpha$, and focus our
investigation on the significance of the BAO feature and the
self-consistency of the radial and transverse wedges.

We allow for $\mu^{\rm tr}$ dependence in the polynomial terms of the
fitting function (\ref{eq:xifit}) since observational biases
such as redshift-space distortions can depend on the separation of the
galaxies relative to the line of sight. We assume that the BAO
feature is independent of $\mu^{\rm tr}$, although asymmetric biases might
enter here as well. However, from our mock investigations (see
section \ref{LCDMmocktest}) we find that we successfully recover the
BAO scale and the distortion parameter $\epsilon$ with the fitting
function (\ref{eq:xifit}), justifying this form at least
for the \LCDM\ model.

We can approximate eq.~(\ref{eq:xifit}) in terms of the model
cosmology through the Alcock-Paczy\'nski scaling explained in section
\ref{APscaling}. Substituting $D^{\rm tr}$ with the approximation
(\ref{eq:DTapproxinverse}) and $\mu^{\rm tr}$ with the approximation
(\ref{eq:muTapproxinverse}), the empirical model (\ref{eq:xifit})
can be written
\begin{align} \label{eq:xifitmodel}
\xi_{Fit}(D^{\rm tr}, \mu^{\rm tr}) & \approx { \xi_{Fit}\left(\tilde{D}^{\rm tr}(D, \mu; \alpha_{\parallel}, \alpha_{\perp}), \tilde{\mu}^{\rm tr}(D, \mu; \alpha_{\parallel}, \alpha_{\perp}) \right) }  \\&= (D)^2 \alpha_{\perp}^2 \left( 1 + \psi \mu^2 \right) A \e^{-\left( D \alpha_{\perp} \sqrt{ 1 + \psi \mu^2} - r_{\text{BAO}} \right)^2\!/(2 \sigma^2)} +\, C_0(\mu) + \frac{C_1(\mu) }{D} + \frac{C_2(\mu )}{(D)^2} , \nonumber
\end{align}
with
\begin{align} \label{eq:xifitmdefs}
\psi \equiv \left( \frac{\alpha_{\parallel}}{\alpha_{\perp}} \right)^2 - 1 = (1 + \epsilon)^6 - 1,
\end{align}
and where $\tilde{D}^{\rm tr}(D, \mu; \alpha_{\parallel}, \alpha_{\perp})$ is the approximation of $D^{\rm tr}$ given by (\ref{eq:DTapproxinverse}) and $\tilde{\mu}^{\rm tr}(D, \mu; \alpha_{\parallel}, \alpha_{\perp})$ is the approximation of $\mu^{\rm tr}$ given by (\ref{eq:muTapproxinverse}).
Thus, when $\xi_{Fit}(D^{\rm tr}, \mu^{\rm tr})$ is expressed in terms of $D$ and $\mu$ through the approximation of the Alcock-Paczy\'nski scaling, it has the form of a Gaussian
in $D$ scaled by $D^2$ plus first and second order polynomial terms in $D^{-1}$.
The coefficients of the Gaussian in the basis of the model cosmology
eq.~(\ref{eq:xifitmodel}) are now dependent on $\mu$.

As discussed in section \ref{2PCFestimatorsmain}, we construct two
wedge correlation functions and the angle-averaged correlation
function, by averaging eq.~(\ref{eq:xifitmodel}) over $\mu$-ranges.
For current galaxy surveys, it is in practice not useful to consider
finer binning in $\mu$, as the noise in the 2-point correlation
function increases with decreasing bin-size, and two wedges already capture the
information on $\alpha$ and $\epsilon$.

In the regime of $\psi D/\sigma \ll 1$, we may expand the Gaussian part of the fitting function (\ref{eq:xifitmodel}) to linear order in $\psi D/\sigma$ before performing the averaging in $\mu$. This has the advantage of providing an analytic expression for the average.
Expanding the Gaussian part of
$\xi_{Fit, \mathcal{N}}$ (\ref{eq:xifitmodel}) to
linear order in $\psi D/\sigma$ we have
\begin{align} \label{eq:xifitmodel2}
\xi_{Fit, \mathcal{N}}(D, \mu) & \approx (D)^2 \alpha_{\perp}^2 \left( 1 + \psi \mu^2 \right) A \e^{-\left( D \alpha_{\perp} - r_{\text{BAO}} \right)^2 /(2 \sigma^2)\,-\, \psi \mu^2 D \alpha_{\perp} \left( D \alpha_{\perp} - r_{\text{BAO}} \right)/(2 \sigma^2) } \\
&\approx (D)^2 \alpha_{\perp}^2 A \e^{-\left( D \alpha_{\perp} - r_{\text{BAO}} \right)^2 /(2 \sigma^2) } \left( 1 - \frac{\psi \mu^2 D \alpha_{\perp} \left( D \alpha_{\perp} - r_{\text{BAO}} \right)}{2 \sigma^2} + \psi \mu^2 \right) \nonumber ,
\end{align}
and taking the average in $\mu$ over the range $[\mu_{1}, \mu_{2}]$ we have
\begin{align} \label{eq:xifitmodelintegral}
&\frac{1}{\mu_{2} - \mu_{1}} \int_{\mu_{1}}^{\mu_{2}} d \mu\, \xi_{Fit, \mathcal{N}}(D, \mu) \\
&\approx (D)^2 \alpha_{\perp}^2 A \e^{-\left( D \alpha_{\perp} - r_{\text{BAO}} \right)^2 /(2 \sigma^2) } \left[ 1 + \frac{1}{3} \psi \frac{\mu^3_{2} - \mu^3_{1}}{\mu_{2}- \mu_{1}} \left( 1 - \frac{D \alpha_{\perp} \left( D \alpha_{\perp} - r_{\text{BAO}} \right)}{2 \sigma^2} \right) \right] \nonumber \\
&\approx (D)^2 \alpha_{\perp}^2 \left( 1 + \kappa \right) A \e^{-\left[ D \alpha_{\perp} \left(1+\half\kappa\right) - r_{\text{BAO}} \right]^2 /(2 \sigma^2)} \nonumber \\
&\approx (D)^2 \widetilde{A} \e^{-\left( D - \tilde{r}_{BAO} \right)^2 /(2 \tilde{\sigma}^2) } , \nonumber
\end{align}
where
\begin{align} \label{eq:kappa1}
& \kappa \equiv \frac{1}{3} \psi \frac{\mu^3_{2} - \mu^3_{1}}{\mu_{2}- \mu_{1}},
\end{align}
we have neglected terms $\orderof(\kappa^2)$ at each step, and in the final line the distorted Gaussian parameters are defined by
\begin{align} \label{eq:gausparams}
&\tilde{r}_{BAO} \equiv \frac{1 - \frac{1}{2} \kappa}{\alpha_{\perp}} r_{\text{BAO}} , \qquad \tilde{\sigma} \equiv \frac{1 - \frac{1}{2} \kappa}{\alpha_{\perp}} \sigma , \qquad \widetilde{A} \equiv \alpha_{\perp}^2 (1 + \kappa) A .
\end{align}
The final wedge fitting function thus yields
\begin{align} \label{eq:xifitwedge}
& \xi_{Fit, [\mu_{1}, \mu_{2}]}(D ) = (D)^2 \widetilde{A}\, \e^{-\left( D - \tilde{r}_{BAO} \right)^2 /(2 \tilde{\sigma}^2) } + \bar{C}_0 + \frac{\bar{C}_1 }{D} + \frac{\bar{C}_2 }{(D)^2} ,
\end{align}
where $\bar{C}_0$, $\bar{C}_1$, and $\bar{C}_2$ are unspecified
coefficients depending on the interval $[\mu_{1}, \mu_{2}]$. In the
following, we investigate some limits of the wedge fitting function
eq.~(\ref{eq:xifitwedge}).

We emphasise that the applicability of the expansion in eq.~(\ref{eq:xifitmodel2}) and the resulting expression for the wedge fitting function (\ref{eq:xifitwedge}) must be checked for a given application. When $\psi D/\sigma \ll 1$ is not satisfied over the fitting range in $D$, one must average the full expression (\ref{eq:xifitmodel}) over $\mu$ in order to obtain the exact expression for the empirical wedge fitting function.
We use the approximation (\ref{eq:xifitwedge}) in our analysis, and confirm its validity by repeating our analysis using the exact expression. (See section \ref{anisotropicfitting} for a discussion of this point.)
\\ \\*
\textbf{\textit{The ideal wedge limit}}.
Let us consider the ideal wedge limit $\mu_{2} \rightarrow\mu_{1}$,
in which the bin width is reduced to zero. In this limit we have
\begin{align} \label{eq:kappaidealwedge}
& \kappa = \psi \mu^2_{1} = \psi \mu^2_{2} .
\end{align}
Working to linear order in the anisotropic distortion parameter, so that by
(\ref{eq:xifitmdefs}) $\alpha_{\parallel}/\alpha_{\perp}\simeq1+3\epsilon$,
the distorted Gaussian parameters (\ref{eq:gausparams}) in this
case read
\begin{align} \label{eq:gausparamsidealwedge}
&\tilde{r}_{BAO} = \frac{r_{\text{BAO}} }{\alpha^{1-\mu^2_{1}}_{\perp} \alpha^{\mu^2_{1}}_{\parallel} } , \qquad \tilde{\sigma} \equiv \frac{\sigma}{\alpha^{1-\mu^2_{1}}_{\perp} \alpha^{\mu^2_{1}}_{\parallel} } , \qquad \widetilde{A} \equiv \alpha^{2-2\mu^2_{1}}_{\perp} \alpha^{2\mu^2_{1}}_{\parallel} A ,
\end{align}
e.g., for the pure transverse wedge $(\mu^2_{1} = \mu^2_{2} = 0)$ and
pure radial wedge $(\mu^2_{1} = \mu^2_{2} = 1)$, one can check that
this expression reduces to the expected scaling by $\alpha_{\perp}$
and $\alpha_{\parallel}$ respectively. For $\mu^2_{1} = \mu^2_{2} =
\frac{1}{2}$, eq.~(\ref{eq:gausparamsidealwedge}) is symmetric in
$\alpha_{\perp}$ and $\alpha_{\parallel}$, as expected.\\ \\*
\textbf{\textit{The observational wedges}}.
In practice we need to make a crude binning in $\mu$ in order to
increase the galaxy counts for each bin. Thus in the further analysis
we shall work with two $\mu$-bins and denote $\mu_{1} = 0, \mu_{2} =
\frac{1}{2}$ the transverse wedge, and $\mu_{1} = \frac{1}{2}, \mu_{2}
= 1$ the radial wedge. For the transverse and
radial wedges we find respectively for $\kappa$
\begin{align} \label{eq:kappa2}
& \kappa_{\perp} = \frac{1}{12} \psi , \qquad \kappa_{\parallel} = \frac{7}{12} \psi ,
\end{align}
which on substitution in eq.~(\ref{eq:gausparams}), to linear order in
$\epsilon$, yield the
distorted Gaussian parameters
\begin{align} \label{eq:gausparamsexpwedgeT}
&\tilde{r}_{BAO \perp} = \frac{r_{\text{BAO}} }{\alpha^{11/12}_{\perp} \alpha^{1/12}_{\parallel} } , \qquad \tilde{\sigma}_{\perp} \equiv \frac{\sigma}{\alpha^{11/12}_{\perp} \alpha^{1/12}_{\parallel} } , \qquad \widetilde{A}_{\perp} \equiv \alpha^{11/6}_{\perp} \alpha^{1/6}_{\parallel} A ,
\end{align}
and
\begin{align} \label{eq:gausparamsexpwedgeR}
&\tilde{r}_{BAO \parallel} = \frac{r_{\text{BAO}} }{\alpha^{5/12}_{\perp} \alpha^{7/12}_{\parallel} } , \qquad \tilde{\sigma}_{\perp} \equiv \frac{\sigma}{\alpha^{5/12}_{\perp} \alpha^{7/12}_{\parallel} } , \qquad \widetilde{A}_{\perp} \equiv \alpha^{5/6}_{\perp} \alpha^{7/6}_{\parallel} A ,
\end{align}
for the transverse and radial wedges, respectively. Note that
eq.~(\ref{eq:gausparamsexpwedgeT}) and (\ref{eq:gausparamsexpwedgeR})
are not symmetric under interchange $\alpha_{\perp} \leftrightarrow
\alpha_{\parallel}$. This asymmetry between the radial and transverse
wedges comes from the fact that we have defined the wedge as an
unweighted average in $\mu$.\\ \\*
\textbf{\textit{The isotropic wedge}}.
For the isotropic wedge $(\mu_{1} = 0, \mu_{2} = 1)$ we have $\kappa =
\psi/3$ which to linear order in $\epsilon$
leads to the \say{isotropically distorted} Gaussian parameters
\begin{align} \label{eq:gausparamsisotropicwedge}
&\tilde{r}_{BAO} = \frac{r_{\text{BAO}} }{\alpha } , \qquad \tilde{\sigma} \equiv \frac{\sigma}{\alpha } , \qquad \widetilde{A} \equiv \alpha^{2} A .
\end{align}
Note that only the isotropic scaling parameter $\alpha$ enters here,
and not the anisotropic distortion parameter $\epsilon$.

\subsection{Testing on \LCDM\ mocks}
\label{LCDMmocktest}

We now apply the fitting function (\ref{eq:xifitwedge}) to
\LCDM\ mocks, to test if we recover the fiducial BAO scale and
distortion parameter. To do this we perform fits to the mean
correlation function of the QPM mocks based on the CMASS NGC and LOWZ
NGC galaxy distributions, assuming a fiducial flat \LCDM\ model
with $\OMn = 0.3$. First we perform a fit to the isotropic
correlation function $\xi(D)$ with the fitting function discussed in
section \ref{empiricalBAO}. Next we perform a joint fit to estimates
of the radial wedge $\xi_{\parallel}(D)$ and transverse wedge
$\xi_{\perp}(D)$ functions. We fit to correlation function
measurements in the range $D \in [50;150]\hm$ with a bin size of
$5\hm$.

For the likelihood function $\mathcal{L}$ of the data given the model,
we assume a Gaussian distribution with mean
$\xi_{\text{Fit}}$
\begin{align} \label{eq:likelihood}
\mathcal{L}\left(\left. \bar{\hat{\xi}} \, \right| \xi_{\text{Fit}} \right) \propto \exp(- \chi^2/2) ,
\end{align}
with
\begin{align} \label{eq:chi2}
\chi^2 = Z^\transpose \doubleunderline{C}_{\bar{\hat{\xi}}}^{-1} Z , \qquad Z = \bar{\hat{\xi}} - \xi_{\text{Fit}} , \qquad
\end{align}
where $\hat{\xi}$ is the binned estimate of the (isotropic or wedge)
2-point correlation function, and $\bar{\hat{\xi}}$ is its average
over the mocks. For the wedge analysis, the transverse and radial
estimates are combined into a single vector $\hat{\xi}$ in order to
perform a combined fit, taking into account the covariance between the
wedges. $\xi_{\text{Fit}}$ is the fitting function prescribed in
eq.~(\ref{eq:xifitwedge}). The covariance matrix of
$\bar{\hat{\xi}}$ is given by the covariance of the individual
measurements $\hat{\xi}$ scaled by the number of mocks over which we
take the mean, $N_{\text{mean}}$
\begin{align} \label{eq:covmat}
\doubleunderline{C}_{\bar{\hat{\xi}}} = \frac{1}{N_{\text{mean}}} \doubleunderline{C}_{\hat{\xi}} , \qquad \doubleunderline{C}_{\hat{\xi}} = \overline{ (\hat{\xi} - \bar{\hat{\xi}}) (\hat{\xi} - \bar{\hat{\xi}})^\transpose } ,
\end{align}
where the overbar represents the averages over the number of mocks,
$N_{\text{mocks}}$. In this analysis we have $N_{\text{mocks}} =
1000$ for both CMASS and LOWZ. $N_{\text{mean}}$ is chosen such that
$\chi^2/N_{\text{dof}} \goesas 1$ in order to not to go beyond
the regime of
applicability of the empirical fitting function ($N_{\text{mean}} = 40$ for CMASS and
$N_{\text{mean}} = 80$ for LOWZ), where $N_{\text{dof}}$ is the number
of independent degrees of freedom.
\begin{figure}[!htb]
\centering
\begin{subfigure}[b]{.47\textwidth}
\includegraphics[width=\textwidth]{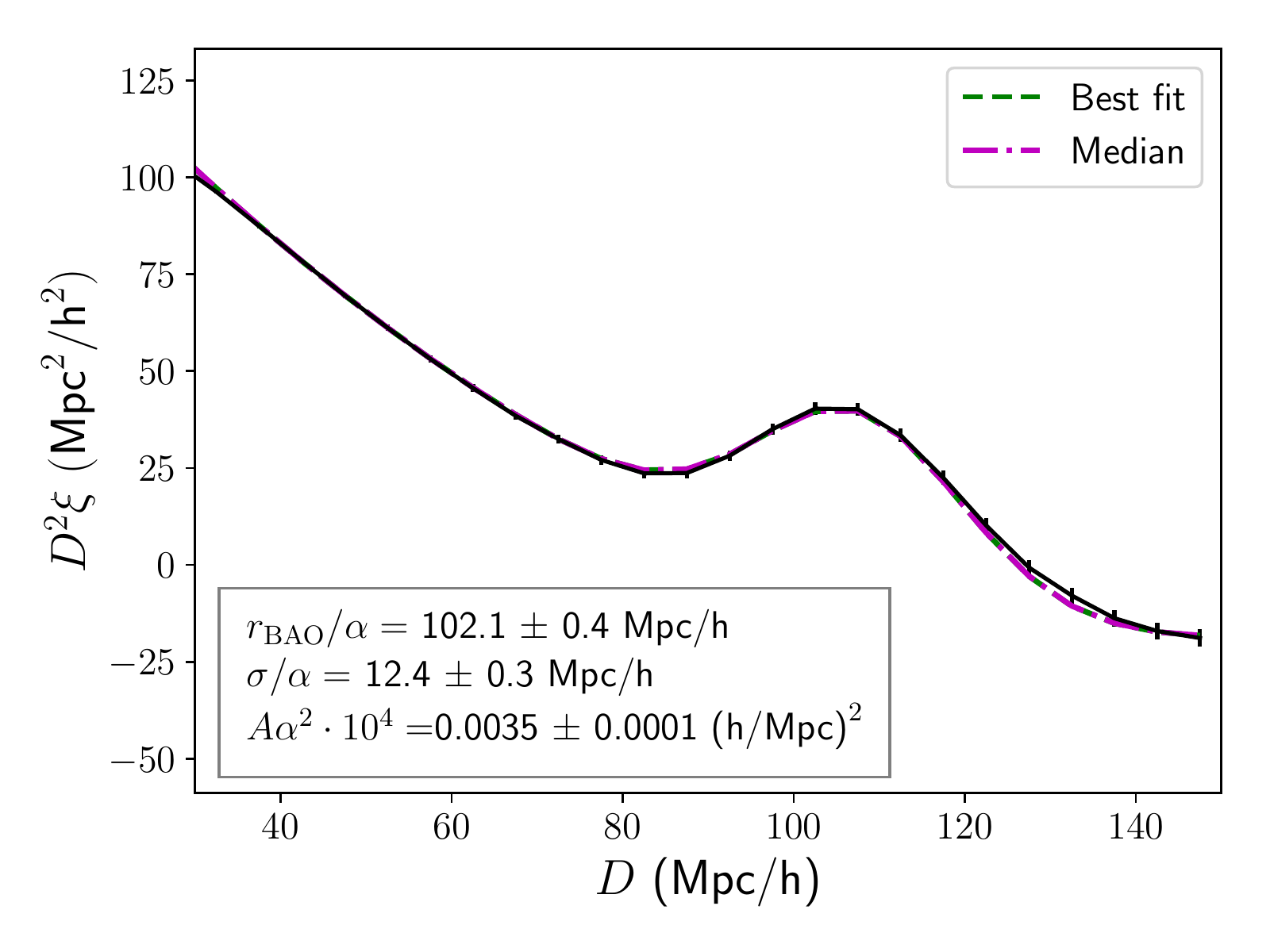}
\caption{$\xi(D_{\Lambda \text{CDM}} )$ CMASS}
\label{fig:mockIsotropicCMASS}
\end{subfigure}
\medskip
\begin{subfigure}[b]{.47\textwidth}
\includegraphics[width=\textwidth]{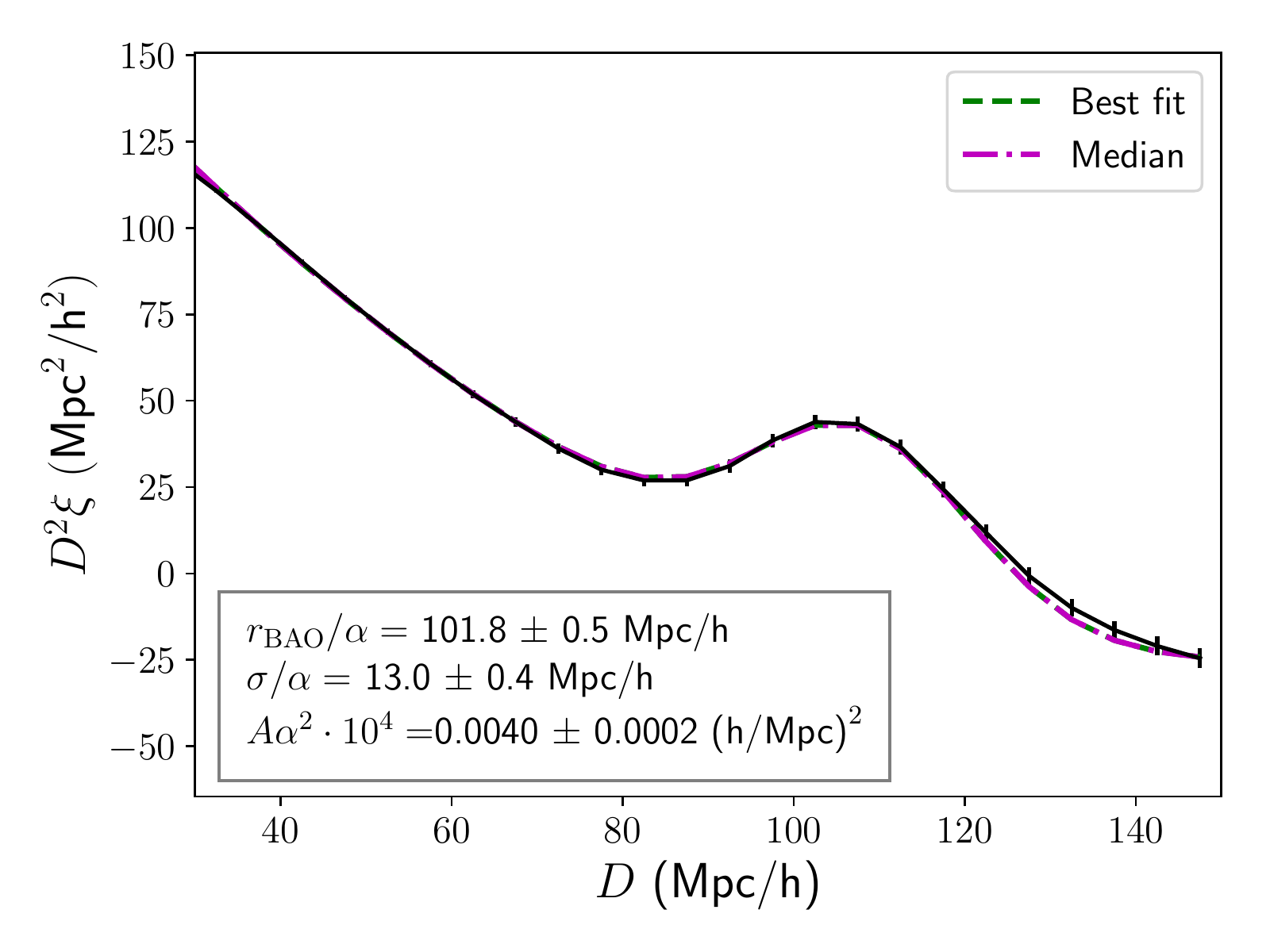}
\caption{$\xi(D_{\Lambda \text{CDM}})$ LOWZ}
\label{fig:mockIsotropicLOWZ}
\end{subfigure}
\caption{Fit to the isotropic wedge $\xi(D_{\Lambda \text{CDM}})$ of
the mean of the CMASS NGC and LOWZ NGC QPM mocks
respectively. $D_{\Lambda \text{CDM}}$ is the Lagrangian distance
evaluated at present times for \LCDM\ with $\OMn =0.3$.
The model fit includes 6 parameters
$\left(\frac{r_{\text{BAO}} }{\alpha } , \frac{\sigma }{\alpha } , A
\alpha^2, C_0, C_1,C_2 \right)$. The best fit (green line) is the
fit that maximises the likelihood function. The median fit
(purple line) is based on the 50\% quantiles of the Bayesian
posterior, resulting from conservative priors (meaning priors that
span the significant volume of the likelihood). Mean values of $\frac{r_{\text{BAO}} }{\alpha }$, $\frac{\sigma }{\alpha }$, and $A
\alpha^2$ with
1$\sigma$ equal tail credible intervals are superimposed on the plots.}
\label{fig:mockMeanIsotropic}
\end{figure}

\begin{figure}[!htb]
\centering
\begin{subfigure}[b]{.47\textwidth}
\includegraphics[width=\textwidth]{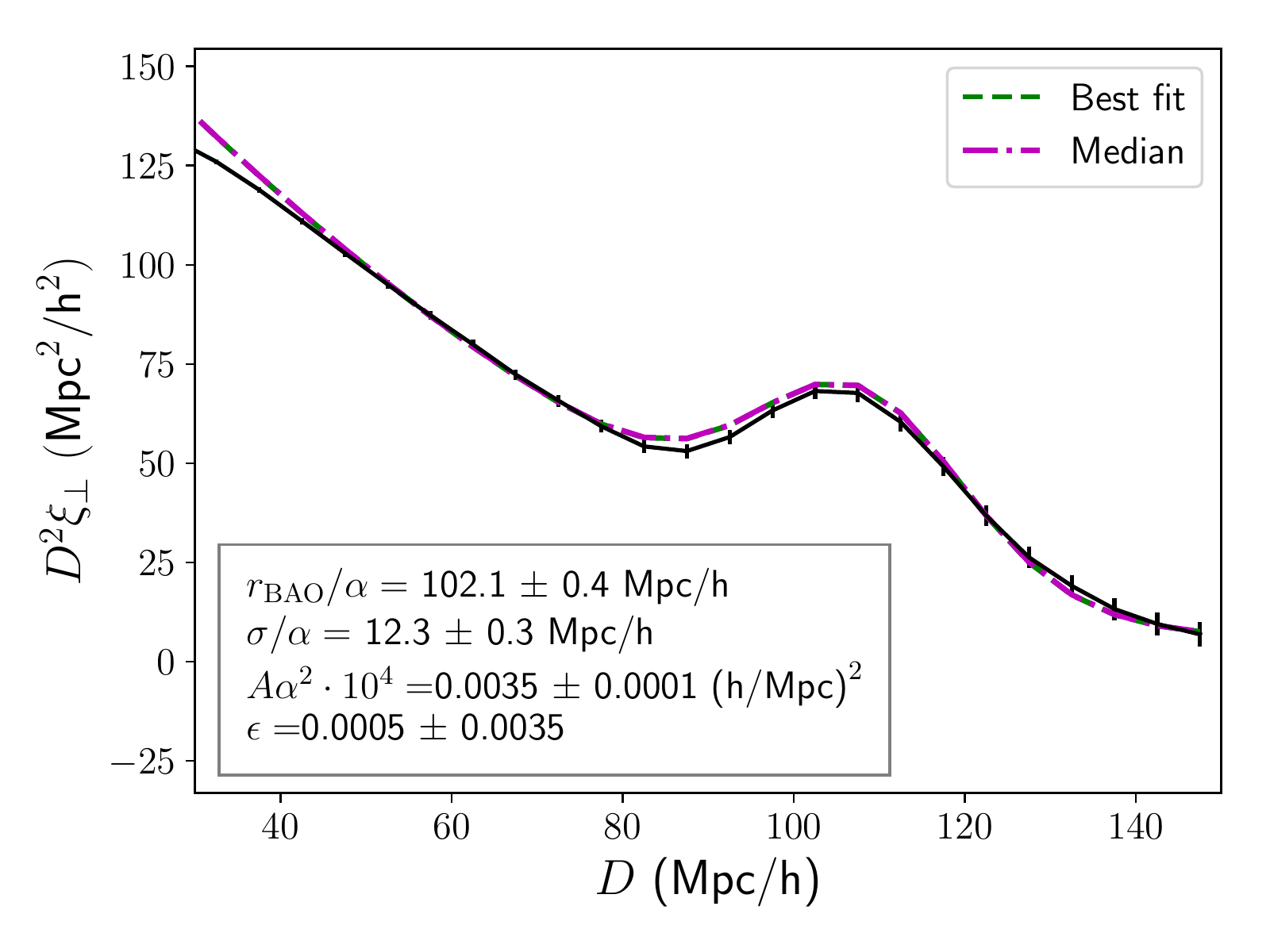}
\caption{$\xi_{\perp}(D_{\Lambda \text{CDM}})$ CMASS NGC}
\label{fig:mocksW0cmass}
\end{subfigure}
\medskip
\begin{subfigure}[b]{.47\textwidth}
\includegraphics[width=\textwidth]{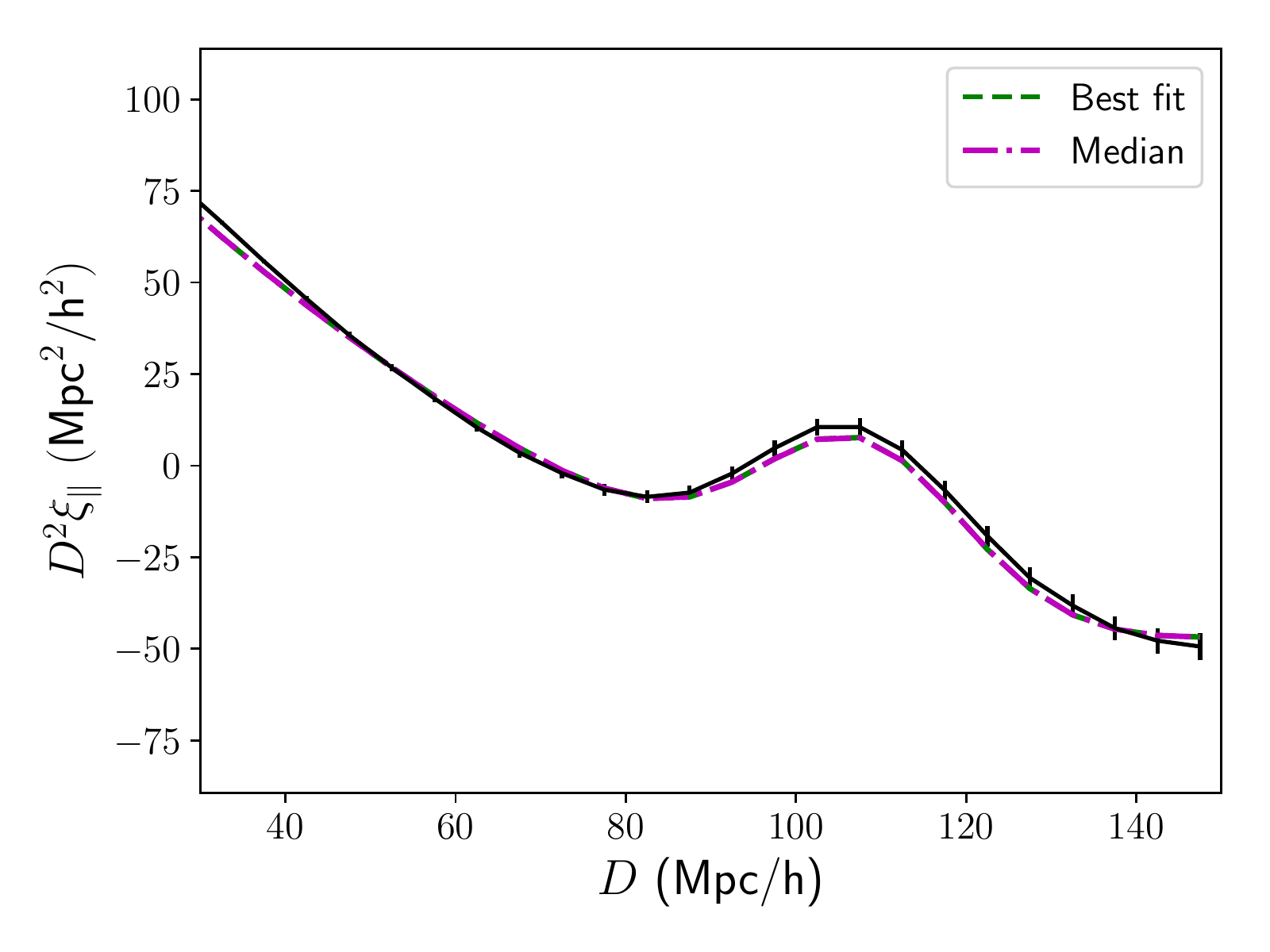}
\caption{$\xi_{\parallel}(D_{\Lambda \text{CDM}})$ CMASS NGC}
\label{fig:mocksW1cmass}
\end{subfigure}
\begin{subfigure}[b]{.47\textwidth}
\includegraphics[width=\textwidth]{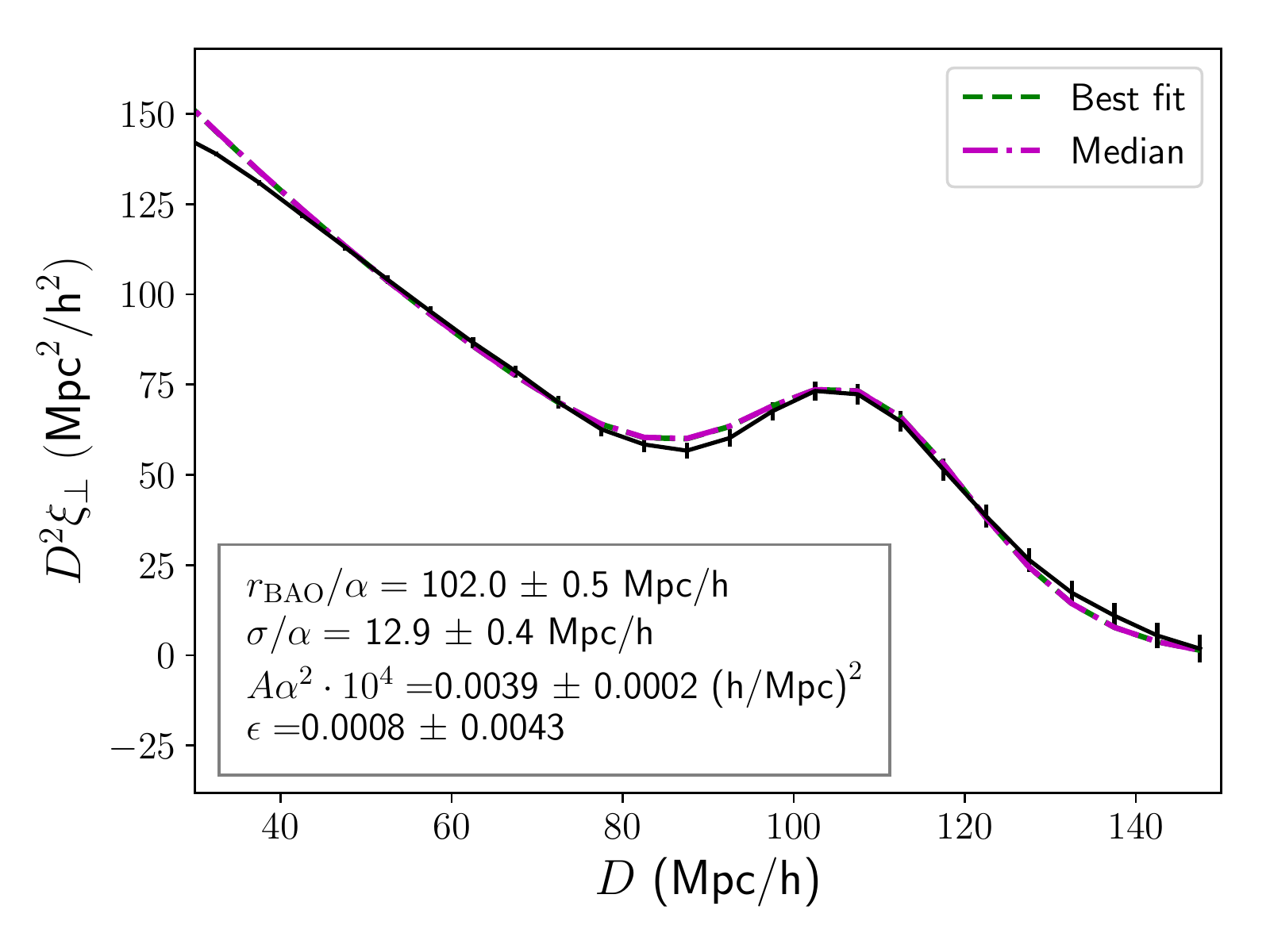}
\caption{$\xi_{\perp}(D_{\Lambda \text{CDM}})$ LOWZ NGC}
\label{fig:mocksW0lowz}
\end{subfigure}
\medskip
\begin{subfigure}[b]{.47\textwidth}
\includegraphics[width=\textwidth]{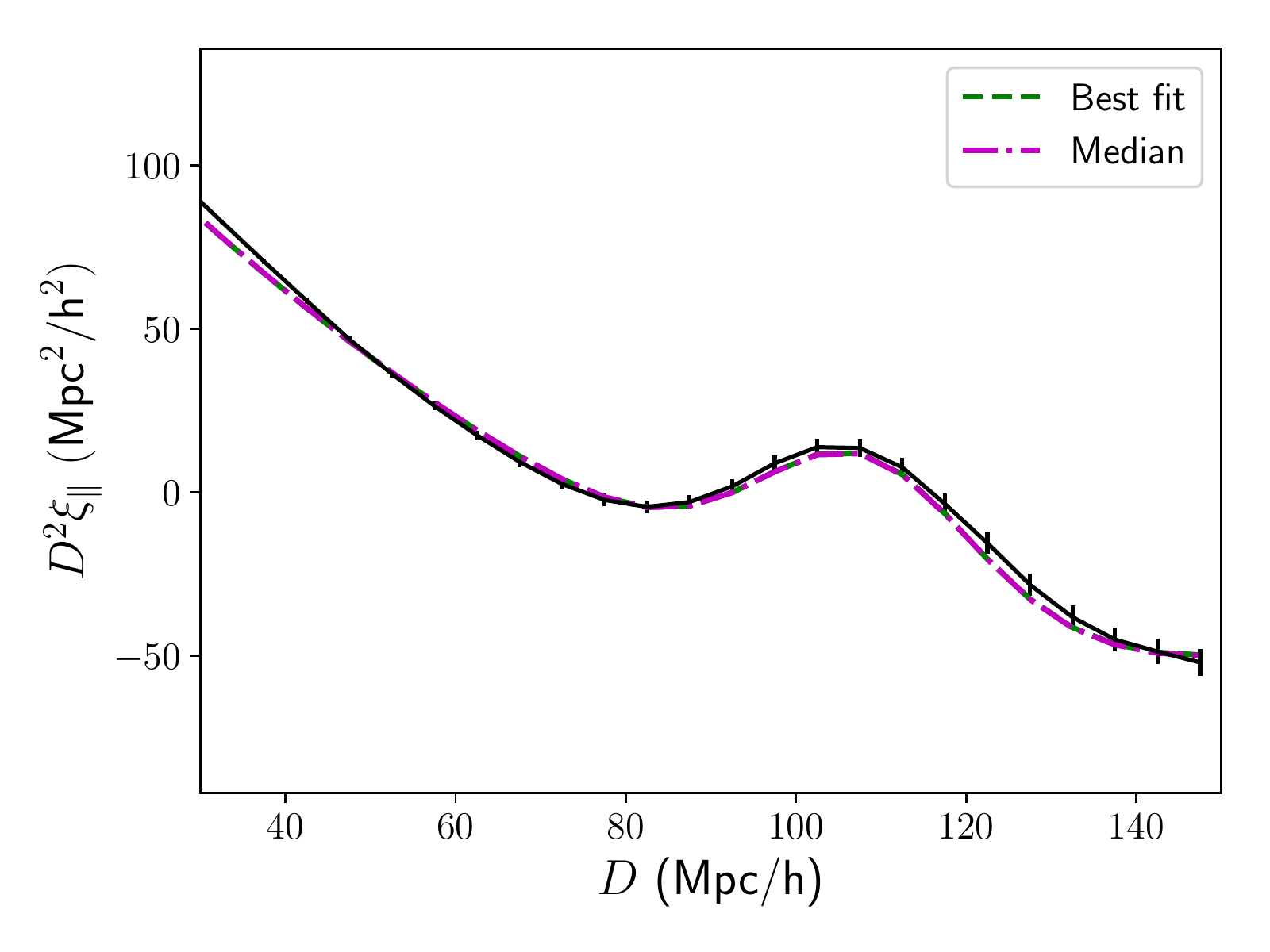}
\caption{$\xi_{\parallel}(D_{\Lambda \text{CDM}})$ LOWZ NGC}
\label{fig:mocksW1lowz}
\end{subfigure}
\caption{Combined fit to the transverse wedge $\xi_{\perp}(D_{\Lambda
\text{CDM}})$ and radial wedge $\xi_{\parallel}(D_{\Lambda
\text{CDM}})$ of the mean of the CMASS NGC and LOWZ NGC QPM
mocks respectively, where $D_{\Lambda \text{CDM}}$ is the Lagrangian
distance evaluated at present times for \LCDM\ with
$\OMn = 0.3$. The model fit includes 10 parameters
$\left(\frac{r_{\text{BAO}} }{\alpha } , \frac{\sigma }{\alpha } , A
\alpha^2, \epsilon, \bar{C}_{0 \perp}, \bar{C}_{1 \perp}, \bar{C}_{2
\perp}, \bar{C}_{0 \parallel}, \bar{C}_{1 \parallel}, \bar{C}_{2
\parallel} \right)$. The best fit (green line) is the fit that
maximises the likelihood function. The median fit (purple line)
is based on the 50\% quantiles of the Bayesian posterior,
resulting from conservative priors (meaning priors that span the
significant volume of the likelihood). The numerical values
superimposed on the plot of $\xi_{\perp}$ are the mean values with
1$\sigma$ equal tail credible intervals.}
\label{fig:mocksWedges}
\end{figure}

We determine the parameters of $\xi_{\text{Fit}}$ in both a
frequentist and Bayesian setting: that is, we find frequentist best
fit parameters as well as Bayesian median parameters with conservative priors.
The results of the fit to the isotropic correlation function for the CMASS and
LOWZ QPM mock mean are shown in
figure \ref{fig:mockMeanIsotropic}, and the results of the fit to the
wedges are shown in figure \ref{fig:mocksWedges}. The estimates of
the parameters describing the isotropic BAO feature
$\left(\frac{r_{\text{BAO}} }{\alpha } , \frac{\sigma }{\alpha } , A
\alpha^2 \right)$ are in good agreement between the isotropic and
wedge analyses. The results for the estimated isotropic BAO scale are
$\frac{r_{\text{BAO}} }{\alpha } = 102.1 \pm 0.4\hm$ for CMASS and
$\frac{r_{\text{BAO}} }{\alpha } = 101.8 \pm 0.5\hm$ for LOWZ, and
the results for the estimated anisotropic distortion parameter are
$\epsilon = 0.0005 \pm 0.0035$ for CMASS and $\epsilon = 0.0008 \pm
0.0043$ for LOWZ.

As noted in section \ref{mocks}, the acoustic scale of the model
underlying the QPM mocks is $r_s = 103.05\hm$. Since the QPM
mocks are generated using $\OMn = 0.29$, and our fiducial model
has $\OMn = 0.30$, we have $\alpha(\bar{z} = 0.55) = 1.005$, and
$\alpha(\bar{z} = 0.32) = 1.003$. We thus compute expected fiducial
values
\begin{align} \label{eq:peakpositiontheory}
\frac{r_s}{\alpha(\bar{z} = 0.55)} &= 102.5 \hm , \qquad \frac{r_s}{\alpha(\bar{z} = 0.32)} = 102.7\hm\\
\epsilon(\bar{z} = 0.55) &= 0.0013 , \qquad \epsilon(\bar{z} = 0.32) =
0.0008 \label{eq:epsilon}
\end{align}
As seen in the isotropic results in figure
\ref{fig:mockMeanIsotropic}, the BAO scale is recovered to a precision of $0.4\% \pm 0.4\%$ for CMASS and $0.9\% \pm 0.5\%$
for LOWZ. The difference between the measured and the model
$\epsilon$-parameter is $\lvert \Delta \epsilon \rvert \lsim 0.0008$,
which is much smaller than typical errors in $\epsilon$ in the context
of \LCDM\ template-fitting approaches to BAO.

We note that $\sim 1\%$ systematic error is significant in
standard BAO template-fitting approaches, where the statistical
errors in the BAO scale measurement from the latest galaxy redshift
surveys are around $1\%$, and the contribution from systematic errors
in
a \LCDM\ model universe are significantly less
than $1\%$ \cite{ShiftPeakBAO}.
Systematic errors
in a empirical fitting procedure will inevitably be
larger, and dependent on the cosmological model.\footnote{{A fiducial \LCDM\ fitting function per construction gives back the correct BAO scale when fitted to mocks generated from that same fiducial \LCDM\ model. Any empirical fitting model, aiming at analysing BAO features for a broader class of models will yield larger systematics in the context of \LCDM\ model simulations than the fitting procedure adapted specifically to \LCDM.
The price to pay for introducing a flexible fitting function adaptable to a large range of cosmologies, is exactly that it is not adapted to a particular cosmology.}} However, the errors
in the underlying calibration of the BAO scale from the CMB are also
larger in models with greater uncertainities in the underlying
physical parameters.
In this paper, we will mainly be
interested in the $\epsilon$ parameter as a consistency check of the
tested fiducial cosmologies, and in comparing the significance of the
BAO feature between the tested models, and do not include a
calibration of the underlying BAO scale.

We experimented with modifications of eq.~(\ref{eq:xifitwedge}),
allowing for a relative scaling of the wedge amplitudes, wedge widths,
or both. The resulting fits were of similar quality to that of
eq.~(\ref{eq:xifitwedge}) from an Akaike Information Criterion
perspective. Thus we had no \LCDM-based motivation for introducing
additional parameters in
the analysis of the galaxy survey. We note, however, that for models
with more complicated curvature evolution than \LCDM, there might be
physical effects equivalent to the \LCDM\ redshift-space
distortions but possibly with stronger magnitude, distorting the
relative amplitude and width of the BAO feature in the two wedges.\footnote{{There is no obvious reason for this to be the case in the timescape model, however, since it implements a ``uniform quasilocal Hubble flow condition'' \cite{clocks,obstimescape}. Calculations of the amplitude of redshift-space distortions require the development of a framework analogous to standard cosmological perturbation theory, which is yet to be done for the timescape cosmology. Estimates of the amplitude of non-kinematic differential expansion \cite{bnw} have been made using the Lema\^{\i}tre-Tolman-Bondi models for local structures on scales of order $10$--$60\,$Mpc \cite{Dthesis}, with the result that differences from the standard model expectation are smaller than current measurement uncertainties in peculiar velocities. Thus we would not expect substantial differences from the amplitude of the standard Kaiser effect \cite{kaiser87}, at least within this class of models.}}

We also experimented with different scaling behaviour of the
Lagrangian distance $D^{\rm tr}$ in eq.~(\ref{eq:xifit}) -- for example,
changing the scaling $(D^{\rm tr})^2$ of the Gaussian function to
$(D^{\rm tr})^n$ with different values of $n$. The inferred peak of the
Gaussian changed as expected, in some cases being significantly
different from the BAO scale. However, $\epsilon$ was consistent with
the expected values in eq.~(\ref{eq:epsilon}) for all investigated
modifications of the fitting function.

\section{Data analysis}
\label{dataanalysis}

The empirical procedure developed in this paper can be applied to a
wide class of cosmological models. In this analysis, we consider two
fiducial model frameworks -- the timescape model and the spatially flat
\LCDM\ model, with $\OMn = 0.3$ in both cases. We note that
both the \LCDM\ and timescape models have a spherically symmetric
effective adapted geometry with a large curvature scale proportional to the
Hubble distance $\sim c/H_0$, which is in both models of order 3
Gpc$/h$. Thus the curvature scale is of order the survey diameter.
The Lagrangian distance introduced in section \ref{comoving} between
two galaxies separated around $\sim 100\hm$ is thus well
approximated by eq.~(\ref{eq:lagrangiandistapprox}) with correction
terms\footnote{See appendix \ref{expgeodesic} for an explicit derivation of
the correction terms.} of order $\lsim 10^{-3}$. We caution that our results
may not be suitable
for extrapolation to other model cosmologies, depending on the
Alcock-Paczy\'nski scaling.

We estimate the 2-point correlation function in each fiducial model
for the CMASS and LOWZ NGC and SGC regions using the LS estimator
described in section \ref{2PCFestimatorsmain}. We compute the
isotropic correlation function estimator $\hat{\xi}(D)$, along with
the radial and transverse wedge correlation function estimators,
$\hat{\xi}_{\parallel}(D)$ and $\hat{\xi}_{\perp}(D)$, defined in eq.~(\ref{eq:xiestimatorLSwedgeTR1}). We use the covariance matrix
$\doubleunderline{C}_{\hat{\xi}}$ formulated in
eq.~(\ref{eq:covmat}), computed from the QPM mocks described in
section \ref{mocks}, to estimate the variance of the correlation
function over realisations of an imagined ensemble of galaxy
catalogues, of which our galaxy catalogue is a single realisation. We
expect different models to \LCDM, with different models for
structure formation and global geometry, to give rise to a different
random process underlying our measured galaxy catalogue. However, we
shall assume that the \LCDM\ estimate provides a reasonable
lowest-order approximation of the covariance.

We combine the estimated correlation functions for the NGC and SGC
regions using the inverse covariance weighting
\cite{BlakeCov,WhiteCov}
\begin{align} \label{eq:InvCovWeighting1}
&\hat{\xi}_{\text{comb}} = \doubleunderline{C}_{\text{comb}} \left( \doubleunderline{C}^{-1}_{\text{NGC}} \,\hat{\xi}_{\text{NGC}} + \doubleunderline{C}^{-1}_{\text{SGC}} \,\hat{\xi}_{\text{SGC}} \right) ,
\end{align}
where
\begin{align} \label{eq:InvCovWeighting2}
\doubleunderline{C}^{-1}_{\text{comb}} = \doubleunderline{C}^{-1}_{\text{NGC}} + \doubleunderline{C}^{-1}_{\text{SGC}} ,
\end{align}
is the inverse covariance of the combined measurement, and where $\xi$
represents either the isotropic correlation functions $\xi(D)$ or the
combined wedge correlation function $(\xi_{\parallel}(D) ,
\xi_{\perp}(D))$. We experimented with different methods of combining
the NGC and SGC measurements, and found that our results were robust
to the exact weighting scheme used.

\subsection{Isotropic fitting analysis}

\begin{figure}[!htb]
\centering
\begin{subfigure}[b]{.47\textwidth}
\includegraphics[width=\textwidth]{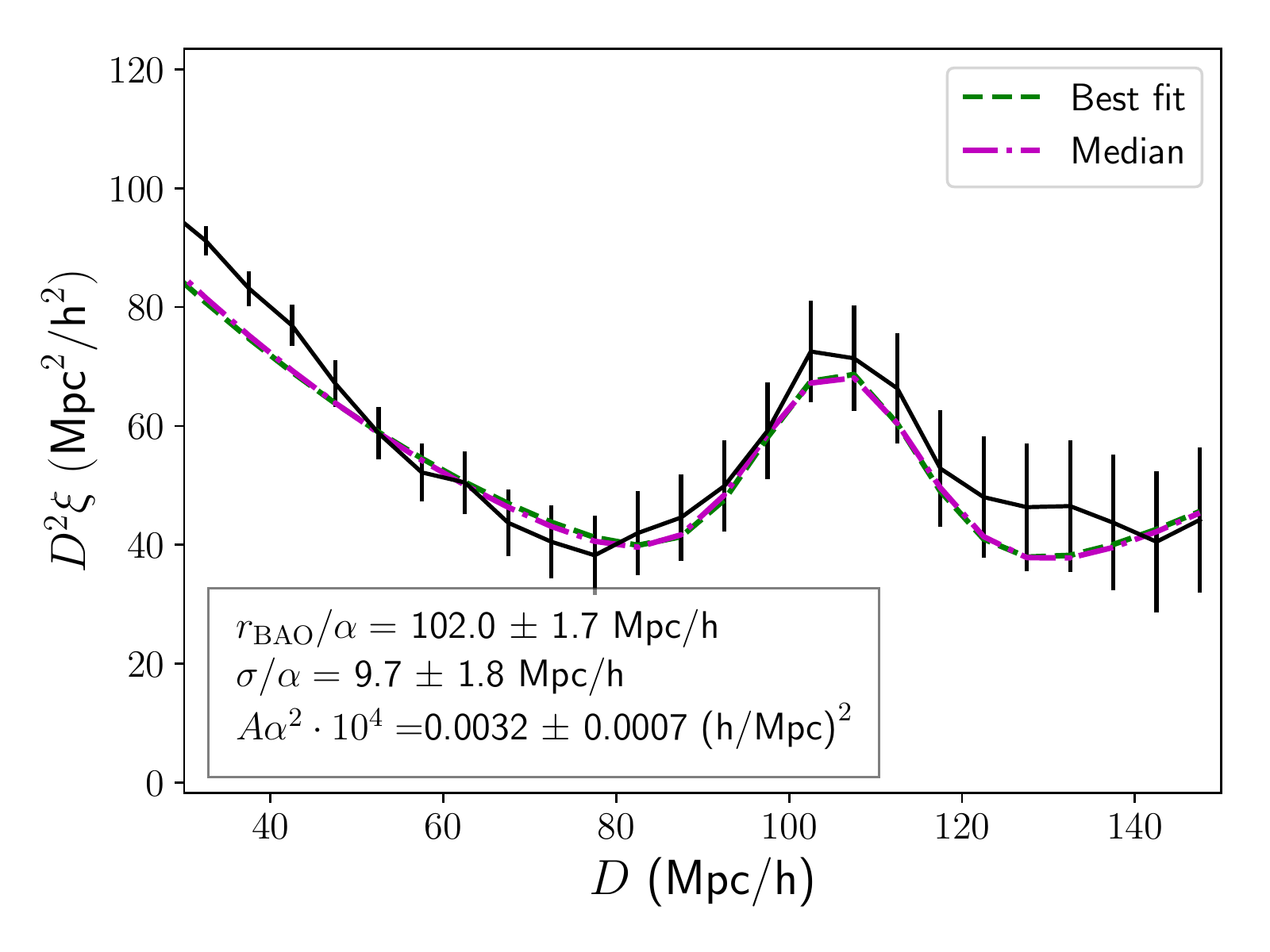}
\caption{$\xi(D_{\Lambda \text{CDM}} )$ CMASS}
\label{fig:LCDMIsotropicCMASS}
\end{subfigure}
\medskip
\begin{subfigure}[b]{.47\textwidth}
\includegraphics[width=\textwidth]{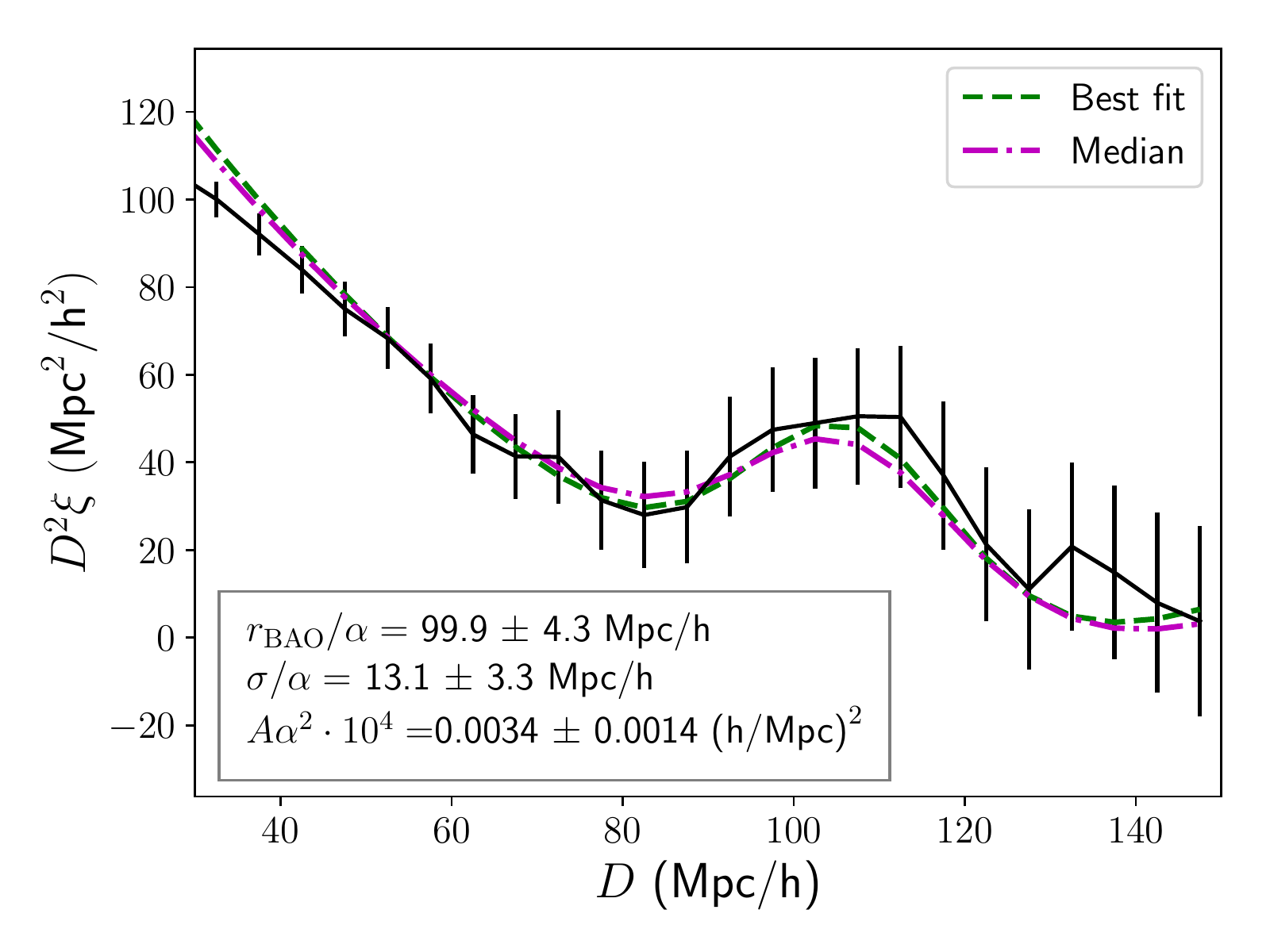}
\caption{$\xi(D_{\Lambda \text{CDM}})$ LOWZ}
\label{fig:LCDMIsotropicLOWZ}
\end{subfigure}
\caption{Fit to the isotropic wedge $\xi(D_{\Lambda \text{CDM}})$ of
the CMASS and LOWZ survey respectively, where $D_{\Lambda
\text{CDM}}$ is the Lagrangian distance evaluated at present times
for \LCDM\ with $\OMn = 0.3$. The model fit includes 6
parameters $\left(\frac{r_{\text{BAO}} }{\alpha } , \frac{\sigma
}{\alpha } , A \alpha^2, C_0, C_1,C_2 \right)$. The best fit (green
line) is the fit that maximises the likelihood function. The median
fit (purple line) is based on the 50\% quantiles of the
Bayesian posterior, resulting from conservative priors (meaning
priors that span the significant volume of the likelihood). Mean values of $\frac{r_{\text{BAO}} }{\alpha }$, $\frac{\sigma }{\alpha }$, and $A
\alpha^2$ with
1$\sigma$ equal tail credible intervals are superimposed on the plots.}
\label{fig:LCDMIsotropic}
\end{figure}

\begin{figure}[!htb]
\centering
\begin{subfigure}[b]{.47\textwidth}
\includegraphics[width=\textwidth]{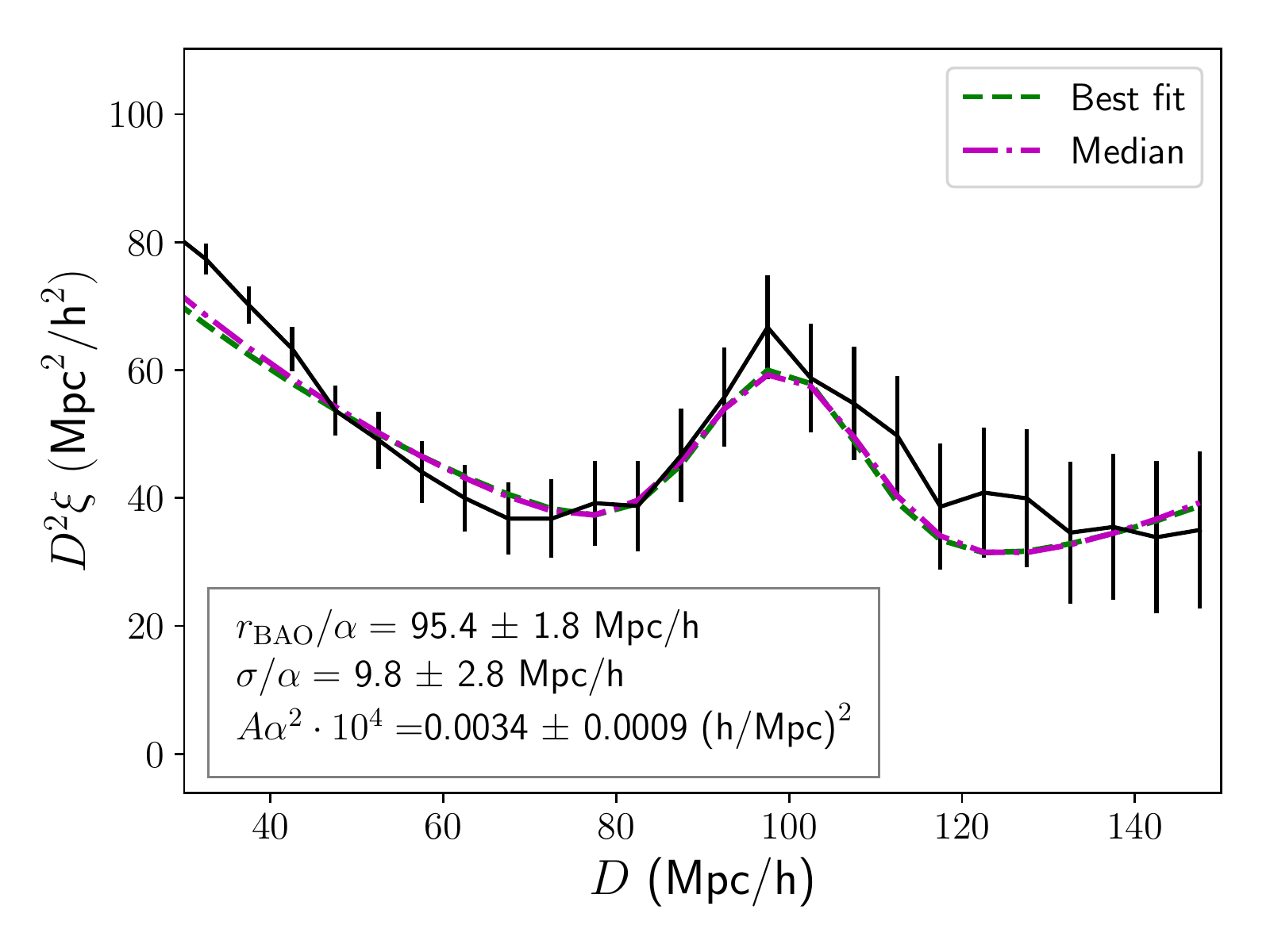}
\caption{$\xi(D_{\text{timescape}} )$ CMASS}
\label{fig:TimescapeIsotropicCMASS}
\end{subfigure}
\medskip
\begin{subfigure}[b]{.47\textwidth}
\includegraphics[width=\textwidth]{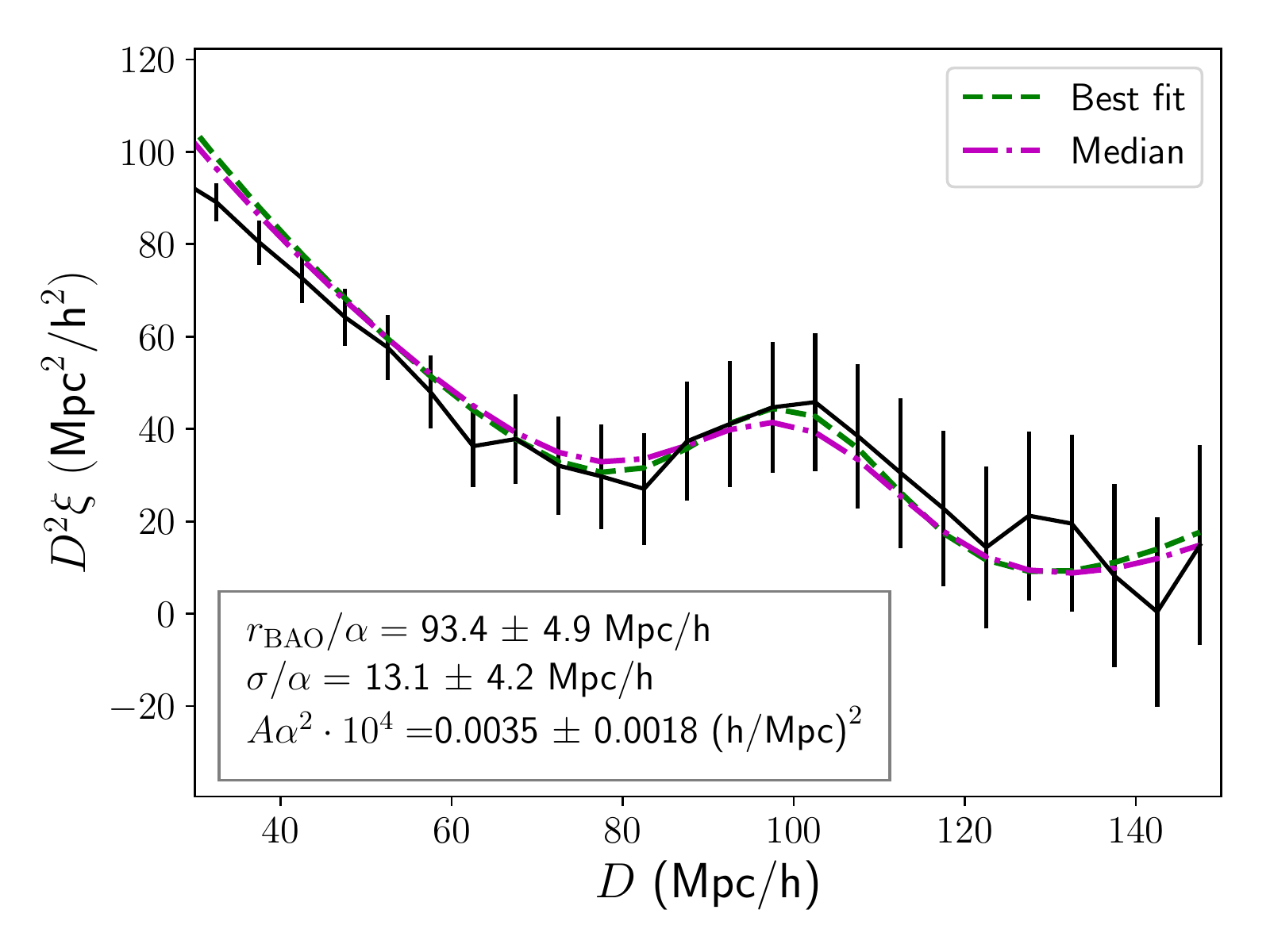}
\caption{$\xi(D_{\text{timescape}} )$ LOWZ}
\label{fig:TimescapeIsotropicLOWZ}
\end{subfigure}
\caption{Fit to the isotropic wedge $\xi(D_{\text{timescape}} )$ of
the CMASS and LOWZ survey respectively, where $D_{\text{timescape}}$
is the Lagrangian distance evaluated at present times for the
timescape model with $\OMn = 0.3$. The model fit includes 6
parameters $\left(\frac{r_{\text{BAO}} }{\alpha } , \frac{\sigma
}{\alpha } , A \alpha^2, C_0, C_1,C_2 \right)$. The best fit (green
line) is the fit that maximises the likelihood function. The median
fit (purple line) is based on the 50\% quantiles of the
Bayesian posterior, resulting from conservative priors (meaning
priors that span the significant volume of the likelihood). Mean values of $\frac{r_{\text{BAO}} }{\alpha }$, $\frac{\sigma }{\alpha }$, and $A
\alpha^2$ with
1$\sigma$ equal tail credible intervals are superimposed on the plots.}
\label{fig:TimescapeIsotropic}
\end{figure}

The estimated isotropic correlation function and best fit and median
models are displayed in figure \ref{fig:LCDMIsotropic} for
\LCDM\ and in figure \ref{fig:TimescapeIsotropic} for the timescape
cosmology, and the results of the fits are summarised in table
\ref{table:CMASSLOWZfit_isotropic}. The Gaussian peak component is
significant in the CMASS isotropic correlation function at the
$4.6\sigma$ level for \LCDM\ and at the $3.8\sigma$ level for
timescape. We quantify the significance of the peak as the posterior
probability of having $\alpha^2 A > 0$.\footnote{We note that this
is different to the typical way of quantifying BAO significance in
\LCDM-based fitting, where a reference power spectrum with no
BAO feature is used to assess the increase of quality in fit when
introducing the BAO feature \cite{BlakeSignificance}.} For the LOWZ
correlation function, the peak is significant at the $2.4\sigma$ level
for \LCDM\ and at the $1.9 \sigma$ level for timescape.

We have used conservative priors for our fits to both timescape and
\LCDM, meaning (log-)uniform priors that span all regions of
parameter space of significant likelihood volume. For the sake of
comparing our \LCDM\ results with the standard fiducial
\LCDM\ analysis of \cite{wedgefit} and \cite{Standardresults},
we have repeated the fit with narrow Gaussian error bars on
$\sigma_{\text{BAO}}/ \alpha$ with mean and standard deviation as
determined by the isotropic mock fit of section \ref{LCDMmocktest}.
For both models, using this prior increases the significance of the
BAO feature and decreases the errors in $r_{\text{BAO}}/ \alpha$. We
only compare model fits when the priors are equally restrictive for
both models and, unless otherwise stated, we comment on the analysis
with conservative priors.

The results for $r_{\text{BAO}}/ \alpha$ and $\sigma_{\text{BAO}}/
\alpha$ are consistent between the LOWZ and CMASS samples for both
timescape and \LCDM. The results for the CMASS BAO peak
positions for the conservative prior analysis are $r_{BAO}/\alpha =
102.0 \pm 1.7\hm$ for \LCDM\ and $95.4 \pm 1.8\hm$ for
timescape. The equivalent results for LOWZ are $99.9 \pm 4.3\hm$
for \LCDM\ and $93.4 \pm 4.9\hm$ for timescape.
The sign and magnitude of the relative peak positions of timescape and \LCDM\ are consistent with figure \ref{fig:LCDMvsTimescape} within the statistical error bars of the analysis. This can be realised by computing the relative isotropic AP-scaling $\alpha$ (\ref{eq:alphaepsilontrans}) between \LCDM\ and timescape based on figure \ref{fig:LCDMvsTimescape} and comparing it to the ratio of the measured peak positions $r_{BAO}/\alpha$ of the models.

Values of the Hubble constant
for the timescape model obtained from CMB constraints can be up to 10\% smaller than for the $\Lambda$CDM model \cite{dnw}. Thus for typical values of $H_0$ the estimated isotropic peak position in units Mpc may in fact be slightly \emph{larger} for timescape than the analogous peak position for $\Lambda$CDM.

The fits are
reasonably good, all with a minimum $\chi^2$ value of reasonable
probability. The most extreme value is $\chi^2 = 22$ for the timescape
LOWZ fit, the probability of $\chi^2>22$ being 8\% for 14
degrees of freedom.

When we include a prior in $\sigma$, our \LCDM\ results for the
isotropic peak position $r_{\text{BAO}}/ \alpha$ are in $\lsim 1\%$
agreement with those found in the fiducial \LCDM\ analysis
considered in e.g., \cite{wedgefit} and \cite{Standardresults}. The
magnitude of the error bars are also comparable to those found in
standard analyses. As an example, we compare our results with the
isotropic pre-reconstruction DR12 results of table 8 in
\cite{Standardresults}, derived assuming the fiducial cosmology
$\OMfid = 0.29$ and $r_{\text{BAO}, fid} = 103.0\hm$.
The isotropic CMASS measurement yields $\tilde{\alpha} =
\alpha(\OMn = 0.29)\, \frac{r_{\text{BAO}, fid} }{r_{\text{BAO}}} =
1.015 \pm 0.013$, which together with the value of $r_{\text{BAO},
fid}$ yields $\frac{r_{\text{BAO}} }{\alpha (\OMn = 0.29) } =
101.4 \pm 1.3\hm$, and finally scaling the result with the
$\alpha$-ratio of \LCDM\ fiducial $\OMfid = 0.29$ and
$\OMn = 0.3$ we have $\frac{r_{\text{BAO}} }{\alpha (\OMn =
0.30)} = 100.9 \pm 1.3\hm$. This result is within 1$\sigma$
agreement with the \LCDM\ results in table
\ref{table:CMASSLOWZfit_isotropic} for both the conservative prior
analysis and for the analysis with a narrow Gaussian prior in
$\sigma_{\text{BAO}}/ \alpha$. The analogous isotropic LOWZ result
from table 8 in \cite{Standardresults} reads $\tilde{\alpha} =
\alpha(\OMn = 0.29)\, \frac{r_{\text{BAO}, fid} }{r_{\text{BAO}}} =
1.009 \pm 0.030$, which gives $\frac{r_{\text{BAO}} }{\alpha (\OMn
= 0.30)} = 101.7 \pm 3.1\hm$, in agreement with the
\LCDM\ results in table \ref{table:CMASSLOWZfit_isotropic} for
the conservative prior analysis and for the narrow Gaussian prior in
$\sigma_{\text{BAO}}/ \alpha$.

\begin{table}[h]
\centering
\begin{tabular}{| llllr |}
\hline
\textbf{Isotropic fit} $ \boldsymbol{\xi}$& $\alpha^2 A \cdot10^4$ & $r_{\text{BAO}}/ \alpha$ & $\sigma_{\text{BAO}}/ \alpha$ & $\chi^2 / N_{\text{dof}}$ \\ [0.5ex]
\LCDM\ CMASS & 0.0032 $\pm$ 0.0007 & 102.0 $\pm$ 1.7 & 9.7 $\pm$ 1.8 & $17/14$\\
\LCDM\ LOWZ & 0.0034 $\pm$ 0.0014 & 99.9 $\pm$ 4.3 & 13.1 $\pm$ 3.3 & $19/14$\\
Timescape CMASS & 0.0034 $\pm$ 0.0009 & 95.4 $\pm$ 1.8 & 9.8 $\pm$ 2.8 & $21/14$ \\
Timescape LOWZ & 0.0035 $\pm$ 0.0018 & 93.4 $\pm$ 4.9 & 13.1 $\pm$ 4.2 & $22/14$ \\
\hline
\LCDM\ CMASS $\mathcal{N}_{\sigma_{\text{BAO}}/ \alpha}$ & 0.0037 $\pm$ 0.0007 & 100.4 $\pm$ 1.5 & 12.2 $\pm$ 0.3 & $17/14$\\
\LCDM\ LOWZ $\mathcal{N}_{\sigma_{\text{BAO}}/ \alpha}$ & 0.0035 $\pm$ 0.0011 & 100.6 $\pm$ 3.0 & 12.2 $\pm$ 0.3 & $19/14$\\
\hline
\end{tabular}
\caption{Results of fitting the isotropic correlation function of
CMASS and LOWZ. The parameter estimates shown are the Bayesian
median with $1\sigma$ equal tail credible intervals. Conservative
priors (meaning priors that span the significant volume of the
likelihood) are used for all parameters in all fits, except for the
\LCDM\ fits labelled $\mathcal{N}_{\sigma_{\text{BAO}}/
\alpha}$, which use a narrow Gaussian prior with mean and width as
determined in the mock analysis of section \ref{LCDMmocktest}. The
minimum $\chi^2$ value divided by number of degrees of freedom
$N_{\text{dof}}$ is also quoted. $r_{\text{BAO}}/ \alpha$ and
$\sigma_{\text{BAO}}/ \alpha$ are in units of $\hm$. $A$ is in
units of $(\!\hm)^2$.}
\label{table:CMASSLOWZfit_isotropic}
\end{table}

\subsection{Anisotropic fitting analysis}
\label{anisotropicfitting}
We now turn to the wedge analysis, which is useful for examining the
consistency of the BAO feature in the transverse and radial separation
of galaxy pairs. The results of fitting the empirical parameters
describing the BAO feature are shown in table
\ref{table:CMASSLOWZfit}. The measurements of the anisotropic
distortion parameter in CMASS are $\epsilon = -0.021 \pm 0.017$ for
\LCDM\ and $\epsilon = 0.021 \pm 0.017$ for timescape, and the
LOWZ results are $\epsilon = -0.022 \pm 0.084$ for \LCDM\ and
$\epsilon = 0.013 \pm 0.110$ for timescape. The CMASS and LOWZ
results for the peak position $r_{\text{BAO}}/ \alpha$ and the width
$\sigma_{\text{BAO}}/ \alpha$ are consistent within 1$\sigma$ for both
\LCDM\ and the timescape model.

The Gaussian peak in the CMASS wedge correlation functions has a
significance of $\sim 4.8\sigma$ for \LCDM\ and $\sim 3.9\sigma$ for
timescape. For LOWZ, the peak has a significance of $\sim
1.4\sigma$ and $\sim 1.3\sigma$ for \LCDM\ and timescape respectively. As above, the significance of the peak is
quantified as the posterior probability of having $\alpha^2 A > 0$.

We note that the values of epsilon gives $\psi \approx 0.1$, for which the expansion in eq.~(\ref{eq:xifitmodel2}) is not guaranteed to hold for the fitting range $D \in [50;150]\hm$. We checked the validity of the approximate fitting model eq.~(\ref{eq:xifitwedge}) by comparing to the exact wedge fitting functions calculated from the average of eq.~(\ref{eq:xifitmodel}) over $\mu$, and found that best fit parameter results derived in our linearised analysis receive corrections of order $\sim 10$\% of the error bars on the same parameters. Since the corrections are an order of magnitude smaller than the error bars, we ignore these corrections here and quote the results from the linearised analysis.

\begin{figure}[!htb]
\centering
\begin{subfigure}[b]{.47\textwidth}
\centering
\includegraphics[width=\textwidth]{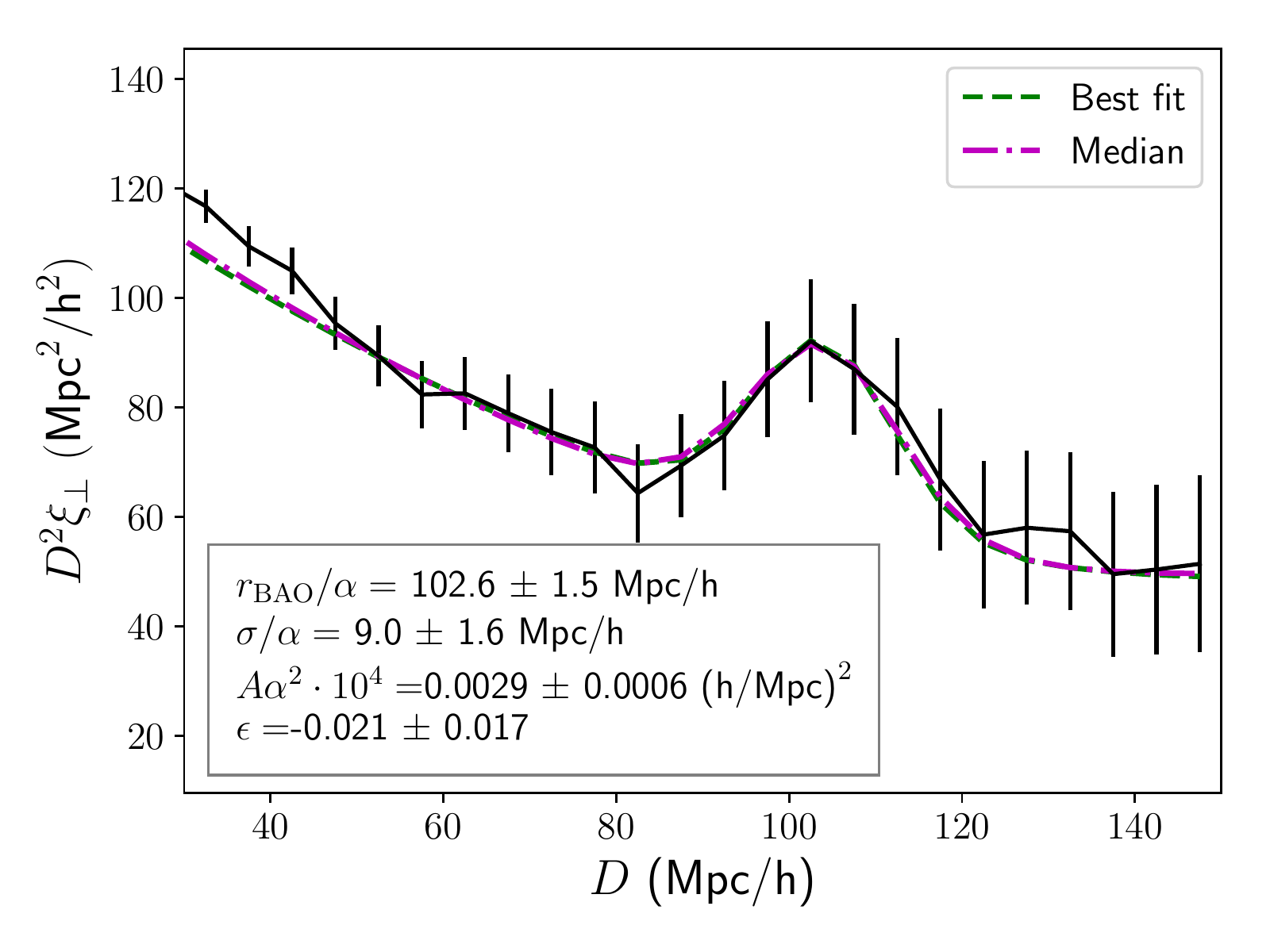}
\caption{$\xi_{\perp}(D_{\Lambda \text{CDM}} )$ CMASS}
\label{fig:LCDMW0CMASS}
\end{subfigure}
\medskip
\begin{subfigure}[b]{.47\textwidth}
\centering
\includegraphics[width=\textwidth]{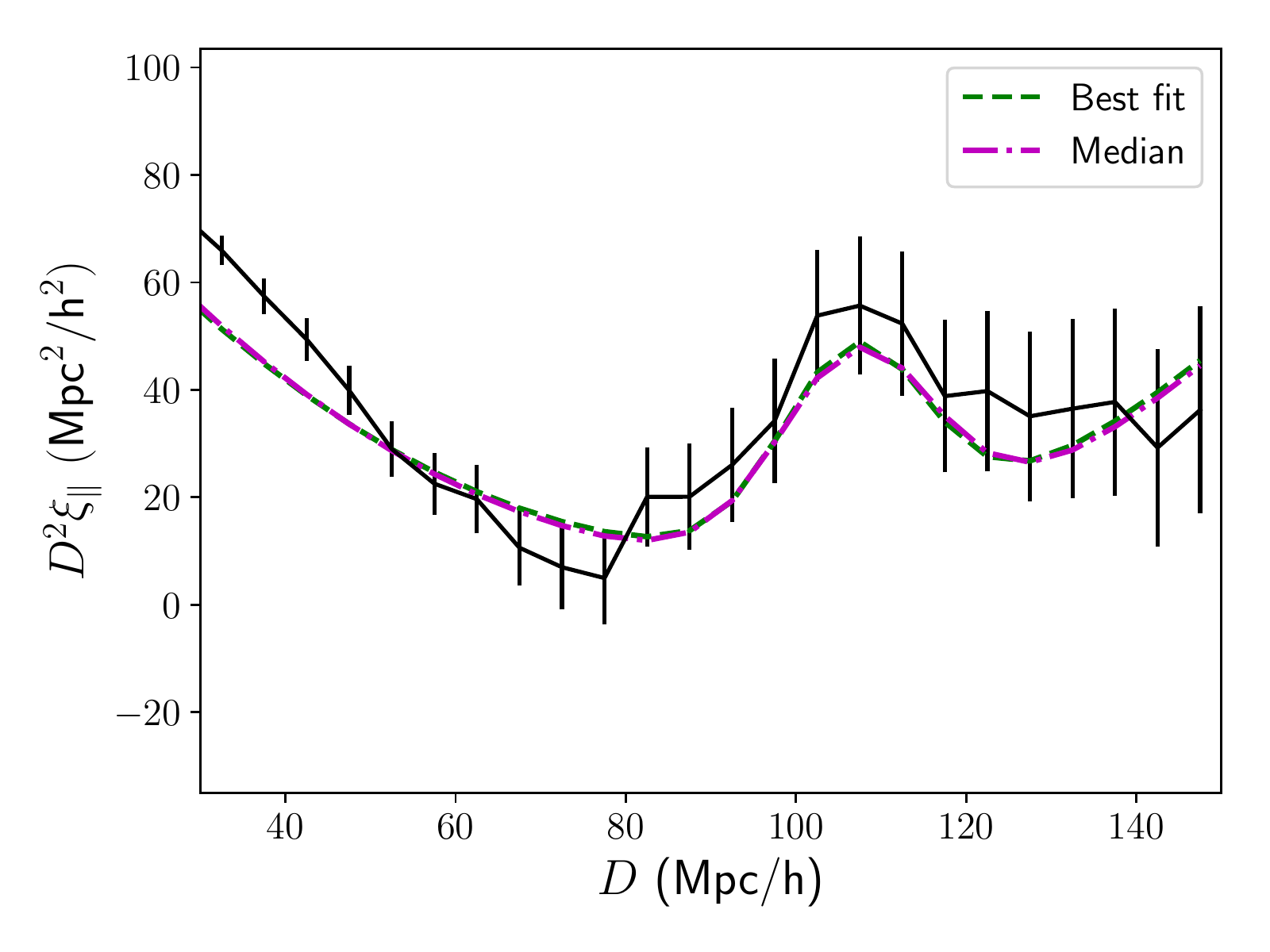}
\caption{$\xi_{\parallel}(D_{\Lambda \text{CDM}})$ CMASS}
\label{fig:LCDMW1CMASS}
\end{subfigure}
\begin{subfigure}[b]{.47\textwidth}
\centering
\includegraphics[width=\textwidth]{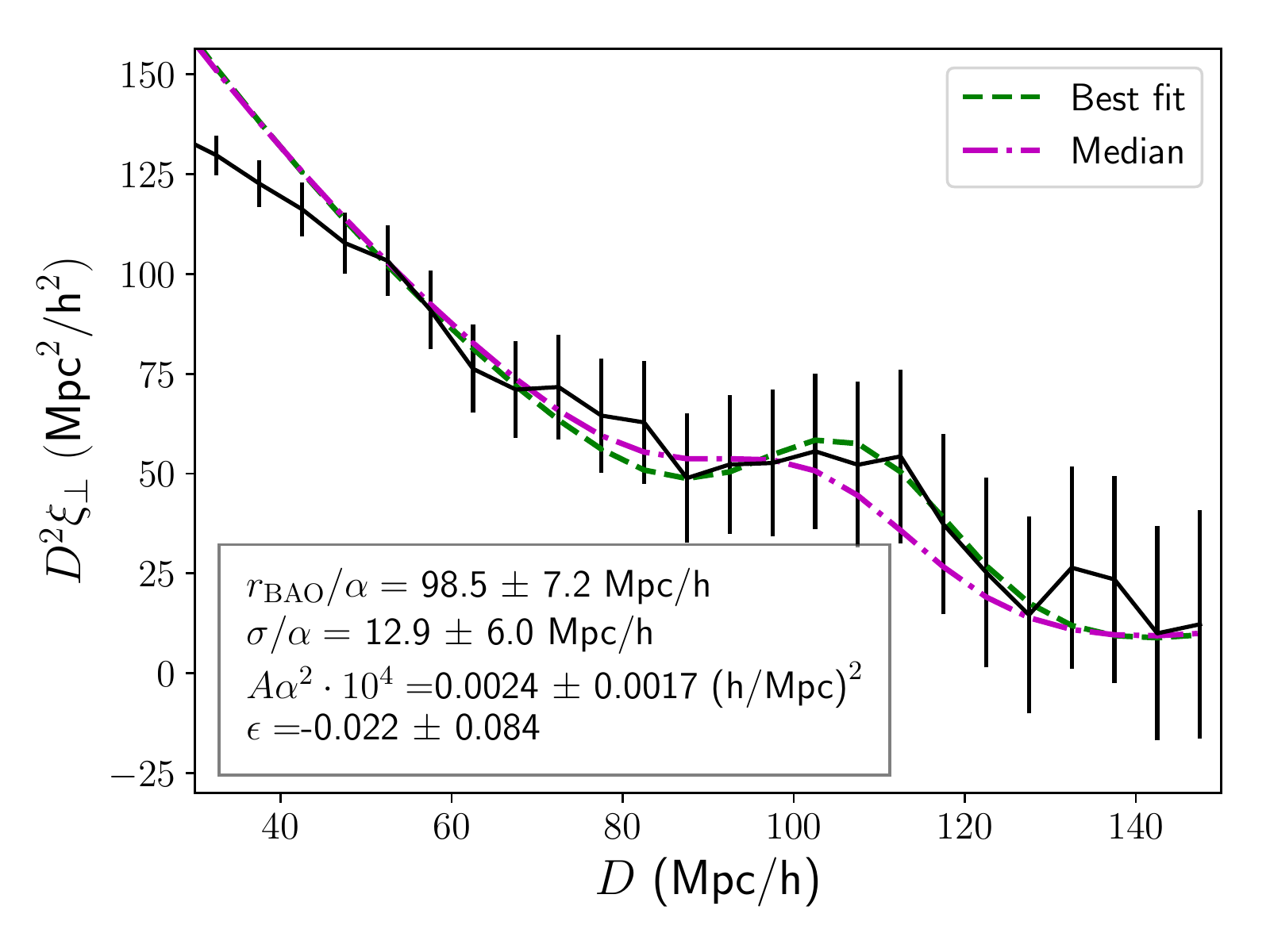}
\caption{$\xi_{\perp}(D_{\Lambda \text{CDM}})$ LOWZ}
\label{fig:LCDMW0LOWZ}
\end{subfigure}
\medskip
\begin{subfigure}[b]{.47\textwidth}
\centering
\includegraphics[width=\textwidth]{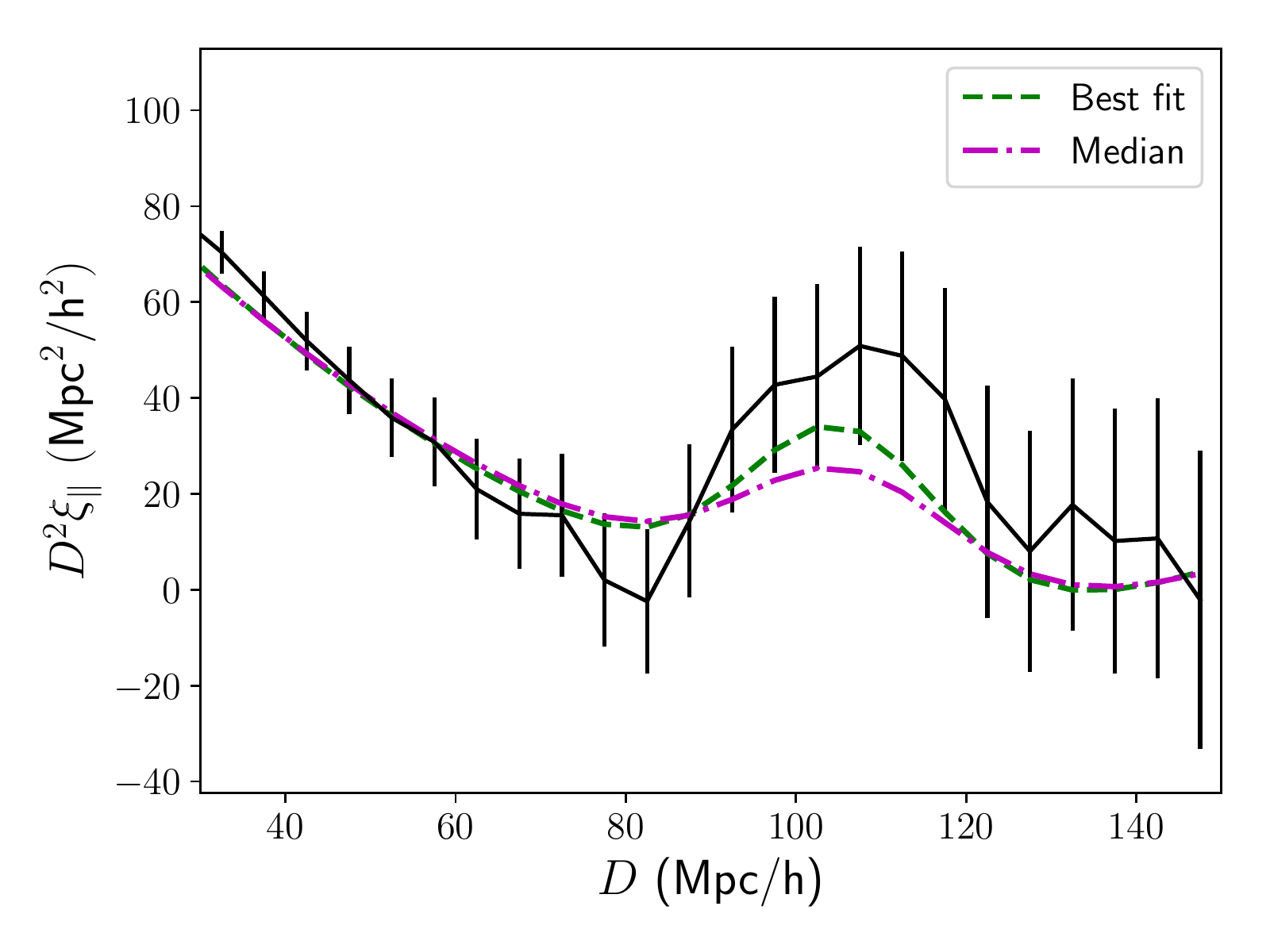}
\caption{$\xi_{\parallel}(D_{\Lambda \text{CDM}})$ LOWZ}
\label{fig:LCDMW1LOWZ}
\end{subfigure}
\caption{Combined fit to the transverse wedge $\xi_{\perp}(D_{\Lambda
\text{CDM}})$ and radial wedge $\xi_{\parallel}(D_{\Lambda
\text{CDM}})$ of the CMASS and LOWZ survey respectively, where
$D_{\Lambda \text{CDM}}$ is the Lagrangian distance evaluated at
present times for \LCDM\ with $\OMn = 0.3$. The model
fit includes 10 parameters $\left(\frac{r_{\text{BAO}} }{\alpha } ,
\frac{\sigma }{\alpha } , A \alpha^2, \epsilon, \bar{C}_{0 \perp},
\bar{C}_{1 \perp}, \bar{C}_{2 \perp}, \bar{C}_{0 \parallel},
\bar{C}_{1 \parallel}, \bar{C}_{2 \parallel} \right)$. The best fit
(green line) is the fit that maximises the likelihood function. The
median fit (purple line) is based on the 50\% quantiles of the
Bayesian posterior, resulting from conservative priors (meaning
priors that span the significant volume of the likelihood). The
numerical values superimposed on the plot of $\xi_{\perp}$ are the
mean values with 1$\sigma$ equal tail credible intervals.}
\label{fig:LCDMWedges}
\end{figure}

\begin{figure}[!htb]
\centering
\begin{subfigure}[b]{.47\textwidth}
\includegraphics[width=\textwidth]{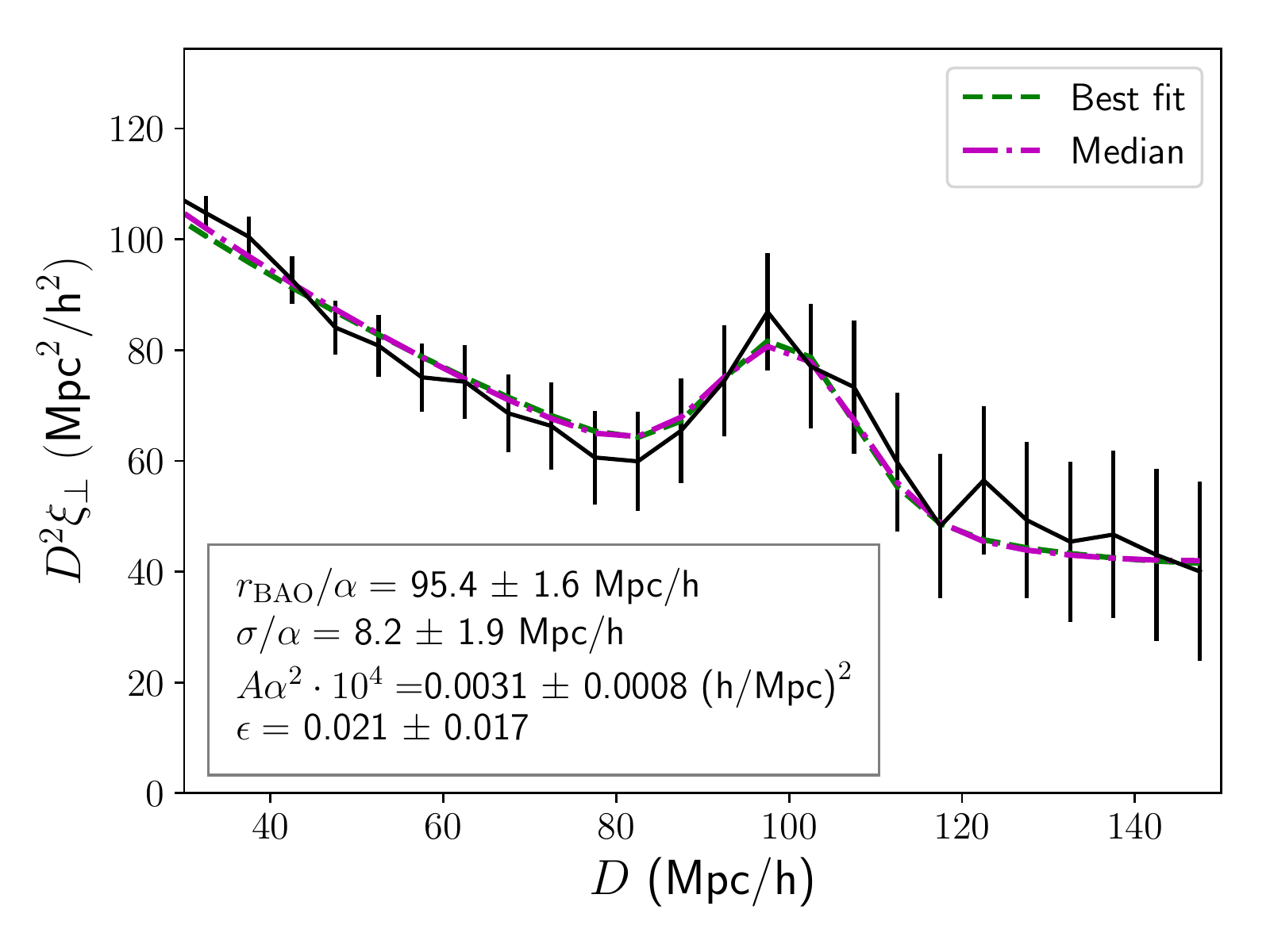}
\caption{$\xi_{\perp}(D_{\text{Timescape}} )$ CMASS}
\label{fig:tsW0CMASS}
\end{subfigure}
\medskip
\begin{subfigure}[b]{.47\textwidth}
\includegraphics[width=\textwidth]{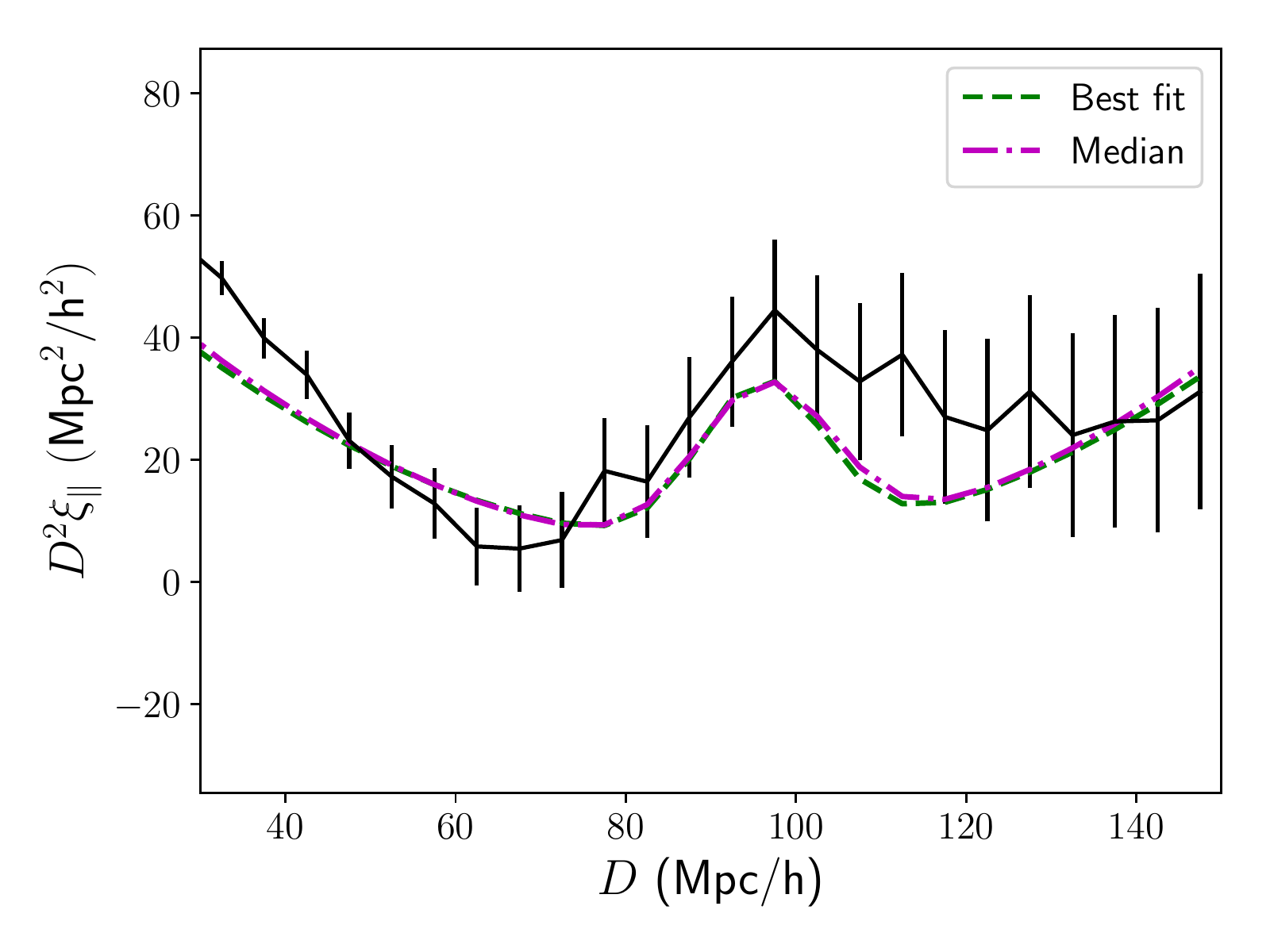}
\caption{$\xi_{\parallel}(D_{\text{Timescape}})$ CMASS}
\label{fig:tsW1CMASS}
\end{subfigure}
\begin{subfigure}[b]{.47\textwidth}
\includegraphics[width=\textwidth]{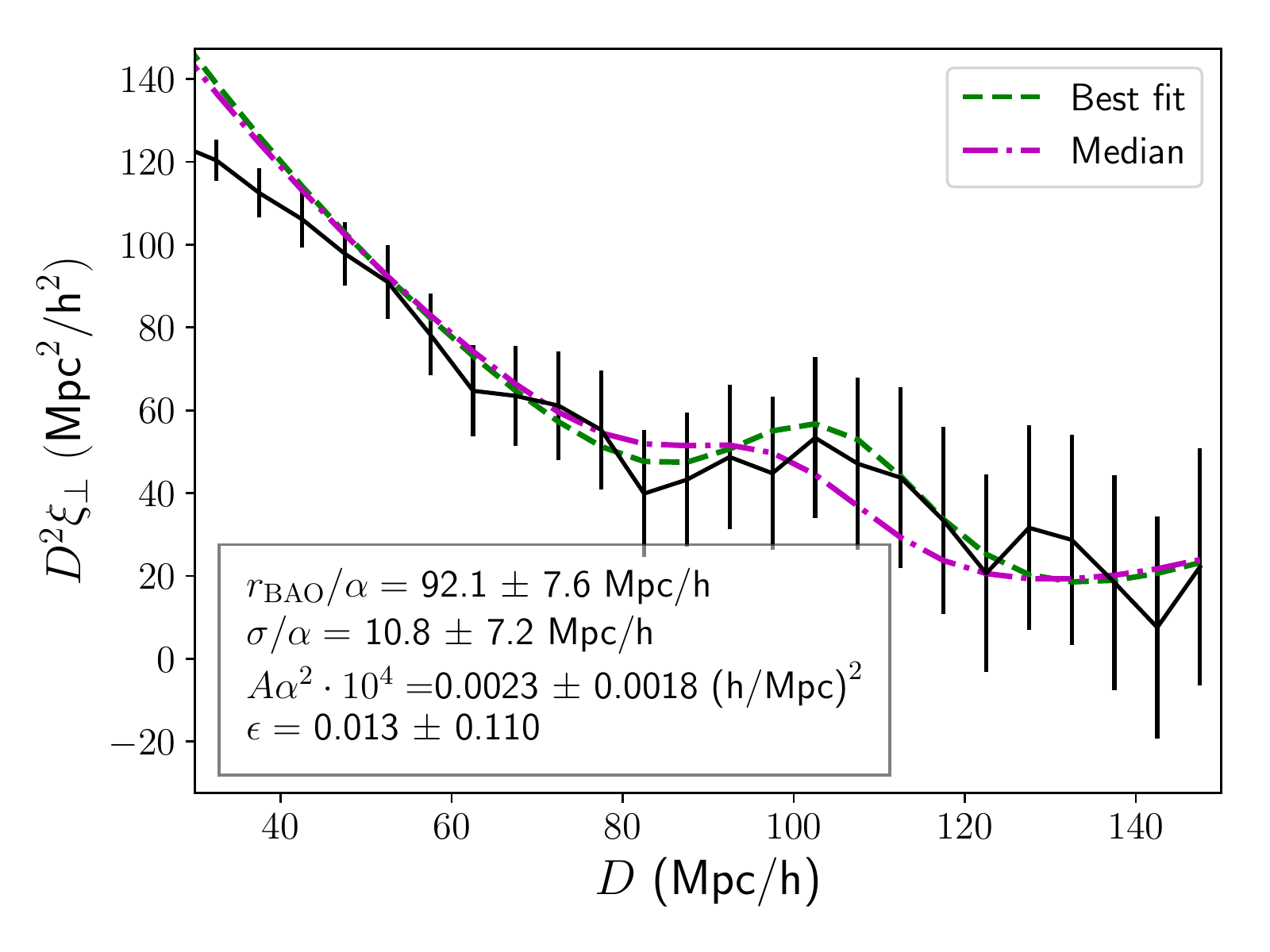}
\caption{$\xi_{\perp}(D_{\text{Timescape}})$ LOWZ}
\label{fig:tsW0LOWZ}
\end{subfigure}
\medskip
\begin{subfigure}[b]{.47\textwidth}
\includegraphics[width=\textwidth]{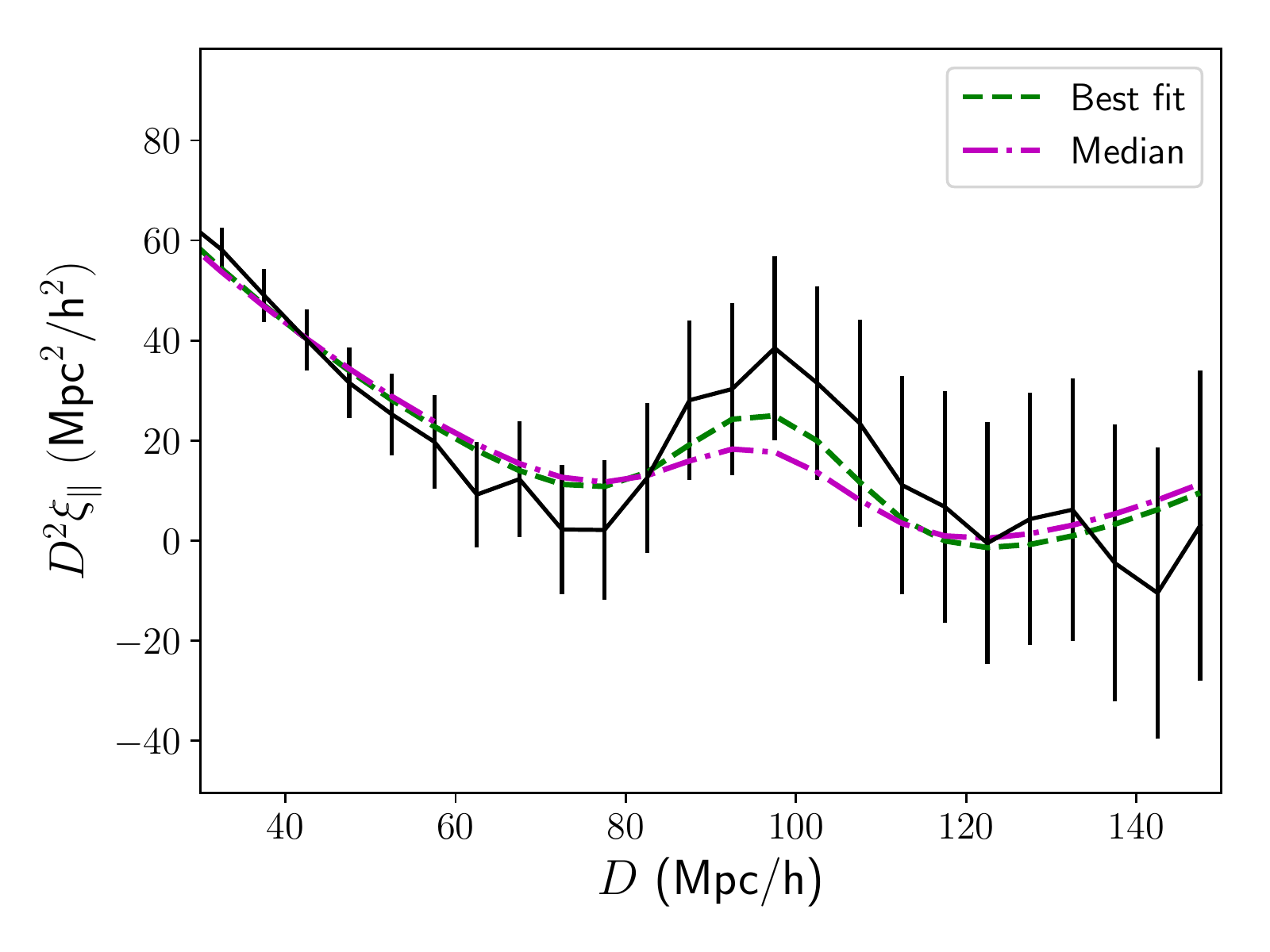}
\caption{$\xi_{\parallel}(D_{\text{Timescape}})$ LOWZ}
\label{fig:tsW1LOWZ}
\end{subfigure}
\caption{Combined fit to the transverse wedge
$\xi_{\perp}(D_{\text{Timescape}})$ and radial wedge
$\xi_{\parallel}(D_{\text{Timescape}})$ of the CMASS and LOWZ survey
respectively, where $D_{\text{Timescape}}$ is the Lagrangian
distance evaluated at present times for the timescape model with
$\OMn = 0.3$. The model fit includes 10 parameters
$\left(\frac{r_{\text{BAO}} }{\alpha } , \frac{\sigma }{\alpha } , A
\alpha^2, \epsilon, \bar{C}_{0 \perp}, \bar{C}_{1 \perp}, \bar{C}_{2
\perp}, \bar{C}_{0 \parallel}, \bar{C}_{1 \parallel}, \bar{C}_{2
\parallel} \right)$. The best fit (green line) is the fit that
maximises the likelihood function. The median fit (purple line) is
based on the 50\% quantiles of the Bayesian posterior,
resulting from conservative priors (meaning priors that span the
significant volume of the likelihood). The numerical values
superimposed on the plot of $\xi_{\perp}$ are the mean values with
1$\sigma$ equal tail credible intervals.}
\label{fig:tsWedges}
\end{figure}

The best fit and median models of eq.~(\ref{eq:xifitwedge}) are shown
superimposed on the $\xi_{\perp}$ and $\xi_{\parallel}$ measurements
for the spatially flat \LCDM\ fiducial cosmology in figure
\ref{fig:LCDMWedges} and for the timescape fiducial cosmology in
figure \ref{fig:tsWedges}. The most extreme $\chi^2$ value is for the
timescape CMASS fit with $\chi^2 = 49$, with 2\% probability
of $\chi^2>49$ for 30 degrees of freedom.

The significance and precision of the acoustic peak in the LOWZ sample
is significantly increased by imposing a prior in
$\sigma_{\text{BAO}}/\alpha$, which is illustrated for the
\LCDM\ case in table \ref{table:CMASSLOWZfit}. Using narrow
Gaussian priors with mean and width determined by the mock analysis of
section \ref{LCDMmocktest}, the significance of the peak goes up to
2$\sigma$ and the errors in $r_{\text{BAO}}/ \alpha$ decrease by $\sim
30$\%. The measurements of $\alpha^2 A$, $r_{\text{BAO}}/
\alpha$, and $\sigma_{\text{BAO}}/ \alpha$ for the wedge analysis are
in good agreement with those of the isotropic analysis in table
\ref{table:CMASSLOWZfit_isotropic} for both timescape and
\LCDM.
We note that the errors on $\alpha^2 A$, $r_{\text{BAO}}/
\alpha$, and $\sigma_{\text{BAO}}/ \alpha$ all decrease when going from the isotropic analysis to the anisotropic analysis for CMASS, whereas they all increase for LOWZ. This might be because of the strong correlation between $\sigma_{\text{BAO}}/ \alpha$ and the remaining parameters of the analysis: a posterior which widens in $\sigma/\alpha$ is likely to widen in the other parameters as well.

The results of our fiducial \LCDM\ analysis displayed in table
\ref{table:CMASSLOWZfit} are in good agreement with previous
measurements reported by \cite{wedgefit} and \cite{Standardresults}.
For example, table 8 of \cite{Standardresults} reports $\epsilon =
-0.016 \pm 0.020$ for a DR12 CMASS pre-reconstruction wedge analysis
which, when transformed through AP-scaling
eq.~(\ref{eq:alphaepsilontrans}) from the fiducial model $\OMn =
0.29$ of \cite{Standardresults} to the fiducial model $\OMn = 0.3$
of this paper, produces $\epsilon = -0.015 \pm 0.020$. This is in
agreement well within 1$\sigma$ of our results listed in table
\ref{table:CMASSLOWZfit}.
The analogous result for the LOWZ sample in table 8 of
\cite{Standardresults} is $0.026 \pm 0.041$, which AP-scaled gives
$\epsilon = 0.025 \pm 0.041$, which is in agreement with our
\LCDM\ results for LOWZ in table \ref{table:CMASSLOWZfit} at the
1$\sigma$ level.

\begin{table}[h]\small
\centering
\begin{tabular}{| lllllr |}
\hline
\textbf{Wedge fit} $\boldsymbol{\xi_{\perp}}$, $\boldsymbol{\xi_{\parallel}}$ & $\alpha^2 A \cdot 10^4$ & $r_{\text{BAO}}/ \alpha$ & $\sigma_{\text{BAO}}/ \alpha$ & $\epsilon$ & $\chi^2 / N_{\text{dof}}$ \\ [0.5ex]
\LCDM\ CMASS & 0.0029 $\pm$ 0.0006 & 102.6 $\pm$ 1.5 & 9.0 $\pm$ 1.6 & -0.021 $\pm$ 0.017 & $48/30$ \\
\LCDM\ LOWZ & 0.0024 $\pm$ 0.0017 & 98.5 $\pm$ 7.2 & 12.9 $\pm$ 6.0 & -0.022 $\pm$ 0.084 & $40/30$ \\
Timescape CMASS & 0.0031 $\pm$ 0.0008 & 95.4 $\pm$ 1.6 & 8.2 $\pm$ 1.9 & 0.021 $\pm$ 0.017 & $49/30$ \\
Timescape LOWZ & 0.0023 $\pm$ 0.0018 & 92.1 $\pm$ 7.6 & 10.8 $\pm$ 7.2 & 0.013 $\pm$ 0.110 & $38/30$ \\
\hline
\LCDM\ CMASS $\mathcal{N}_{\sigma_{\text{BAO}}/ \alpha}$ & 0.0035 $\pm$ 0.0006 & 100.9 $\pm$ 1.8 & 12.3 $\pm$ 0.2 & -0.022 $\pm$ 0.023 & $48/30$ \\
\LCDM\ LOWZ $\mathcal{N}_{\sigma_{\text{BAO}}/ \alpha}$ & 0.0027 $\pm$ 0.0012 & 100.3 $\pm$ 4.6 & 12.3 $\pm$ 0.3 & -0.008 $\pm$ 0.060 & $40/30$ \\
\hline
\end{tabular}
\caption{Results of the combined fit to the transverse and radial
wedge for CMASS and LOWZ. The parameter estimates shown are the
Bayesian median with $1\sigma$ equal tail credible
intervals. Conservative priors (meaning priors that span the
significant volume of the likelihood) are used for all parameters in
all fits, except for the \LCDM\ fits labelled
$\mathcal{N}_{\sigma_{\text{BAO}}/ \alpha}$, where a narrow Gaussian
prior is used with mean and width as determined in the mock analysis
of section \ref{LCDMmocktest}. The minimum $\chi^2$ value divided
by number of degrees of freedom $N_{\text{dof}}$ is also
quoted. $r_{\text{BAO}}/ \alpha$ and $\sigma_{\text{BAO}}/ \alpha$
are in units of $\hm$. $A$ is in units of $(\!\hm)^2$.}
\label{table:CMASSLOWZfit}
\end{table}

\begin{figure}[!htb]
\centering
\includegraphics[scale=0.55]{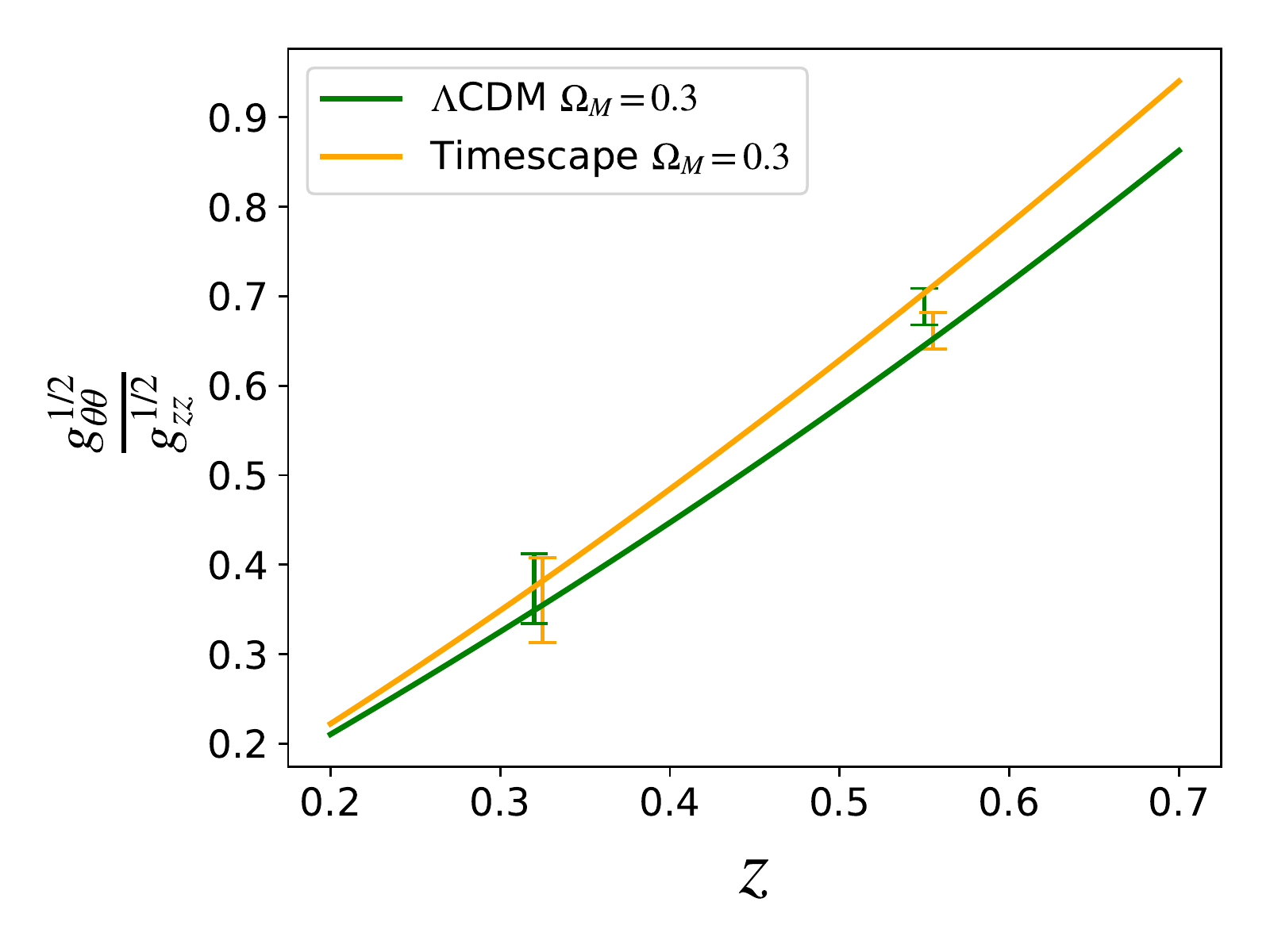}
\caption{Effective measurement of the metric combination $g_{\theta
\theta}^{1/2} / g_{z z}^{1/2}$ for LOWZ and CMASS within the timescape model and
\LCDM\ respectively, with the fiducial \LCDM\ and
timescape $\OMn = 0.3$ predictions
superimposed. The timescape measurements are artificially shifted slightly in redshift relative to the mean LOWZ and CMASS redshifts $z=0.32$ and $z=0.55$, in order to see the measurements and their comparison more clearly.}
\label{fig:EffectivedAH}
\end{figure}

The anisotropic distortion parameter $\epsilon$ describes how the
fiducial model is distorted in a relative angular and radial sense
compared to the \say{true} underlying cosmology, to lowest order.
Since $\epsilon$ is consistent with zero at the $< 2 \sigma$ level for
both timescape and \LCDM\ in the above data analysis, both
models are in agreement with no anisotropic distortion. We can
formulate the $\epsilon = 0$ consistency test in terms of the
effective metric combination $g_{\theta \theta}^{1/2} / g_{z z}^{1/2}$
(equal to $d_A H / c$ in \LCDM, where $d_A$ is the angular
diameter distance, and $H$ is the Hubble parameter), which from the
AP-scaling of our results can be formulated as
\begin{align} \label{eq:effDAH}
& \frac{g_{\theta \theta}^{1/2} }{ g_{z z}^{1/2}} \approx \frac{ \alpha_{\perp} }{\alpha_{\parallel}} \, \frac{ g_{\fid, \theta \theta}^{1/2}}{ g_{\fid, z z}^{1/2} } = (1 + \epsilon)^{-3} \,\frac{ g_{\fid, \theta \theta}^{1/2}}{ g_{\fid, z z}^{1/2} }
\end{align}
where $g_{\fid}$ corresponds to the fiducial adapted metric of either
\LCDM\ or timescape, and where $\epsilon$ is the estimate quoted
in table \ref{table:CMASSLOWZfit} for the respective fiducial
cosmologies. The results of the effective measurement of the metric
combinations (\ref{eq:effDAH}) for CMASS and LOWZ are shown in figure
\ref{fig:EffectivedAH}.
We see that both effective measurements are consistent with the respective fiducial lines, as expected since the estimated $\epsilon$-parameter is consistent with zero within both models. The precision in the measurements of the metric combination is comparable to the difference between the fiducial metric combination of the two cosmologies for the CMASS survey, potentially making this metric combination a useful discriminator between the \LCDM\ and timescape model for future surveys.
We also note that the systematics in the measurement arising from the choice of fiducial cosmology is of order the distance between the cosmologies, indicating that a careful analysis of the regime of application of the AP-scaling is needed.

\section{Discussion}
\label{discussion}

In this paper we have developed methods for examining BAO features in
the 2-point correlation function for cosmological models with
non-trivial curvature: models that are not necessarily spatially
flat, close to spatially
flat, nor with constant spatial curvature. The methods outlined in
section \ref{theory} and \ref{empiricalBAO} are applicable for a broad
class of large-scale cosmological models. (See section \ref{models} for
precise statements about the regime of applicability.)
Our assumptions on the model cosmology can be summarised as follows:
\begin{itemize}
\item We assume global hyperbolicity of the average space-time, and that the galaxies can to a good approximation be described as particles in a non-caustic, vorticity-free fluid description. These assumptions are made in order to formulate the reduced 2-point correlation function descriptive statistic in terms of the Lagrangian distance definition given in section \ref{comoving}, generalising the comoving distance definition of FLRW cosmology.
\item We further impose the assumptions outlined in section \ref{models}, such that the Lagrangian distance definition can be approximated as in eq.~(\ref{eq:lagrangiandistapprox}). The approximation (\ref{eq:lagrangiandistapprox}) is needed to: (i) define the \say{radial fraction} of the separation $\mu_T$ in (\ref{eq:mu}); and (ii) make sense of the generalised AP-scalings $\alpha_{\parallel}, \alpha_{\perp}$ of the \say{radial} and \say{transverse} component of the metric introduced in section \ref{APscaling}. \item Finally, we assume that the empirical fitting function described in section \ref{empiricalBAO} is appropriate for extracting the isotropic BAO characteristic scale and the anisotropic distortion between the radial and transverse scale. (This assumption is tested and confirmed for a fiducial \LCDM\ model using mocks catalogues, but is left as an ansatz for other cosmologies.)\end{itemize}
Our methods allow us to explicitly formulate the 2-point correlation
function in the context of a broad class of cosmologies and hence
analyse the clustering statistics for those cosmologies in detail,
instead of relying on results extrapolated from \LCDM. The
only \LCDM\ estimate used in this paper enters when estimating
errors in the observed 2-point correlation function, where we use
mocks generated from a fiducial \LCDM\ model to give a rough
estimate for the variance over ensembles of our sky.

When testing our methods on \LCDM\ mocks we recover the
isotropic peak position to within one per cent of the fiducial value.
This $\lsim 1\%$ discrepancy is due to a calibration issue between the
characteristic scale extracted in the fitting procedure and the
underlying BAO scale discussed in section \ref{empiricalBAOmain}.
It should be
noted that, while this level of systematic error is somewhat higher
than obtained by \LCDM\ fitting procedures, it can be considered
low in (semi-)model independent analysis. Removing cosmology
dependence in data reduction necessarily comes at the price of
increasing uncertainties. The systematics related to the BAO scale
extraction in the context of other models must be assessed for each
cosmology of interest. The anisotropic scaling parameter $\epsilon$
is recovered to high precision; the systematics in our mock analysis
on the determination of $\epsilon$ are much smaller than the usual
statistical errors in \LCDM\ fitting procedures. The estimation
of the $\epsilon$ parameter is robust to the exact form of the fitting
function assumed, and is not associated with the calibration issues of
the statistical BAO scale.

A shortcoming of this analysis is that a fiducial cosmology of choice
is still needed in order to reduce data into a 2-point correlation
function. Model-independent analysis has been proposed in,
e.g., \cite{sanchezradial} and \cite{sancheztransverse}. While such
procedures are certainly relevant for next-generation surveys, the
signal strength is greatly reduced due to the split of the fiducial spatial scale to a range of
angular and redshift separations.

Another shortcoming of this paper is the approximations of section \ref{models}, implying that only effective cosmic metric theories which are averaged on scales of the order of the BAO scale can be tested in our framework. While testing more complicated models with a hierarchy of curvature scales, describing different scales of structure in our universe, would be of interest, this is beyond the scope of this paper.

We apply our fitting methods to the BOSS DR12 CMASS and LOWZ galaxy
surveys using two fiducial cosmologies: a spatially flat \LCDM\
model and the timescape model, which at the present epoch has a marginal
apparent acceleration with a recent expansion history closer to the empty
FLRW universe.
We recover the pre-reconstruction results for the BAO peak position and the anisotropic distortion parameter $\epsilon$ based on \LCDM\ template-fitting obtained in \cite{Standardresults}.

It should be noted, that
since the parameter estimates of our empirical procedure and of the standard \LCDM\ template-fitting are based on the same
datasets, any difference in the results can be attributed to
systematic differences in the parameter extraction procedures. For
procedures with small systematic differences as compared to the
statistical errors, we would thus expect differences in parameter
estimates much smaller than $1\sigma$. The systematic differences
between the present procedure and the standard \LCDM\ procedure
are smaller, but of order, the statistical errors. The main
difference between the estimated BAO peak position of the present
procedure and of the standard \LCDM\ power spectrum fitting
procedure can be ascribed to the systematics related to the
calibration of the BAO scale in the empirical fit (see discussion in
section \ref{empiricalBAOmain}). Other examples of systematics
between the procedures that can lead to differences in the parameter
estimates are: choice of statistical framework, choice of priors, and
galaxy weights. For example, the $\sigma_{\text{BAO}}/ \alpha$ prior
in the present analysis has a $\sim 1\%$ effect on the peak position,
which is comparable to the differences in our inferred scale as
compared to the results of \cite{Standardresults}.

Based on our empirical model for the shape of the
2-point correlation function, we find that the BAO feature of the
models is detected at a similar level of significance in the two
cosmologies, and that the distortion between the radial and transverse
directions, quantified by the $\epsilon$ parameter, is consistent with
zero for both fiducial models within 2$\sigma$.
Thus, both models are consistent with no anisotropic distortion with respect to the \say{true} cosmological model, and thus provide self-consistent fits to the BAO-data.
This finding is
interesting in light of the significant difference between the
timescape model and the \LCDM\ model distance measures (see
figure \ref{fig:LCDMvsTimescape}).

Our analysis suggests that a wide class of cosmological models can
yield a statistically isotropic BAO feature with $\epsilon = 0$,
consistent with the expectation of statistical homogeneity and
isotropy of our universe. In future work, we will combine these BAO
measurements with estimations of the standard ruler scale in timescape
cosmology to perform a full model comparison.

\acknowledgments
We wish to thank Thomas Buchert
for hospitality at the ENS, Lyon, France. This work was supported by
Catalyst grant CSG-UOC1603 administered by the Royal Society of New Zealand.
AH is grateful for the support given by the funds: `Knud H{\o}jgaards Fond', `Torben og Alice Frimodts Fond', and `Max N{\o}rgaard og Hustru Magda N{\o}rgaards Fond'.

\begin{appendices}

\section{Taylor expansion of geodesic distances}
\label{expgeodesic}

In the present analysis we make use of a Taylor expansion of the
spatial geodesic distance between two points on a spatial
hypersurface. Such an expansion is convenient when the spatial
geodesic equation (defined on spatial hypersurfaces of interest) of
the model under investigation has no analytic solution, and applicable
when the curvature scale of the model is much larger than the particle
separation of interest.

We consider a metric on the form eq.~(\ref{eq:metriclagrangian})
\begin{align} \label{eq:metriclagrangian3}
&ds^2 = - \alpha^2 c^2 dt^2 + g_{ij} dx^{i} dx^{j} ,
\end{align}
where $t$ defines a spatial foliation of interest. (In the context of
this analysis, $t =$ constant slices are taken to coincide with the
matter frame, which can be done in the absence of vorticity.)

Consider a geodesic spatial line between two points $P_{1}$ and
$P_{2}$ on the hypersurface $t = T$, such that the line is required to
lie in the $t=T$ plane everywhere. The geodesic distance between the
points is given by
\begin{align} \label{eq:exactgeodesicD}
d_{T}(P_{1}, P_{2}) &\equiv \int_{l_{1}}^{l_{2}} dl \sqrt{g_{ij} \frac{dx^{i}}{dl} \frac{dx^{j}}{dl} } = l_2 - l_1
\end{align}
where $l$ is the affine parameter along a spatial geodesic connecting
$P_{1}$ and $P_{2}$, $\frac{dx^{i}}{dl}$ is the tangent to the
geodesic with $g_{ij}\frac{dx^{i}}{dl} \frac{dx^{j}}{dl} = 1$, $l_1 =
l(P_1)$ and $l_2 = l(P_2)$ is the affine parameter evaluated at the
endpoints. The function $d_{T}$ coincides with the Lagrangian
distance $D_T$ defined in section (\ref{comoving}), when the points
$P_{1}$ and $P_{2}$ represent the intersection of two particle
worldlines with the surface $t = T$.

We expand the coordinate functions on the line in the affine parameter
$l$
\begin{align} \label{eq:xiexpand}
x^{i}_2 &= x^{i}_1 + \left. \frac{dx^{i}}{dl} \right|_{l = l_1} (l_2 - l_1) + f^{i} ,\\
f^{i} &= \sum^{\infty}_{n = 2} f^{i}_{n} , \qquad f^{i}_{n} = \frac{1}{n!} \left. \frac{d^n x^{i}}{dl^n} \right|_{l = l_1} (l_2 - l_1)^n \nonumber
\end{align}
where $x^i_1 = x^i(P_1)$ and $x^i_2 = x^i(P_2)$ are the coordinate
labels of the end points. The higher-order terms $f^{i}_{n}$ can be
expressed in terms of $\Delta x^i = x^i_2- x^i_1$ up to a given
order. Here we shall keep terms up to $\orderof \left( f_3^i
\right)$, where we assume $\Delta x^j f_2^k\goesas\orderof \left( f_3^i
\right)$ etc. The second order term yields
\begin{align} \label{eq:secondorder}
f^{i}_{2} &= \frac{1}{2}\, \frac{d^2 x^{i}}{dl^2}\, (l_2 - l_1)^2 = - \frac{1}{2} \Gamma^{i}_{\; jk}\, \frac{d x^{j}}{dl} \,\frac{d x^{k}}{dl} (l_2 - l_1)^2\\
&= - \frac{1}{2} \Gamma^{i}_{\; jk} (\Delta x^j - f^j) (\Delta x^k - f^k) = - \frac{1}{2} \Gamma^{i}_{\; jk} \,\Delta x^j\, \Delta x^k + \Gamma^{i}_{\; jk}\, \Delta x^j\, f_2^k + \orderof \left( f_4^i \right) \nonumber \\
&= - \frac{1}{2} \Gamma^{i}_{\; jk} \,\Delta x^j \,\Delta x^k - \frac{1}{2} \Gamma^{i}_{\; jk} \Gamma^{k}_{\; st}\, \Delta x^j\, \Delta x^s\, \Delta x^t + \orderof \left( f_4^i \right) , \nonumber
\end{align}
where the first line follows from the affine geodesic equation, the
second line follows from applying (\ref{eq:xiexpand}) and keeping
terms up to $\orderof \left( f_3^i \right)$. The third line comes from
recursively plugging (\ref{eq:secondorder}) into itself and again
keeping terms up to $\orderof \left( f_3^i \right)$. The evaluation
at $l=l_1$ is implicit.

With a similar derivation, the third-order term of the expansion
eq.~(\ref{eq:xiexpand}) yields
\begin{align} \label{eq:thirdorder}
f^{i}_{3} &= \left( \frac{1}{3} \Gamma^{i}_{\; jk} \Gamma^{k}_{\; st} - \frac{1}{6} \partial_{s} \Gamma^{i}_{\; jt} \right) \Delta x^j \Delta x^s \Delta x^t + \orderof \left( f_4^i \right) .
\end{align}
We can now expand the geodesic distance (\ref{eq:exactgeodesicD})
in an adapted coordinate system $x^{i}$ of choice. Keeping terms up to
$\orderof \left( f_4^i \right)$ we have
\begin{align} \label{eq:exactgeodesicDthird}
d_{T}(P_{1}, P_{2})& = l_2 - l_1 = \sqrt{g_{ij} \frac{dx^{i}}{dl} \frac{dx^{j}}{dl} (l_2 - l_1)^2} \\
&= \sqrt{g_{ij} (\Delta x^i - f^i) (\Delta x^j - f^j) } \nonumber \\
&= \sqrt{g_{ij} \Delta x^i \Delta x^j - 2g_{ij} \Delta x^i (f_2^j + f_3^j ) + g_{ij} f_2^i f_2^j + \orderof \left( f_5^i \right) } \nonumber \\
&= \sqrt{{}^{(0)}g + {}^{(1)}g + {}^{(2)}g + \orderof \left( f_5^i \right) } \nonumber ,
\end{align}
where all terms are evaluated at $l = l_1$. The first line follows
from a convenient multiplication by $1 = \sqrt{g_{ij}
\frac{dx^{i}}{dl} \frac{dx^{j}}{dl} }$. The following lines come
from applying the expansion (\ref{eq:xiexpand}) and truncating
the resulting terms at $\orderof \left( f_4^i \right)$. In the last
line we have used the definitions
\begin{align} \label{eq:expdef}
{}^{(0)}g & \equiv g_{ij}\, \Delta x^i \Delta x^j , \qquad {}^{(1)}g \equiv g_{ij} \, \Gamma^{j}_{\; st} \,\Delta x^i \Delta x^s \Delta x^t\\
{}^{(2)}g & \equiv \left[\frac{1}{3} g_{ij} \left(\partial_{s} \Gamma^{j}_{\; tb} + \Gamma^{j}_{\; ab} \Gamma^{a}_{\; st}\right) + \frac{1}{4} g_{kj} \,\Gamma^{k}_{\; st} \Gamma^{j}_{\; ib} \right] \Delta x^i \Delta x^s \Delta x^t \Delta x^b \nonumber
\end{align}
The extent to which the coordinate expansion
(\ref{eq:exactgeodesicDthird}) is accurate at a given truncation
of the series depends on the space-time metric and the chosen events
$P_1$ and $P_2$, but also on the adapted coordinates used in the
expansion. The convergence of the expansion
(\ref{eq:exactgeodesicDthird}) must be examined for the
particular problem at hand.

\subsection{Spherically symmetric metrics}
\label{sphericalsymmDist}

As a special case relevant for this paper we consider the
spherically-symmetric metric (\ref{eq:metriclagrangian2}) of
section \ref{models}. The adapted metric on the spatial
hypersurfaces given by
eq.~(\ref{eq:metriclagrangiansphericalsymmetry})
\begin{align} \label{eq:metriclagrangiansphericalsymmetry2}
&ds_T^2 = g_{rr}(t=T,r) dr^2 + g_{\theta \theta}(t=T,r) \left(d\theta^2 + \cos^2(\theta) d \phi^2 \right) .
\end{align}
In this case we have for the lowest-order term of
eq.~(\ref{eq:exactgeodesicDthird})
\begin{align} \label{eq:geodesicfirstordersphericals}
&{}^{(0)}g = { g_{rr} ( \Delta r )^2 + g_{\theta \theta} \left( ( \Delta \theta )^2 + \cos^2(\theta) ( \Delta \phi )^2 \right) } .
\end{align}
The first order correction yields
\begin{align} \label{eq:secondorderspherical}
{}^{(1)}g& = \frac{1}{2} {\frac{\partial}{\partial x^{k}}} (g_{sm}) \Delta x^m \Delta x^k \Delta x^s \\
&= \frac{1}{2} \Delta g_{rr} (\Delta r)^2 + \frac{1}{2} \Delta g_{\theta \theta} \left( (\Delta \theta)^2 + \cos^2(\theta) (\Delta \phi)^2 \right) + \frac{1}{2} g_{\theta \theta} [\Delta \cos^2(\theta)] ( \Delta \phi)^2 , \nonumber
\end{align}
where we have defined
\begin{align} \label{eq:secondordersphericaldef}
& \Delta g_{rr} \equiv \frac{d g_{rr} }{dr} \Delta r , \qquad \Delta g_{\theta \theta} \equiv \frac{d g_{\theta \theta} }{dr} \Delta r , \qquad \Delta \cos^2(\theta) \equiv \frac{d ( \cos^2(\theta)) }{d\theta} \Delta \theta .
\end{align}
Combining the lowest order term and the first order correction we thus have for
eq.~(\ref{eq:exactgeodesicDthird}) up to $\orderof \left( {}^{(2)}g
\right)$
\begin{align} \label{eq:geodesicDexpandsphericals}
&d_{T}(P_{1}, P_{2}) = \sqrt{\overline{g}_{rr} ( \Delta r )^2 + \overline{g}_{\theta \theta} \left( ( \Delta \theta )^2 + \overline{\cos^2(\theta)} ( \Delta \phi )^2 \right) + \orderof \left( {}^{(2)}g \right) } ,
\end{align}
where we have used the definition
\begin{align} \label{eq:bargdef}
\overline{g}_{rr} \equiv g_{rr} + \half \Delta g_{rr} , \qquad \overline{g}_{\theta \theta} \equiv g_{\theta \theta} + \half \Delta g_{\theta \theta}, \qquad
\overline{\cos^2(\theta)} \equiv \cos^2(\theta) + \half \Delta \cos^2(\theta),
\end{align}
and a term $-\half\Delta g_{\theta \theta} [\Delta \cos^2(\theta)]
(\Delta \phi)^2$ is subsumed in the $\orderof \left( {}^{(2)}g \right)$
terms in (\ref{eq:geodesicDexpandsphericals}).
Hence the first order correction represents a shift of evaluation at
$x^i_1$ to the mean coordinate point $\bar{x}^i = x^i_1 + \frac{1}{2}
\Delta x^{i}$.

We can examine the accuracy of the approximation
(\ref{eq:geodesicDexpandsphericals}), truncated at first order,
by evaluating the second order correction terms of ${}^{(2)}g$, which
will contain terms of order $\sim \frac{\left({}^{(1)}g \right)^2 }{
{}^{(0)}g }$ and terms involving the second derivatives of the
metric.
All of these terms should be evaluated in the model of interest and in
the desired coordinate system, in order to examine the approximation
(\ref{eq:geodesicDexpandsphericals}). For observational
coordinates $(z,\theta,\phi)$ in both the FLRW and timescape model
with realistic model parameters, we find $\frac{{}^{(2)}g }{{}^{(0)}g
} \lsim 10^{-3}$, for separation distances $\Delta z$, $\Delta
\theta$, $\Delta \phi$ around the BAO scale.

\section{The 2-point correlation function}
\label{2PCF}
The spatial 2-point correlation function $\xi$ describes the excess
probability of finding two galaxies at two given points on a
spatial surface, relative to an uncorrelated sample. The typical
formulation of the 2-point correlation function in standard cosmology
is tightly linked to the assumption of symmetries of the \say{background} FLRW space-time, and the ergodic assumptions on the density perturbation field on top of the background, which leads to the modelling of the galaxy distribution as a stationary and ergodic point process.

Thus if we revisit the \say{background} cosmology, or do cosmology without imposing a background, we should also revisit the theory underlying the 2-point correlation function.
Here we seek to provide a more general introduction
to the 2-point correlation function, valid for models with no exact
symmetries in the pointwise ensemble average of the galaxy counts.

Consider a spatial domain of a hypersurface $\domain$. We view the
position of the galaxies within this domain as random variables, and
fix the total number of galaxies $N$ within the domain $\domain$. We
use adapted coordinates $x^{i}$ on the hypersurface,
and denote the random position of the $a$'th particle $x_{a}^{i}$.
The scaled probability (ensemble average number count) of finding two
galaxies located in the infinitesimal volume elements $dV_X$
and $dV_Y$ centred at the points $x^{i} = X^i$ and $x^{i} = Y^i$ can
be written as
\begin{align} \label{eq:excessprob}
&f(X,Y) dV_X \,dV_Y \equiv \braket{N(dV_{X}) N(dV_{Y})} ,
\end{align}
where $f(X,Y)$ is the number count density and where
\begin{align} \label{eq:NEnsemble}
N(dV_{X}) \equiv \sum_{a}^{N} \mathbb{1}_{dV_{X}}(x^i_a) .
\end{align}
is the number count in the volume element $dV_{X}$ in a given realisation,
where
\bea
\mathbb{1}_{dV_{X}}(x^i_a)=\begin{cases}
1,&x^i_a\in dV_X,\\ 0,&x^i_a\notin dV_X,
\end{cases}
\eea
is the indicator function. (If the volume $dV_{X}$ is made small enough,
this is zero or one in practice.)
The spatial volume elements $dV$ are given by\footnote{We could
alternatively absorb any non-zero function into the number count density $f(X,Y)$ and make the redefinition $f(X,Y) \rightarrow \det(g_{ij}) f(X,Y) , \quad dV \rightarrow dV/\sqrt{ \det(g_{ij})} = dx^1 \wedge dx^2 \wedge dx^3$ if we prefer to work in terms of coordinate volumes instead of physical volumes.}
the adapted metric
(\ref{eq:metriclagrangian})
\begin{align} \label{eq:volumeelement}
dV = \sqrt{ \det(g_{ij})} dx^1 \wedge dx^2 \wedge dx^3 .
\end{align}
The integral of (\ref{eq:excessprob}) over two arbitrary domains $\domain_X \in \domain$ and $\domain_Y \in \domain$ is
\begin{align} \label{eq:excessprob2}
&\int_{X \in \domain_X} \int_{Y \in \domain_Y} f(X,Y) dV_X \,dV_Y = \braket{N(\domain_{X}) N(\domain_{Y})}
\end{align}
following from the property $\mathbb{1}_{A \cup B}(y) = \mathbb{1}_{A}(y) + \mathbb{1}_{B}(y)$ of the indicator function, where $A$ and $B$ are disjoint sets.

The scaled probability of finding a galaxy in the small volume $dV_X$ (ensemble average number count) can be expressed as an integral over (\ref{eq:excessprob2})
\begin{align} \label{eq:probsinglegalaxy}
f(X)dV_X &\equiv \braket{N(dV_{X})} = \frac{1}{N} \int_{Y \in \domain} f(X,Y) dV_X \,dV_Y .
\end{align}
We shall be interested in writing the probability
(\ref{eq:excessprob}) in terms of the excess probability of the
uncorrelated process
\begin{align} \label{eq:excessprobUncor}
f_{\mathrm{Poisson}}(X,Y) = f(X) f(Y) , \qquad X \neq Y .
\end{align}
Assuming that $f(X)\neq 0$ over the domain $\domain$ we can write
\begin{align} \label{eq:excessprobxi}
f(X,Y)\, dV_X \,dV_Y &= f(X)f(Y) \left(1 + \xi(X,Y) \right) dV_X \,dV_Y
\end{align}
where we have defined
\begin{align} \label{eq:xidef}
\xi(X,Y)= \frac{f(X,Y)}{f(X)f(Y) } - 1 .
\end{align}
This correlation function,
$\xi$, is zero for $X \neq Y$ for a Poisson point process per construction, and
measures the departure from an uncorrelated distribution of galaxies.

The correlation function
(\ref{eq:xidef}) is a function of all $6$ variables
$(X^{i},Y^{i})$ in a general inhomogeneous universe. In practice, in
BAO analysis, we are interested in integrating out some of these
degrees of freedom, to isolate a characteristic statistical scale. We
can make the substitution $(X^{i},Y^{i}) \rightarrow (X^{i}, \hat{n}_X^{i}, D)$ in (\ref{eq:excessprob}),
where $\hat{n}_X^{i}$ is a unit vector at $X^{i}$ defining a geodesic starting at $X^{i}$ and intersecting $Y^{i}$ and $D$ is the geodesic distance from $X^{i}$ to $Y^{i}$
\begin{align} \label{eq:excessprobtrans}
&f(X,Y)\, dV_X \,dV_Y = f(X, \hat{n}_X, D)\, dV_X \,d\hat{n}_X \, dD = f(X, \hat{n}_X, D)\, J\, dV_X \,dV_Y
\end{align}
with
\begin{align} \label{eq:Jacobian}
&J\equiv \det\left(\frac{{\partial} (X,\hat{n}_X,D)}{{\partial} (X,Y)}\right)
\end{align}
being the Jacobian of the transformation. It follows that (\ref{eq:xidef}) reads
\begin{align} \label{eq:xideftrans}
\xi(X,\hat{n}_X, D)= \frac{f(X,\hat{n}_X, D)}{f_{\mathrm{Poisson}}(X,\hat{n}_X, D) } - 1 ,
\end{align}
where the Jacobian $J$ of the transformation $(X^{i},Y^{i}) \rightarrow (X^{i}, \hat{n}_X^{i}, D)$ cancels in (\ref{eq:xidef}), since $f$ and $f_{\mathrm{Poisson}}$ have identical transformations.
We denote the random process underlying the ensemble homogeneous and isotropic if $f_{HI}(X + \alpha) = f_{HI}(X)$, $f_{HI}(X+ \alpha, R \hat{n}_X , D) = f_{HI}(X,\hat{n}_X, D)$ are satisfied,
where $\alpha$ is an arbitrary translation, $R$ is an arbitrary rotation of the unit vector $\hat{n}_X$, and where the subscript $HI$ stands for homogenous and isotropic.
In this case (\ref{eq:xideftrans}) becomes the so-called reduced 2-point correlation function
\begin{align} \label{eq:xideftranshom}
\xi_{HI}(D)= \frac{f_{HI}(D) }{ f_{HI \; \mathrm{Poisson}}(D) } - 1 .
\end{align}
In the general case where the random process underlying the ensemble is not associated with any exact symmetries, we can still create a reduced version of the correlation function (\ref{eq:xideftrans}) by marginalising over the direction and position degrees of freedom $\hat{n}_X , \, X$.
This can be done as follows. We define the marginalised number count density over a subdomain $\domain_S \in \domain$ as
\begin{align} \label{eq:marginalisedf}
f(D, \domain_S) &\equiv \int_{X \in \domain_S} dV_{X} \int dn_{X} f(X,\hat{n}_X, D) .
\end{align}
The marginalised ensemble number count in a small range of affine distance $dD$ is given by
\begin{align} \label{eq:marginalisedfdD}
f(D, \domain_S) dD = \Braket{\sum_{a,b}^{N} \mathbb{1}_{[D, D + dD]}(D(x^i_a,x^i_b)) \mathbb{1}_{\domain_S}(x^i_a) } ,
\end{align}
where we have used the fact that we can rewrite the number count in terms of the new coordinates $X,\hat{n}_X, D$,
\begin{align} \label{eq:marginalisedfNN}
N(dV_{X}) N(dV_{Y}) &= \sum_{a,b}^{N} \mathbb{1}_{dV_{X}}(x^i_a) \mathbb{1}_{dV_{Y}}(x^i_b) \nonumber \\
&= \sum_{a,b}^{N} \mathbb{1}_{dV_{X}}(x^i_a) \mathbb{1}_{[\hat{n}_X, \hat{n}_X + d\hat{n}_X]}(\hat{n}_X(x^i_a,x^i_b)) \, \mathbb{1}_{[D, D + dD]}(D(x^i_a,x^i_b)) ,
\end{align}
and that by (\ref{eq:excessprobtrans}) $f(X, \hat{n}_X, D)\, dV_X \,d\hat{n}_X \, dD = \braket{N(dV_{X}) N(dV_{Y})}$.
We can write (\ref{eq:marginalisedf}) in terms of $f_{\mathrm{Poisson}}(D, \domain_S)$ defined through the integral over $f_{\mathrm{Poisson}}(X,\hat{n}_X, D)$ analogous to (\ref{eq:marginalisedf}).
\begin{align} \label{eq:marginalisedfxi}
f(D, \domain_S)\, dD = f_{\mathrm{Poisson}}(D, \domain_S) \left(1 + \xi(D, \domain_S) \right) dD ,
\end{align}
with
\begin{align} \label{eq:marginalisedxidef}
\xi(D, \domain_S) = \frac{f(D, \domain_S)}{f_{\mathrm{Poisson}}(D, \domain_S)} - 1 ,
\end{align}
which we denote the \say{marginalised} two point correlation function.

Note that eq.~(\ref{eq:marginalisedxidef}) has the form of the
conventional reduced 2-point correlation function of a homogenous and isotropic cosmology.
However, the interpretation is different here, as the reduction does not
follow from symmetry assumptions on the probability distribution of
the density field, but rather follows from marginalisation over the
position and direction degrees of freedom (and hence depends on scale through $\domain_S$). Eq.~(\ref{eq:marginalisedxidef})
coincides with the conventional 2-point correlation function (\ref{eq:xideftranshom}) when the galaxy distribution
is assumed to be represented by a homogeneous and isotropic point
process. We can thus view eq.~(\ref{eq:marginalisedxidef}) as a generalisation
of the 2-point correlation function to inhomogeneous space-times.

For models of the form outlined in section \ref{models} we can decompose $\hat{n}_X$ into $\mu$, $\sgn(\delta z)$, and the normalised angular separation vector
$\delta \hat{\Theta} = \frac{1}{ \left| \delta \Theta \right| }(\delta
\theta, \cos(\theta) \delta \phi)$.
In this case we can write
\begin{align} \label{eq:excessprobtransmu}
&f(X,Y) dV_X \,dV_Y = f(X, \mu, \sgn(\delta z) , \delta \hat{\Theta}, D) dV_X \,d\mu \,d\delta \hat{\Theta} \, dD
\end{align}
and we can construct a marginalised number count density in $D$ analogous to (\ref{eq:marginalisedf}) by marginalising over the remaining variables.
We shall sometimes be interested in keeping $\mu$ as a variable, and construct the following marginalised number count density
\begin{align} \label{eq:marginalisedfmu}
f(D,\mu, \domain_S) &\equiv \sum_{\sgn(\delta z)=\pm 1} \int_{X \in \domain_S} dV_{X} \int d\delta \hat{\Theta}\; f(X, \mu, \sgn(\delta z) , \delta \hat{\Theta}, D) ,
\end{align}
for which we can define the marginalised $\mu$-dependent 2-point correlation function
\begin{align} \label{eq:marginalisedxidefmu}
\xi(D, \mu, \domain_S) = \frac{f(D,\mu, \domain_S)}{f_{\mathrm{Poisson}}(D,\mu, \domain_S)} - 1 .
\end{align}
Integrating out the $\mu$-dependence in (\ref{eq:marginalisedfmu}) we arrive at the marginalised isotropic number count density $f(D, \domain_S)$
from which we can construct
 the isotropic marginalised 2-point correlation function of (\ref{eq:marginalisedxidef}).\footnote{For the estimate of (\ref{eq:marginalisedxidefmu}) or (\ref{eq:marginalisedxidef}) based on number counts in a subdomain $\domain_{S'}$ of a single realisation of the ensemble to be representative for the ensemble average, we must invoke the approximate convergence condition $\hat{\xi}(D,\domain_S')_{\lim V(\domain_{S'}) \rightarrow \infty} \approx \xi(D, \domain_S)$ for
some choice of scale $V(\domain_S)$, with fast enough convergence of the estimate. In practice $\domain_{S'}$ will correspond to a given survey domain.}

We define the \say{wedge} as the mean of
eq.~(\ref{eq:marginalisedxidefmu}) over a given $\mu$ range $[\mu_{1},
\mu_{2}]$.
\begin{align} \label{eq:xiDmudefwedge}
\xi_{[\mu_{1}, \mu_{2}]} (D) \equiv \frac{1}{\mu_{2}- \mu_{1}} \int_{\mu_{1}}^{\mu_{2}} d \mu\, \xi(D, \mu) ,
\end{align}
where the dependence on $\domain_S$ is implicit in (\ref{eq:xiDmudefwedge}) and in the following.
It can be viewed as the mean excess of probability of finding two
galaxies a distance $D$ apart over the given $\mu$ range. We define
the transverse and the radial wedge as respectively
\begin{align} \label{eq:xitransradial}
\xi_{\perp} (D) \equiv \xi_{ [0, 0.5]} (D) , \qquad \xi_{\parallel} (D) \equiv \xi_{[0.5, 1]} (D) .
\end{align}
When $f(D,\mu)$ is mainly depending on $D$ such that
\begin{align} \label{eq:Pmuapprox}
f(D, \mu) &= f(D)(1 + h(D,\mu)) , \qquad h(D,\mu) \ll 1\\
f_{\mathrm{Poisson}}(D, \mu) &= f_{\mathrm{Poisson}}(D)(1 + h_{\mathrm{Poisson}}(D,\mu)) , \qquad h_{\mathrm{Poisson}}(D,\mu) \ll 1, \nonumber
\end{align}
we have the useful approximation
\begin{align} \label{eq:xiapprox}
\int_{0}^{1} d\mu\, \xi(D, \mu) &= \int_{0}^{1} d\mu \frac{f(D)(1 + h(D,\mu)) }{f_{\mathrm{Poisson}}(D)(1 + h_{\mathrm{Poisson}}(D,\mu))} - 1 \\
&\approx \int_{0}^{1} d\mu \frac{f(D) }{f_{\mathrm{Poisson}}(D)} (1 + h(D,\mu) - h_{\mathrm{Poisson}}(D,\mu)) - 1 \nonumber \\
& = \frac{f(D) }{f_{\mathrm{Poisson}}(D)} - 1 = \xi(D) \nonumber
\end{align}
where we have used $\int_{0}^{1} d \mu\, h(D,\mu) = 0$ and $\int_{0}^{1}
d \mu\, h_{\mathrm{Poisson}}(D,\mu) = 0$ by construction. Note that
corrections to eq.~(\ref{eq:xiapprox}) are \emph{second order} in $h$
and $h_{\mathrm{Poisson}}$. A similar approximation can be formulated
for the wedges (\ref{eq:xiDmudefwedge})
\begin{align} \label{eq:xiapproxwedge}
\xi_{[\mu_{1}, \mu_{2}]} (D) &\approx \xi(D, \mu_1 \leq \mu \leq \mu_2) \equiv \frac{f(D, \mu_1 \leq \mu \leq \mu_2)}{f_{\mathrm{Poisson}}(D, \mu_1 \leq \mu \leq \mu_2) } - 1 .
\end{align}
\end{appendices}

\end{document}